\documentclass[aos,preprint,reqno]{imsart}

\RequirePackage{xr}
\RequirePackage[OT1]{fontenc}
\RequirePackage{amsthm,amsmath,amssymb,dsfont}
\RequirePackage[numbers]{natbib}
\RequirePackage{algorithm}
\RequirePackage{algpseudocode}
\RequirePackage{graphicx} 
\RequirePackage{capt-of}
\RequirePackage[section]{placeins}
\RequirePackage{color}
\RequirePackage{chngcntr}
\RequirePackage{enumerate}
\RequirePackage{longtable}

\startlocaldefs

\counterwithin{table}{section}

\algnewcommand\algorithmicinput{\textbf{Input:}}
\algnewcommand\Input{\item[\algorithmicinput]}

\newcounter{rrr}
\newcounter{aaa}
\newcounter{iii}

\newtheorem{lem1}{Lemma} \numberwithin{lem1}{section} 
\newtheorem{cor1}[lem1]{Corollary}
\newtheorem{theo1}[lem1]{Theorem}

\theoremstyle{definition}
\newtheorem{df1}[lem1]{Definition}
\newtheorem{example1}[lem1]{Example}

\theoremstyle{remark}
\newtheorem{remark1}[lem1]{Remark}

\newenvironment{rr}{\refstepcounter{rrr} \par \begin{itemize} \item[\textbf{R \therrr.}]}{ \end{itemize}}

\newenvironment{model}{\par\vspace*{0.1cm}\noindent\textit{The SBSSR-model }}{}

\newenvironment{exampleon}{\begin{example1}}{\end{example1}}  

\newenvironment{remark}{\begin{remark1}}{\end{remark1}}
\newenvironment{remarkmn}[1]{\begin{remark1}[#1]}{\end{remark1}}

\newenvironment{theo}{\begin{theo1}}{\end{theo1}}
\newenvironment{theomn}[1]{\begin{theo1}[#1]}{\end{theo1}}

\newenvironment{coron}{\begin{cor1}}{\end{cor1}} 
\newenvironment{dfon}{\begin{df1}}{\end{df1}}

\newcommand{\N}{\ensuremath{\mathds{N}}}  
\newcommand{\Q}{\ensuremath{\mathds{Q}}}  
\newcommand{\R}{\ensuremath{\mathds{R}}}  
\newcommand{\B}{\ensuremath{\mathds{B}}}  
\newcommand{\Pp}{\ensuremath{\textbf{P}}}  

\newcommand{\cC}{\mathcal{C}}
\newcommand{\cD}{\mathcal{D}}

\newcommand{\cH}{\mathcal{H}}
\newcommand{\cI}{\mathcal{I}}
\newcommand{\cJ}{\mathcal{J}}

\newcommand{\cM}{\mathcal{M}}
\newcommand{\cN}{\mathcal{N}}
\newcommand{\cO}{\mathcal{O}}
\newcommand{\co}{\mbox{\scriptsize $\mathcal{O}$}}

\newcommand{\cS}{\mathcal{S}}

\newcommand{\cU}{\mathcal{U}}

\newcommand{\fA}{\ensuremath{\mathfrak A}}
\newcommand{\fB}{\ensuremath{\mathfrak B}}

\newcommand{\fH}{\ensuremath{\mathfrak H}}

\newcommand{\fL}{\ensuremath{\mathfrak L}}

\newcommand{\fT}{\ensuremath{\mathfrak T}}

\newcommand{\abbr}[3]{ #1 : \; #2 \; \rightarrow \; #3 }  
		
\newcommand{\norm}[1]{\|#1\|} 
\newcommand{\abs}[1]{\ensuremath{\left\vert#1\right\vert}}
\newcommand\ZuWeis{\mathrel{\mathop:\!\!=}} 
\newcommand\WeisZu{\mathrel{=\!\!\mathop:}} 
\newcommand{\indE}{\mathds{1}} 
\newcommand{\imag}{\operatorname{Im}} 
\newcommand{\dist}{\overline{\operatorname{dist}}}
\newcommand{\argmin}{\operatorname{argmin}}
\newcommand{\argmax}{\operatorname{argmax}}

\newcommand{\proj}{\operatorname{proj}}

\newcommand{\mae}{\operatorname{MAE}}
\newcommand{\miae}{\operatorname{MIAE}}
\newcommand{\meanNcp}{\operatorname{Mean}(\hat{K})}
\newcommand{\medNcp}{\operatorname{Med}(\hat{K})}
\newcommand{\meanKK}{\operatorname{Mean}(\hat{K} = K )}
\newcommand{\vm}{\operatorname{V_1}}
\newcommand{\fpsle}{\operatorname{FPSLE}}
\newcommand{\fnsle}{\operatorname{FNSLE}}
\newcommand{\mean}{\operatorname{Mean}}
\newcommand{\E}{\operatorname{E}}

\endlocaldefs
\begin{document}

\begin{frontmatter}
\title{Multiscale Blind Source Separation}
\runtitle{Multiscale Blind Source Separation}

\begin{aug}
\author{\fnms{Merle} \snm{Behr}\thanksref{m1, t1}\ead[label=e1]{behr@math.uni-goettingen.de}},
\author{\fnms{Chris} \snm{Holmes}\thanksref{m2, t2}\ead[label=e3]{cholmes@stats.ox.ac.uk}},
\and
\author{\fnms{Axel} \snm{Munk}\thanksref{m1,m3,t3}\ead[label=e2]{munk@math.uni-goettingen.de}}

\thankstext{t1}{Merle Behr acknowledges support of DFG RTG 2088 and CRC 803 Z.}
\thankstext{t3}{Chris Holmes is supported by the Oxford-Man Institute, an EPSRC programme grant i-like, the Medical Research Council UK, and the Alan Turing Institute.}
\thankstext{t2}{Axel Munk is supported by DFG CRC 803 Z and FOR 916 B1, B3, Z.}

\runauthor{M. Behr et al.}

\affiliation{University of Goettingen \thanksmark{m1}, University of Oxford \thanksmark{m2},\\ and {Max Planck Institute for Biophysical Chemistry \thanksmark{m3}}}

\address{University of Goettingen \\
Institute for Mathematical Stochastics\\
Goldschmidtstr. 7\\
37077 G\"ottingen\\
Germany\\
\printead{e1}\\
\printead{e2}}

\address{University of Oxford\\
Department of Statistics\\
24-29 St Giles'\\
Oxford. OX1 3LB\\
United Kingdom\\
\printead{e3}}

\end{aug}

\begin{abstract}
We provide a new methodology for statistical recovery of single linear mixtures of piecewise constant signals (sources) with unknown mixing weights and change points in a multiscale fashion. We show exact recovery within an $\epsilon$-neighborhood of the mixture when the sources take only values in a known finite alphabet. 
Based on this we provide the SLAM (Separates Linear Alphabet Mixtures) estimators for the mixing weights and sources. For Gaussian error, we obtain uniform confidence sets and optimal rates (up to log-factors) for all quantities. 
SLAM is efficiently computed as a nonconvex optimization problem by a dynamic program tailored to the finite alphabet assumption. Its performance is investigated in a simulation study. Finally, it is applied to assign copy-number aberrations from genetic sequencing data to different clones and to estimate their proportions. 
\end{abstract}



\end{frontmatter}

\counterwithin{figure}{section}

\section{Introduction}\label{sec: intro}


As the presented methodology requires a quite broad range of techniques we will briefly introduce them in this section for explanatory purposes.
Details are given in subsequent sections and a supplement.

\subsection{The statistical blind source separation problem}\label{subsec: SBSSR}

We will start by introducing a particular kind of the blind source separation (BSS) problem which will be considered throughout this paper. 
More generally, in BSS problems (for a review see Section \ref{subsec: relW}) one observes a mixture of signals (sources) and aims to recover these sources from the available observations, usually corrupted by  noise. The blindness refers to the fact that neither the sources nor the mixing weights are known. Of course, without any additional information on the sources the BSS problem is unsolvable as the weights and sources cannot be separated, in general. However, under the additional assumption that the sources take values in a known finite alphabet, we will show that estimation of all quantities and inference for these is indeed possible.

Motivated by several applications mainly from digital communications (e.g., the recovery of mixtures of multi-level PAM signals (see \cite{verdu, proakis})) and cancer genetics (see Section \ref{subsec: intapp}), we assume, from now on, that the $m$ source functions $f^i,$ $i = 1,\ldots,m$, consist of arrays of constant segments, that is, step functions with unknown jump sizes, numbers, and locations of change points (c.p.'s), respectively. More specifically, let for a finite (known) ordered alphabet $\fA= \{a_1,\ldots,a_k\}\subset\R$, with $a_1<\ldots <a_k$, each source function be in the class of step functions on $[0,1)$ 
\begin{equation}\label{def:SA}
\small
\cS(\fA) \ZuWeis \Big\{ \sum_{j=0}^{K} \theta_j \indE_{[\tau_j,\tau_{j+1})}: \theta_j\in\fA,  0=\tau_0<...<\tau_{K}<\tau_{K+1} = 1, K\in\N \Big\}.
\end{equation} 
Note that this implies that for each source function the number $K(f^i)$ of c.p.'s is assumed to be finite, possibly different, and unknown.
We will assume $\theta_j \neq \theta_{j+1}$ for $j = 0,...,K$ to ensure identifiability of the c.p.'s $\tau_j$. Note that without further specification $\cS \ZuWeis \cS(\fA)$ is an extremely flexible class of functions, including any discretized source function taking values in $\fA$.
Moreover, we define the set of all possible (linear) mixtures with $m$ components each in $\cS$ as
\begin{equation}\label{MA}
\cM \ZuWeis \cM(\fA,m)= \Big\{\omega^{\top}f = \sum_{i=1}^{m}\omega_i f^i : \omega \in \Omega(m) \text{ and } f \in \cS(\fA)^m \Big \},
\end{equation}
with mixing weights $\omega$ in the $m$-simplex
\begin{equation}\label{def: Omega}
\Omega(m)\ZuWeis \big\{\omega\in\R^m: 0 \leq \omega_1 \leq \ldots \leq \omega_m \text{ and }\sum_{i=1}^{m}\omega_i =1\big\}.
\end{equation}
For a set $\tilde{\Omega}\subset \Omega(m)$ we define $\cM(\fA,\tilde{\Omega})$ analogously.
Throughout the following we assume that $m$ is known. Extension to unknown $m$ is akin to a model selection type of problem and beyond the scope of this paper.

In summary, in this paper we will be concerned with the \textit{statistical blind source separation regression} model.
\begin{model}
For a given finite alphabet $\fA$ and a given number of mixture components $m\in \N$ let $g = \sum_{i = 1}^m \omega_i f^i \in \cM$ be an arbitrary mixture of $m$ piecewise constant source functions $f^i \in \cS$. Suppose we observe
\begin{equation}\label{def:Y}
Y_j= g(x_j)+\sigma \epsilon_j,\qquad  j=1,\ldots,n,
\end{equation}
at sampling points $x_{j} \ZuWeis (j-1)/n $, s.t. the error $ (\epsilon_1,..., \epsilon_n)^{\top} \sim \cN (0,I_n)$, $\sigma > 0$, that is, i.i.d. centered normal random variables with variance $\sigma^2$. 
\end{model}

\begin{figure}[ht]
\includegraphics[width=\textwidth]{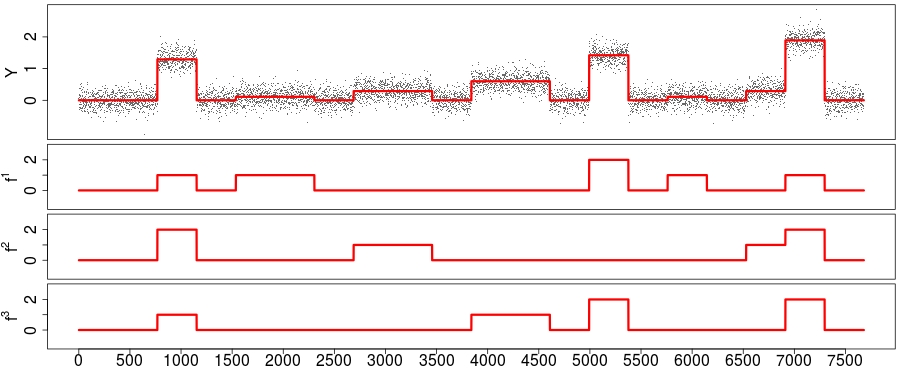}
\caption{The mixture $g = 0.11 f^1 + 0.29 f^2 + 0.6 f^3$, together with the observations $Y$ (gray dots), and the sources $f^1$, $f^2$, $f^3$ from Example \ref{example1} (from top to bottom).}\label{fig: example1}
\centering
\end{figure}

\begin{exampleon}\label{example1}
In Figure \ref{fig: example1}, a mixture $g$ of $m=3$ source functions $f^1$, $f^2$, $f^3$, taking values in the alphabet $\fA=\{0,1,2\}$, is displayed. The mixing weights are given by $\omega^{\top}=(0.11,0.29,0.6)$. Normal noise with standard deviation $\sigma=0.22$ is added according to the SBSSR-model, $n = 7680$. Both, $n$ and $\sigma$ were chosen close to our data example in Section \ref{sec:appgen}.
\end{exampleon}

In summary, the unknowns in the SBSSR-model are
\begin{enumerate}
\item the weights $\omega = (\omega_1,\ldots,\omega_m)^\top$ and \label{p: estw}
\item the source functions $f^i$, $i = 1,\ldots,m$, i.e. their \label{p: estf}
\begin{enumerate}
\item number of c.p.'s $K(f^i)$,
\item c.p. locations $\tau_j^i$, $j = 1,\ldots, K(f^i)$, and
\item function values $f^i(x)$ ($\in \fA$) at locations $x \in [0,1)$.
\end{enumerate}
\end{enumerate}

In this paper we will address estimation of all the quantities in \ref{p: estw}. and \ref{p: estf}. and, in addition, we will construct under further assumptions
\begin{enumerate}
\setcounter{enumi}{2}
\item a uniform (i.e., honest) confidence region $\cC_{1-\alpha}$ for the weights $\omega$ and\label{p: estcw}
\item asymptotically uniform multivariate confidence bands for the source functions $f=(f^1,\ldots,f^m)^{\top}$.\label{p: estcf}
\end{enumerate} 
\begin{remark}
\quad 

\begin{enumerate}[a)]
\item For simplicity, we assume throughout the following that $g$ in (\ref{def:Y}) is sampled equidistantly at $x_j = (j-1) / n$, $j = 1,\ldots,n$ and that all functions are defined on the domain $[0,1)$. We stress that extensions to more general domains $\subseteq \R$ and sampling designs are straightforward under suitable assumptions (see, e.g., \cite{boysen}) but will be suppressed to ease notation. 
\item Further, for sake of brevity, we will assume that in (\ref{def:Y}) the variance $\sigma^2$ is known, otherwise one may pre-estimate it $\sqrt{n}$-consistently by standard methods, see, for example, \cite{muller, hall, dette, davies} and Section \ref{sec:appgen}.
\end{enumerate}
\end{remark}

\subsection{Identifiability and exact recovery} \label{subsec: intI}
Before we introduce estimators for $\omega$ and $f$, we need to discuss identifiability of these parameters in the SBSSR-model, that is, conditions when $g$ determines them uniquely via $g = \sum_{i=1}^m \omega_i f^i$. 

Although, deterministic finite alphabet instantaneous (linear) mixtures, i.e., $\sigma = 0 $ in the SBSSR-model (\ref{def:Y}), received a lot of attention in the literature \cite{telwar,pajunen,yuanqing,diamantaras2006,diamantras,gu,rostami}, a complete characterization of identifiability remained elusive and has been recently provided in \cite{behr}, which will be briefly reviewed here as far as it is required for our purposes.
Obviously, not every mixture $g \in \cM$ in (\ref{MA}) is identifiable. Consider, for example, $\omega \in \Omega(m)$ in (\ref{def: Omega}) such that $\omega_1 = \omega_2$. Then a jump in the source function $f^1$ has the same effect on the mixture $g$ as a jump in $f^2$ and hence, $f^1$ and $f^2$ cannot be distinguished from the mixture $g$. Likewise, when $\omega_1$ and $\omega_2$ are close, i.e., $\omega_2 - \omega_1 \rightarrow 0  $, it becomes arbitrarily difficult to separate $f^1$ and $f^2$ from the observations $Y$ in the SBSSR-model. For statistical estimation, it is therefore necessary that different source function values $f(x) = (f^1(x),\ldots, f^m(x)) \in \fA^m $ are sufficiently well separated by the mixing weights $\omega$. This is quantified by the \textit{alphabet separation boundary} \cite{behr}
\begin{align}\label{def: ASB}
ASB(\omega) = ASB(\omega, \fA) \ZuWeis \min _{a \neq a^{\prime}\in \fA^m } \abs{\omega^{\top}a - \omega^{\top}a^{\prime}}.
\end{align}
A necessary identifiability condition in the SBSSR-model is $ASB(\omega) > 0$ (see \cite[Section 3.A]{behr}), where the size of $ASB(\omega)$ can be understood as a conditioning number for the difficulty of separating the sources in the SBSSR-model, that is, the smaller $ASB(\omega)$, the more difficult separation of sources. Therefore, to quantify the estimation error of any method which serves the purposes in \ref{p: estw}.\ - \ref{p: estcf}.\ we must restrict to submodels of mixing weights which sufficiently separate different alphabet values in $\fA^m$, that is, for given $\delta > 0$ we introduce
\begin{equation}\label{def: OmegaD}
\Omega^\delta = \Omega^\delta(\fA, m) \ZuWeis \big\{\omega\in\Omega(m): ASB(\omega) \geq \delta \}.
\end{equation}

Note further that $ASB(\omega) > 0$ implies that any jump in the source vector $f$ (i.e., at least one source $f^i$ jumps) occurs as well in the mixture $g = \omega^{\top}f$ and that $ASB(\omega)$ coincides with the minimal possible jump height of $g$.

Just as we have restricted the possible $\omega$'s in (\ref{def: OmegaD}), it is necessary to further restrict the set of possible source functions $f \in \cS(\fA)^m$ in (\ref{def:SA}). Consider for example the case of two sources, $m = 2$, such that $f^1 = f^2$. Then $g = \omega_1 f^1 + \omega_2 f^2 = f^1$, independently of $\omega$, and hence, $\omega$ cannot be determined from $g$. Therefore, a certain kind of variability of the sources $f^i$ is necessary to ensure identifiability of the mixing weights $\omega$. 
We employ from \cite{behr} the following simple sufficient identifiability condition.


\begin{dfon}\label{def:separable}
A vector of source functions $f = (f^1,\ldots,f^m)^\top \in \cS(\fA)^m$ is \textit{separable} if there exit intervals $I_1,\ldots, I_m \subset [0,1)$ such that $f$ is constant on $I_r$  with function values
\begin{equation}\label{separable}
f|_{I_r} \equiv [A]_r, \qquad r = 1,\ldots,m,
\end{equation}
with
\begin{equation}
A \ZuWeis a_{1} E_m + (a_{2} - a_{1}) I_m = \begin{pmatrix}
a_{2} & a_{1} & a_{1} & \ldots & a_{1} \\
a_{1} & a_{2} & a_{1} & \ldots & a_{1} \\
\vdots &&&& \vdots \\
a_{1} & a_{1} & \ldots & a_{1} & a_{2}
\end{pmatrix} \in \fA^{m\times m}, \label{Amatrix}
\end{equation}
where $E_m$ denotes the matrix of ones, $I_m$ the identity matrix, and $[A]_r$ the $r$-th row of $A$.
\end{dfon}


The notation ``separable'' is borrowed from identifiability conditions for nonnegative matrix factorization \cite{donoho2003, arora, recht}, see Section \ref{subsec: relW} for details.
Separability in Definition \ref{def:separable} means that for each of the $m$ sources $f^i$ there is a region where only this source function is ``active'' (taking the second smallest alphabet value $a_{2}$) and all the other sources are ``silent'' (taking the smallest alphabet value $a_{1}$). For example, if we have an alphabet of the form $\fA = \{0,1,a_3,...,a_k\}$, $A$ becomes the identity matrix and separability means that each of the mixing weights $\omega_i$ appears at least once in the mixture $g = \omega^\top f$.
Note that separability in Definition \ref{def:separable} only requires that the values $[A]_r \in \fA^m$ are attained \textit{somewhere} by the source functions $f^1,\ldots,f^m$ and does not specify the location.
For specific situations it is possible to replace the matrix $A$ in (\ref{Amatrix}) by a different invertible (as a function from $\Omega(m)$ to $\R^m$) matrix if this matrix induces enough variability in the sources for the weights to be identifiable from their mixture (see \cite{behr}). Here, however, we consider arbitrary alphabets and number of sources and the separability condition in Definition \ref{def:separable} ensures identifiability for arbitrary $\fA$ and $m$, in general.
Note that when the source functions $f = (f^1,\ldots,f^m)^{\top}$ attain all $k^m$ possible function values in $\fA^m$ somewhere in $[0,1)$, the case of maximal variation, then, in particular, $f$ is separable (see \cite{behr} for further examples). 
We stress that the above assumption (\ref{separable}) on the variability of  $f$ is close to being necessary for identifiability (see \cite[Theorem 3.1]{behr}). Hence, without such an assumption no method can provide a unique decomposition of $g$ into the $f^i$'s and its weights $\omega_i$, $i=1,\ldots,m$, even in the noiseless case.
Summing up, we will, in the following, restrict to those mixtures $g$ in the SBSSR-model, which are in
\begin{equation}\label{def: MA}
\cM^{\delta} \ZuWeis \Big\{\omega^{\top}f = \sum_{i=1}^{m}\omega_i f^i : \omega \in \Omega^\delta \text{ and } f \in \cS(\fA)^m \text{ is separable} \Big \}.
\end{equation}
For instance, in Example \ref{example1} $f $ is separable and $\omega \in \Omega^{0.02}$, i.e., $g \in \cM^{0.02}$.
 
The following simple but fundamental result will guide us later on to derive estimators for all quantities in 1. and 2. in the statistical setting (\ref{def:Y}) (see Section \ref{subsec: intSESAME}).

\begin{theomn}{Stable recovery of weights and source functions}\label{lem: epsilon}
Let $g = \omega^{\top} f, \tilde{g} = \tilde{\omega}^{\top} \tilde{f}$ be two mixtures in $\cM^\delta$ for some $\delta > 0$ and let $\epsilon$ be such that $0<\epsilon< \delta (a_2 - a_1)/(2m(a_k - a_1))$. If
\begin{equation}\label{supghg}
\sup_{x\in[0,1)}\abs{g(x)-\tilde{g}(x)} < \epsilon,
\end{equation}
\begin{enumerate}
\item then the weights satisfy the stable approximate recovery (SAR) property $\max_{i=1,\ldots,m}\abs{\omega_i - \tilde{\omega}_i} < \epsilon/(a_2 - a_1)$ and \label{a}
\item the sources satisfy the stable exact recovery (SER) property $f = \tilde{f}$.\label{b}
\end{enumerate}
\end{theomn}

For a proof see Section \ref{subsec: appExactR} in the supplement.

\subsection{Methodology: first approaches} \label{subsec:intMeth}
\begin{figure}[h]
\includegraphics[width=\textwidth]{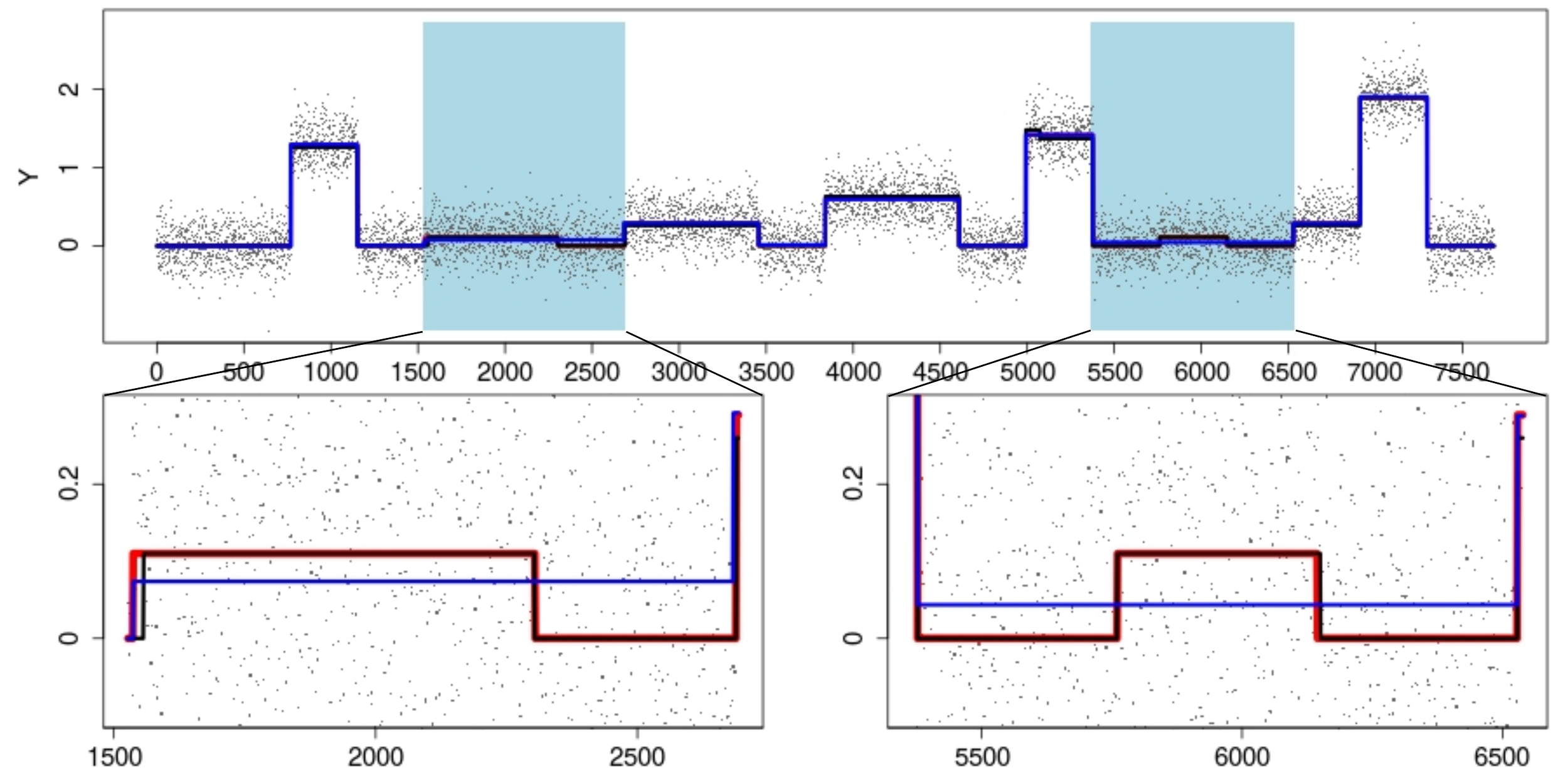}
\caption{Observations $Y$ from Example \ref{example1} (gray dots), together with the true underlying mixture $g$ (red line). The blue line shows the c.p. estimate from \cite{frick}, which does not incorporate the mixture structure. The red line shows the estimate with the proposed method (see Figure \ref{fig:fgest} for the underlying recovery of $\omega$ and the sources $f$). The blue areas display a region where $g$ has a small jump (red line), which is not detected by the c.p. estimator \cite{frick} (blue line), but by the proposed method (black line). The bottom plots show a zoom in of the blue regions.}
\label{fig: SmuceVsSlam}
\centering
\end{figure}

In order to motivate our (quite involved) methodology, let us discuss briefly some attempts which may come to mind at a first glance.
As a first approach to estimate $\omega$ and $f$ from the data $Y$ in the SBSSR-model one might pre-estimate the mixture $g$ with some standard c.p. procedure, ignoring its underlying mixture structure, and then try to reconstruct $\omega$ and $f$ afterwards. One problem is that the resulting step function cannot be decomposed into mixing weights $\omega \in \Omega(m)$ and source function $f\in \cS^m(\fA)$, in general, as the given alphabet $\fA$ leads to restrictions on the function values of $g$. 
But already for the initial step of reconstructing the mixture $g$ itself, a standard c.p. estimation procedure (which does ignore the mixture structure) is unfavorable as it discards important information on the possible function values of $g$ (induced by $\fA$). For example, if $g$ has a small jump in some region, this might be easily missed (see Figure \ref{fig: SmuceVsSlam} for an example). Consequently, subsequent estimation of $f$ and $\omega$ will fail as well. In contrast, a procedure which takes the mixture structure explicitly into account right from its beginning is expected to have better detection power for a jump.
As a conclusion, considering the SBSSR-model as a standard c.p. model discards important information and does not allow for demixing, in general. 

\begin{figure}[h]
\includegraphics[width=\textwidth, height= 4cm]{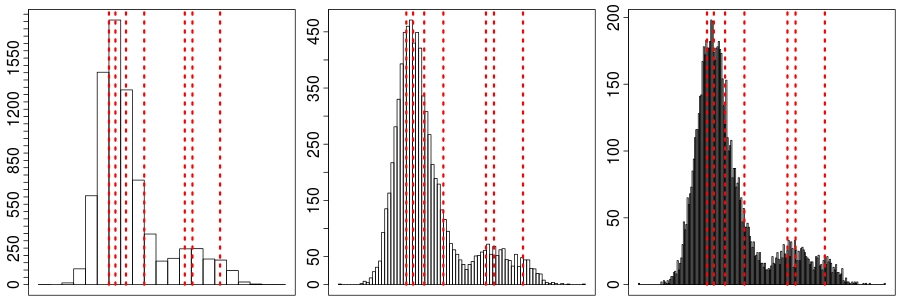}
\caption{Histogram of the data from Example \ref{example1} with $20$, $100$, and $200$ equidistant bins, respectively (from left to right). The vertical red lines indicate the true function values (modes) of $g$ which have to be identified.}
\label{fig: hist_example1}
\centering
\end{figure}

A second approach which comes to mind is to first use some clustering algorithm to pre-estimate the function values of $g$, ignoring its serial c.p. structure, and infer the mixing weights $\omega$ from this. This pre-clustering approach has been pursued in several papers \cite{diamantaras2006,yuanqing,gu} for the particular case of a binary alphabet, that is, $k = 2$.
However, as the number of possible function values of $g$ equals $k^m$ (recall that $k$ is the size of the alphabet and $m$ is the number of sources), recovery of these values in a statistical context by clustering is a difficult task in general, as it amounts to estimate the location of (at most) $k^m$ modes correctly from the marginal distributions of the observations $Y_j$. 
In fact, this corresponds to mode hunting (see, e.g., \cite{polonik1998, cheng1999, tibshirani2001, li2007, dumbgen2008, ooi}) with potentially large number of modes which is known to be a hard problem.
We illustrate the difficulty of this in Figure \ref{fig: hist_example1} employing histograms of the $Y_j$'s in Example \ref{example1} with different bin widths. From this, it becomes obvious that a pre-clustering approach is not feasible for the present data.

Summing up, ignoring either of both, the c.p. and the finite alphabet mixture structure, in a first pre-estimation step discards important information which is indispensable for statistically efficient recovery. We emphasize that we are not aware of any existing method taking both aspects into account, in contrast to the method presented in this paper (SLAM), which will be briefly described now.

\subsection{Separate Linear Alphabet Mixtures (SLAM)}\label{subsec: intSESAME}
In a first step, we will construct a confidence region $\cC_{1-\alpha}$ for the weights $\omega$ which can be characterized by the acceptance region of a specific multiscale test with test statistic $T_n$, which is particularly well suited to capture both, the c.p. and the mixture structure, of $g$. The confidence level is determined by a threshold $q_n(\alpha)$ such that for any $g = \sum_{i=1}^m\omega_i f^i \in \cM^\delta$
\begin{equation}\label{feasible}
\{\omega \in \cC_{1-\alpha}(Y)\}\supseteq \{T_n \leq q_n(\alpha) \}.
\end{equation}
In a second step we estimate $f$ based on a multiscale constraint again. In the following section we will introduce this procedure in more detail.
We stress that the multiscale approach underlying SLAM is crucial for valid recovery of sources and mixing weights as the jumps potentially can occur at any location and any scale (i.e., interval length of neighboring sampling points). 

\subsubsection{Multiscale statistic and confidence boxes underlying SLAM}\label{subsubsec: Tn}
As the jump locations may occur at any place, a well established way for inferring the function values of $g$ is to use local log-likelihood ratio test statistics in a multiscale fashion (see e.g., \cite{ siegmund, dumbgen2001, davies, dumbgen2008, frick}). 
Let $g|_{[x_i,x_j]} \equiv  g_{ij}$ denote that $g$ is constant on $[x_i, x_j]$ with function value $g_{ij}$.
For the local testing problem on the interval $[x_i,x_j]\subset [0,1)$ with some given value $g_{ij}\in \R$
\begin{equation}\label{def:locH}
H_0 : g|_{[x_i,x_j]} \equiv  g_{ij} \quad \text{vs.}\quad H_1: g|_{[x_i,x_j]}\not \equiv g_{ij}
\end{equation}
the local log-likelihood ratio test statistic is
\begin{equation}\label{def: Tij}
T_i^j(Y_i,\ldots,Y_j,g_{ij})= \ln \left( \frac{\sup _{\theta \in \R}\prod_{l=i}^j \phi_\theta(Y_l)}{\prod_{l=i}^j \phi_{g_{ij}}(Y_l)}\right)= \frac{(\sum_{l=i}^j Y_l - g_{ij})^2}{2 \sigma^2 (j-i+1)},
\end{equation} 
where $\phi_{\theta}$ denotes the density of the normal distribution with mean $\theta$ and variance $\sigma^2$.
We then combine the local testing problems in (\ref{def:locH}) and define in our context the multiscale statistic $T_n$ for some candidate function $\tilde{g}$ (which may depend on $Y$) as
\begin{equation}
T_n(Y,\tilde{g})\ZuWeis \max_{\substack{1\leq i\leq j\leq n \\ \tilde{g}|_{[x_i, x_j]} \equiv \tilde{g}_{ij}}}\frac{|\sum_{l=i}^{j}Y_l - \tilde{g}_{ij}|}{\sigma \sqrt{j-i+1}}-pen(j-i+1),\label{def: Tn}
\end{equation}
where $pen(j-i+1)\ZuWeis \sqrt{2 \left( \ln\left( n / (j-i+1)\right) + 1 \right)}$. The maximum in (\ref{def: Tn}) is understood to be taken only over those intervals $[x_i, x_j]$ on which $\tilde{g}$ is constant with value $\tilde{g}_{ij} = \tilde{g}(x_i)$. The function values of $\tilde{g}$ determine the local testing problems (the value $g_{ij}$ in (\ref{def:locH})) on the single scales $[x_i, x_j]$.
The calibration term $pen(\cdot)$ serves as a balancing of the different scales in a way that the maximum in (\ref{def: Tn}) is equally likely attained on all scales (see \cite{dumbgen2001,frick}). Other scale penalizations can be employed as well (see e.g., \citep{walther2010}), but, for the ease of brevity, will not be discussed here.
With the notation 
$\bar{Y}_i^j\ZuWeis \sum_{l=i}^j Y_l / (j-i+1)$,
the statistic $T_n(Y,\tilde{g})$ in (\ref{def: Tn}) has the following geometric interpretation:
\begin{equation} \label{geoInt}
T_n(Y,\tilde{g})\leq q \; \Leftrightarrow 
\tilde{g}_{ij}  \in B(i,j) \; \forall 1\leq i \leq j \leq n  \text{ with } \tilde{g}|_{[x_i, x_j]} \equiv \tilde{g}_{ij},
\end{equation}
for $q \in \R$, with intervals
\begin{equation}\label{def: box}
B(i,j)\ZuWeis \bigg\lbrack \overline{Y}_{i}^{j}-\frac{q+ pen(j-i+1)}{\sqrt{j-i+1}/\sigma},\overline{Y}_{i}^{j}+\frac{q+ pen(j-i+1)}{\sqrt{j-i+1}/\sigma}\bigg\rbrack.
\end{equation}
In the following we will make use of the fact that the distribution of $T_n(Y,g)$, with $g\in\cM^\delta$ (see (\ref{def: MA})) the true signal from the SBSSR-model, can be bounded from above with that of $T_n = T_n(Y,0)$. It is known that $T_n \overset{\cD}{\Rightarrow} L(\B) < \infty$ a.s. as $n\rightarrow \infty$, a certain functional of the Brownian motion $\B$ (see \cite{dumbgen2001, dumbgen2006}). Note that the distribution of $T_n(Y,0)$ does not depend on the (unknown) $f$ and $\omega$ anymore. As this distribution is not explicitly accessible and to be more accurate for small $n$ ($\leq 5000$ say) the finite sample distribution of $T_n$ can be easily obtained by Monte Carlo simulations. From this one obtains $q_n(\alpha)$, $\alpha \in (0,1)$, the $1-\alpha$ quantile of $T_n$. We then obtain
\begin{equation}\label{def: qalpha}
\inf_{g \in \cM^\delta} \Pp(T_n(Y,g)\leq q_n(\alpha))\geq 1-\alpha.
\end{equation}
Hence, for the intervals in (\ref{def: box}) with $q = q_n(\alpha)$ it follows that for all $g \in \cM^\delta$
\begin{equation}\label{Pbox}
\Pp(g_{ij} \in B(i,j)\;\forall 1\leq i \leq j \leq n \text{ with }g|_{[x_i, x_j]} \equiv g_{ij} )\geq 1-\alpha.
\end{equation}

In the following, we use the notation $B(i,j)$ for both, the intervals in (\ref{def: box}) and the corresponding boxes $[i,j]\times B(i,j)$.

\subsubsection{Inference about the weights}\label{subsec: introWeights}

We will use now the system of boxes $\fB \ZuWeis \{B(i,j): 1\leq i \leq j \leq n\}$ from (\ref{def: box}) with $q = q_n(\alpha)$ as in (\ref{def: qalpha}) to construct a confidence region $\cC_{1-\alpha}$ for $\omega$ such that (\ref{feasible}) holds, which ensures 
\begin{equation}\label{covComega}
\inf_{g \in \cM^\delta} \Pp(\omega \in \cC_{1-\alpha})\geq 1-\alpha.
\end{equation}
More precisely, we will show that a certain element $B^{\star} \in \fB^m$ (denoted as the space of $m$-boxes) directly provides a confidence set $\cC_{1-\alpha}^{\star} = A^{-1} B^{\star}$ for $\omega$, with $A$ as in (\ref{Amatrix}). As $B^{\star}$ cannot be determined directly, we will construct a covering, $\fB^{\star}\ni B^{\star}  $, of it such that the resulting confidence set
\begin{equation}\label{Comega}
\cC_{1-\alpha} = \bigcup_{B \in \fB^{\star}} A^{-1} B
\end{equation}
has minimal volume (up to a log-factor) (see Section \ref{subsec: consi}). 
The construction of $\fB^{\star}$ is done by applying certain reduction rules on the set $\fB^m$ reducing it to a smaller set $\fB^{\star} \subset \fB^m$ with $B^{\star} \in \fB^{\star}$. This is summarized in the CRW (confidence region for the weights) algorithm in Section \ref{subsec: CO} (and Section \ref{subsec: psCRW} in the supplement, respectively), which constitutes the first part of SLAM.

In Example \ref{example1} for $\alpha = 0.1$ this gives 
$\cC_{0.9}= [0.00,0.33]\times [0.07,0.41]\times [0.39,0.71]$ 
as a confidence box for 
$\omega = (\omega_1, \omega_2, \omega_3)^{\top}$ which covers the true value $\omega = (0.11, 0.29, 0.60)^{\top}$ in this case.

As the boxes $B(i,j)$ from (\ref{def: box}) are constructed in a symmetric way, SLAM now simply estimates $\omega$ by
\begin{equation}\label{omes}
\hat{\omega}= \frac{1}{\sum_{i=1}^m (\underline{\omega}_i+\overline{\omega}_i)}(\underline{\omega}_1+\overline{\omega}_1,\ldots,\underline{\omega}_m+\overline{\omega}_m),
\end{equation}
with $\cC_{1-\alpha}\WeisZu [\underline{\omega}_1,\overline{\omega}_1]\times \ldots \times [\underline{\omega}_m,\overline{\omega}_m]$. In Example \ref{example1}, (\ref{omes}) gives for $\alpha = 0.1$ $\hat{\omega}= (0.17, 0.25, 0.58)^{\top}$.

For $D\subset \R^m$ and $d\in \R^m$ define the maximal distance
\begin{equation}\label{def: dist}
\dist(d,D)\ZuWeis \sup_{\tilde{d}\in D}\norm{d-\tilde{d}}_{\infty}.
\end{equation}
Further, and for all following considerations, define 
\begin{equation}\label{alphaN}
\alpha_n = \exp(- c_1 \ln^2(n)) \text{ and } \beta_n = \exp\left(- 75 m^2 \left(\frac{a_k - a_1}{a_2 - a_1} \right)^2 c_1 \ln^2(n)\right),
\end{equation}
for some constant $c_1$, to be specified later, see (\ref{c0c1c2}).
Denote the minimal distance between any two jumps of $g \in \cM^\delta$ (and hence of the $f^i$'s, recall the discussion in Section \ref{subsec: intI}) as $\lambda$. Then, in addition to uniform coverage in (\ref{covComega}) for $\alpha = \alpha_n$ in (\ref{alphaN}), we will show that the confidence region $\cC_{1-\alpha}$ from (\ref{Comega}) covers the unknown weight vector $\omega$ with maximal distance shrinking of order $\ln(n)/\sqrt{n}$ with probability tending to one at a superpolynomial rate,
\begin{equation*}
\Pp\left(\dist(\omega, C_{1-\alpha_n}(Y)) < \frac{c_2}{a_2 - a_1} \frac{\ln(n)}{\sqrt{n}}\right) \geq 1 - \exp(- c_1 \ln^2(n))
\end{equation*}
for all $n\geq N^{\star}$, for some constants $c_1 = c_1(\delta)$, $c_2 = c_2 (\lambda, \delta)$ and some explicit $N^{\star} =  N^{\star}(\lambda, \delta) \in \N$  (see Corollary \ref{cor: omegaconsi}).

\subsubsection{Inference about the source functions}\label{subsec: intSources}

Once the mixing weights $\omega$ have been estimated by $\hat{\omega}$ (see (\ref{omes})), SLAM estimates $f^1,\ldots,f^m$ in two steps. First, the number of c.p.'s $K(g)$ of $g = \omega^{\top}f \in \cM^\delta$ will be estimated by solving the constrained optimization problem 
\begin{equation}\label{Khat}
\hat{K} \ZuWeis \min_{\tilde{g} \in \; \cM(\fA, \hat{\omega}) } K(\tilde{g}) \quad \text{s.t.} \quad T_n(Y,\tilde{g})\leq q_n(\beta).
\end{equation}
Here, the multiscale constraint on the r.h.s. of (\ref{Khat}) is the same as for $\cC_{1-\alpha}(Y)$ in (\ref{feasible}), but with a possibly different confidence level $1-\beta$. 
Finally, we estimate $f^1,\ldots,f^m$ as the constrained maximum likelihood estimator 
\begin{equation}\label{def: gEstS}
\hat{f} = (\hat{f}^1,\ldots, \hat{f}^m)^{\top} \ZuWeis \argmax_{\tilde{f} \in \cH(\beta)} \sum_{i = 1}^n \ln\left(\phi_{\hat{\omega}^{\top}\tilde{f}(x_i)}(Y_i)\right),
\end{equation}
with (see Section \ref{subsec: estfF})
\begin{equation}\label{Hn}
\cH(\beta)\ZuWeis \Big\{\tilde{f}\in \cS(\fA)^m : T_n\left(Y, \hat{\omega}^{\top}\tilde{f}\right) \leq q_n(\beta) \text{ and } K\left(\hat{\omega}^{\top}\tilde{f}\right) = \hat{K}\Big\}.
\end{equation}

\begin{figure}[h]
\includegraphics[width=\textwidth]{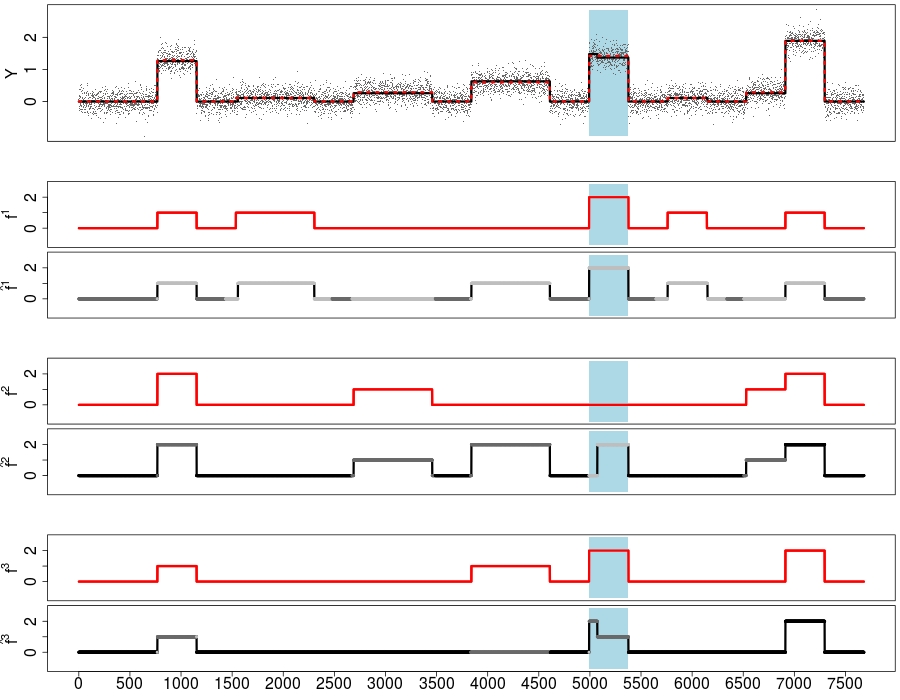}
\caption{First row: $g$ (red dotted line), $\hat{g}$ (black line) with $\hat{\omega} = (0.11, 0.26, 0.63)^{\top}$, and data $Y$ (gray) from Example \ref{example1}. Subsequent rows: $f^i$ (red line) and SLAM's estimate $\hat{f}^i$ (gray/black line) for $q_n(\alpha) = 0.2$ and $q_n(\beta) = 2.1 $ (see Section \ref{subsec: chQ}). 
Gray shades for segments of $\hat{f}^i$ indicate the confidence for the given segment:
a maximal deviation of two (light gray), one (gray), and no deviation (black) at confidence level $\beta = 0.01$. The blue area displays a constant region of $g$ where $\hat{g}$ includes a (wrong) jump and  its effect on estimation of the sources.}
\label{fig:fgest}
\centering
\end{figure}

Choosing $\alpha = \alpha_n$ and $\beta = \beta_n$ as in (\ref{alphaN}), in Section \ref{subsec: consi} (see Theorem \ref{consimain}) we show that with probability at least $1-\alpha_n$, for $n$ large enough, the SLAM estimator $\hat{f}$ in (\ref{def: gEstS}) estimates for all $i = 1,\ldots,m$ 
\begin{enumerate}
\item the respective number of c.p.'s $K(f^i)$ correctly, 
\item all c.p. locations with rate $\ln^2(n)/n$ simultaneously, and \label{intoOptEq2}
\item the function values of $f^i$ exactly (up to the uncertainty in the c.p. locations).
\end{enumerate}

Obviously, the rate in \ref{intoOptEq2}.\ is optimal up to possible log-factors as the sampling rate is $1/n$. From Theorem \ref{consimain} it follows further (see Remark \ref{remarkOptRates}) that the minimax detection rates are even achieved (again up to possible log-factors) when $\delta, \lambda \to 0$ (as $n\to \infty$).

Further, we will show that a slight modification $\tilde{\cH}(\beta)$ of $\cH(\beta)$ in (\ref{Hn}) constitutes an asymptotically uniform (for given ASB $\delta$ and $\lambda$) multivariate confidence band for the source functions $(f^1,\ldots,f^m)$ (see Section \ref{subsec: estfC}).

To illustrate, Figure \ref{fig:fgest} depicts SLAM's estimates of the mixture $\hat{g} = \hat{\omega}^{\top}\hat{f}$, with $\hat{\omega} = (0.11, 0.26, 0.63)^{\top}$, and the source functions $\hat{f}^1$, $\hat{f}^2$, $\hat{f}^3$ from (\ref{def: gEstS}) with $Y$ as in Example \ref{example1}, $\beta = 0.01$ (corresponding to $q_n(\beta) = 2.1$), and an automatic choice of $\alpha$, the MVT-selection method explained in Section \ref{subsec: chQ}.
In order to visualize $\tilde{\cH}(\beta)$, we illustrate the provided confidence in gray scale encoding the projections of $\tilde{\cH}(\beta)$ (recall the alphabet $\fA = \{0,1,2\}$).

\subsection{Algorithms and software}

SLAM's estimate for $\omega$ (see (\ref{omes}) and Algorithm CRW, in Section \ref{subsec: psCRW} in the supplement) can be computed with polynomial complexity between $\cO(n^m)$ and $\cO(n^{2m})$ (see Section \ref{subsec: Imple}). Using dynamic programming, the final estimate of sources can then be computed with a complexity ranging from  $\cO(n)$ and $\cO(n^2)$ depending on the final solution (see Section \ref{subsec: Imple} for details).
An R-package including an implementation of SLAM is available on request.

\subsection{Simulations}

The performance of SLAM is investigated in a simulation study in Section \ref{subsec: Appsim}. We first investigate accuracy of $\hat{\omega}$ and the confidence region $\cC_{1-\alpha}(Y)$ as in (\ref{omes}) and (\ref{Comega}). We found always higher coverage of $\cC_{1-\alpha}(Y)$ than the nominal confidence level $1-\alpha$. In line with this,  $\hat{\omega}$ appeared to be very stable under the choice of the confidence level $\alpha$.
Second, we investigate SLAM's estimates $\hat{f}$. A major conclusion is that if $g$ is not well estimated in a certain region, this typically will influence the quality of the estimates of $f^i$ in this region but not beyond (see the marked lightblue region in Figure \ref{fig:fgest} where the estimator $\hat{g}$ includes a wrong jump in a constant region of $g$ but this error does not propagate serially). This may be explained by the flexible c.p. model $\cM^\delta$ together with the multiscale nature of SLAM, which locally "repairs" estimation errors.
Finally, in Section \ref{subsec: chQ} we comment on practical choices for $\alpha$ and $\beta$ complementing the theoretically motivated choices in (\ref{alphaN}). To this end, we suggest a data driven selection method for $\alpha$ when it is considered as tuning parameter for the accuracy of the estimate $\hat{\omega}$ and $\hat{f}$ rather than a confidence level for the coverage of $\omega$. 

\subsection{Application to cancer genetics}\label{subsec: intapp}

Blind source separation in the context of the SBSSR-model occurs in different areas, for example in digital communications and signal transmission.
The main motivation for our work comes from cancer genetics, in particular from the problem to assign copy-number aberrations (CNAs) in cell samples taken from tumors (see \cite{liu}) to its clones. CNAs refer to stretches of DNA in the genomes of cancer cells which are under copy-number variation involving deletion or duplication of stretches of DNA relative to the inherited (germline) state present in normal tissues. CNAs are known to be key drivers of tumor progression through the deletion of ``tumor suppressing'' genes and the duplication of genes involved in processes such as cell signaling and division. Understanding where, when and how CNAs occur during tumourgenesis, and their consequences, is a highly active and important area of cancer research (see e.g., \cite{beroukhim}).
Modern high-throughput technologies allow for routine whole genome DNA sequencing of cancer samples and major international efforts are underway to characterize the genetic make up of all cancers, for example The Cancer Genome Atlas, {\tt{http://cancergenome.nih.gov/}}. 

A key component of complexity in cancer genetics is the ``clonal'' structure of many tumors, which relates to the fact that tumors usually contain distinct cell populations of genetic sub-types (clones) each with a distinct CNA profile
(see e.g., \cite{greaves, shah}). High-throughput sequencing technologies act by bulk measurement of large numbers of pooled cells in a single sample, extracted by a micro-dissection biopsy or blood sample for haematological cancers. 

The copy-number, that is, number of copies of DNA stretches at a certain locus, of a single clone's genome is a step function mapping chromosomal loci to a value $i \in \{0,\ldots,k\}$ corresponding to $i$ copies of DNA at a locus, with reasonable biological knowledge of $k$ (in  our example $k=5$, see Section \ref{sec:appgen}). 

From the linear properties of the measurement technologies the relative amount of DNA measured at any loci is therefore a mixture of step functions, with mixture weights given by the relative proportion of the clone's DNA in the pool. The estimation of the mixed function, that is, estimating the locations of varying overall copy numbers, has perceived considerable interest in the past (see \cite{olshen, mBIC, tibshirani, jeng, chen, niu2012, frick, du}).
However, the corresponding demixing problem, that is, jointly estimating the number of clones, their proportion, and their CNAs, has only perceived more recently as an important issue and hence received very little attention in a statistical content so far and is the main motivation for this work. 

In Section \ref{sec:appgen}, we illustrate SLAM's ability to recover the CNA's of such clones by utilizing it on real genetic sequencing data. On hand of a special data set, with measurements not only for the mixture but also for the underlying source functions (clones) and with knowledge about the mixing weights, we are able to report on the accuracy of SLAM's estimates of the corresponding CNA profile and the mixing proportion of the clones.

\subsection{Related work}\label{subsec: relW}
Each, BSS of finite alphabet sources (see, e.g., 
\cite{pajunen, lee, bofill, yuanqing,  diamantaras2006, amari, diamantras, gu, rostami}) 
and the estimation of step functions, with unknown number and location of c.p.'s (see, e.g., \cite{carlstein,olshen, fearnhead,  friedrich, tibshirani, spokoiny,jeng, killick, zhang2012, niu2012, siegmund20132, frick, matteson, fryzlewicz, harchaoui, du}), are widely discussed problems.
However, the combination of both, as discussed in this paper, is not. 
Rigorous statistical methodology and theory for finite alphabet BSS problems is entirely lacking to best of our knowledge and we are not aware of any other method which provides estimates for and confidence statements in the SBSSR-model in such a rigorous and general way. There are, however, related problems, discussed in the following.

Rewriting the SBSSR-model (\ref{def:Y}) in matrix form
$Y = F \omega  + \epsilon$
with $F = (f^i(x_j))_{1\leq j \leq n, 1 \leq i \leq m}$ shows some commonality to signal recovery in linear models.
In fact, our Theorem \ref{lem: epsilon} reveals some analogy to exact and stable recovery results in compressive sensing and related problems (see \cite{donoho2006, candes}).
We stress, however, that there are fundamental differences.
There typically the systems matrix $F$ is known and $\omega$ is a sparse vector to be recovered, having only a very few non null coefficients. Under an additional finite alphabet assumption (for known $F$) recovery of $\omega$ is, for example, addressed in \cite{draper, das, bioglio, aissa}. In our setting both, $F$ and $\omega$ are unknown.

Another related problem is non-negative matrix factorization (NMF) (see e.g., \cite{lee1999, donoho2003, arora, recht}), where one assumes a multivariate signal $Y \in \R^{n \times M}$ resulting from $M$ different (unknown) mixing vectors, that is, $\omega \in \R_+^{m \times M}$, and an (unknown) non-negative source matrix $F \in \R_+^{n \times m}$. There, a fundamental assumption is that $m \ll \min(n,M)$, which obviously does not hold in our case where $M = 1$. Instead we employ the additional assumption of a known finite alphabet, i.e., $F \in \fA^{n \times m}$. Indeed, techniques and algorithms for NMF are quite different from the ones derived here, as our methodology explicitly takes advantage of the one dimensional (i.e., ordered) c.p. structure under the finite alphabet assumption. 

However, the identifiability conditions (\ref{def: OmegaD}) and (\ref{separable}) from Section \ref{subsec: intI} are similar in nature to identifiability conditions for the NMF problem \cite{donoho2003, arora}, from where the notation ``separable'' originates. In order to ensure identifiability in the NMF problem, the ``$\alpha$-robust simplicial'' condition (see e.g., \cite[Definition 2.1]{recht}) on the mixing matrix $\omega \in \R_+^{m \times M}$ and the ``separability'' condition (see e.g. \cite[Definition 2.2]{recht}) on the source matrix $F \in \R_+^{n \times m}$ are well established \cite{donoho2003, arora, recht}. 

There, the ``$\alpha$-robust simplicial'' condition assumes that the mixing vectors $\omega_{1\cdot},\ldots, \omega_{m\cdot} \in \R^M_+$ constitute vertices of an $m$-simplex with minimal diameter (distance between any vertex and the convex hull of the remaining vertices) $\alpha$. This means that different source values $F_{i\cdot} \in \R^m$ are mapped to different mixture values $F_{i \cdot}\omega \in \R_+^M$ by the mixing matrix $\omega \in \R_+^{m \times M}$. This condition is analog to the condition $ASB(\omega) \geq \delta$ in (\ref{def: OmegaD}), which also ensures that different source values $f(x) \in \fA^m$ are mapped to different mixture values $\omega^\top f(x) \in \R$ via the mixing weights $\omega \in \Omega(m)$, with minimal distance $\delta$ between different mixture values.

The ``separability'' condition in NMF  is the same as in Definition \ref{def:separable} but with $A$ replaced by the identitiy matrix (recall that in NMF the sources can take any positive value in $\R_+$, in contrast to the SBSSR-model where the sources can only take values in a given alphabet $\fA$) and the intervals $I_r \subset (0,1]$ are replaced by measurement points $i_r \in \{1,\ldots,n\}$ (recall that the SBSSR-model considers a change-point regression setting, in contrast to NMF where observations do not necessarily come from discrete measurements of an underlying regression function). In both models (NMF and SBSSR) the separability condition ensures a certain variability of the sources in order to guarantee identitfiability of the mixing matrix and vector, respectively, from their mixture.

Besides NMF, there are many other matrix-factorization problems, which aim to decompose a multivariate signal $Y \in \R^{n \times M}$ in two matrices of dimension $n \times m$ and $m \times M$, respectively. A popular example is independent component analysis (ICA) (see e.g., \cite{comon, belkin, arora2015}), which exploits statistical independence of the $m$ different sources. We stress that this approach becomes infeasible in our setting where $M=1$ as the error terms then sum up to a single error term and ICA would treat this as one observation. Other matrix-factorization methods assume a certain sparsity of the mixing-matrix \cite{spielman}. Similar to NMF methods, in general all these methods, however, again rely on the assumption that $M>1$ (most of them even require $M \geq m$) as otherwise the signal is not even identifiable, in contrast to our situation again due to the finite alphabet. 

Minimization of the $\ell_0$ norm using dynamic programming (which has a long history in c.p. analysis, see e.g., \cite{bai1998, fearnhead, friedrich, killick}) for segment estimation under a multiscale constraint has been introduced in \cite{boysen} (see also \cite{davies2012} and \cite{frick}) and here we extend this to mixtures of segment signals and in particular to a finite alphabet restriction. 

The SBSSR problem becomes tractable as we assume that our signals occur with sufficiently many alphabet combinations which may be present already on small scales on the one hand, and on the other hand we also observe  long enough segments (large scales) in order to estimate reliably the corresponding mixing weights on these (see the identifiability condition in (\ref{separable})). Both assumptions seem to be satisfied in our motivating application, the separation of clonal copy numbers in a tumor. 

To best of our knowledge, the way we treat the problem of clonal separation is new, see, however, \cite{yau,carter,liu,roth,titan,ding}. Methods suggested there, all rely on specific prior information about the sources $f$ and cannot be applied to the general SBSSR-model. Moreover, most of them treat the problem from a Bayesian perspective.

\section{Method and theory}  \label{sec: theory}

\subsection{Confidence region for the weights} \label{subsec: CO}
Let $Y$ and $g = \omega^{\top}f \in \cM^\delta$ be as in the SBSSR-model (\ref{def:Y}).
Our starting point for the recovery of the weights $\omega$ and the sources $f$ is the construction of proper confidence sets for $\omega$ which is also of statistical relevance by its own as the source functions are unknown which hinders direct inversion of a confidence set for $g$.

Consider the system of boxes $\fB=\{B(i,j): 1\leq i \leq j \leq n\}$ from (\ref{def: box}) with $q = q_n(\alpha)$ as in (\ref{def: qalpha}) for some given $\alpha \in (0,1)$, as described in Section \ref{subsubsec: Tn}. 

As the underlying sources $f$ are assumed to be separable (see Definition \ref{def:separable} and (\ref{def: MA})) there exist intervals $[x_{i^{\star}_r},x_{j^{\star}_r}] \subset (0,1]$, for $r = 1,\ldots,m$, such that
\begin{equation}\label{iStar}
f|_{[x_{i^{\star}_r},x_{j^{\star}_r}]}\equiv [A]_r,
\end{equation}
with $A$ as in (\ref{Amatrix}). 
Assume for the moment that these intervals would be known and let $B^{\star} \ZuWeis B(i^{\star}_1,j^{\star}_1)\times \ldots \times B(i^{\star}_m,j^{\star}_m) \in \fB^m$ be the corresponding $m$-box.
Then a $1-\alpha$ confidence region for $\omega$ is given as
\begin{equation}\label{ciomegacan}
\cC_{1-\alpha} (i^{\star}_1,j^{\star}_1,...,i^{\star}_m,j^{\star}_m) \ZuWeis A^{-1}B^{\star}.
\end{equation}

To see that (\ref{ciomegacan}) is, indeed, a $1-\alpha$ confidence region for $\omega$, note that
\begin{equation*}
\{\omega \in \cC_{1-\alpha} (i^{\star}_1,j^{\star}_1,...,i^{\star}_m,j^{\star}_m)\} \supset
\bigcap_{1 \leq r \leq m}\{g|_{[x_{i^{\star}_{r}}, x_{j^{\star}_{r}}]} \equiv \omega^{\top}[A]_r \in B(i^{\star}_{r},j^{\star}_{r})\}
\end{equation*}
and
\begin{equation*}
\{T_n (Y,g)\leq q_n(\alpha)\} = \bigcap_{\substack{1 \leq i \leq j \leq n \\ g|_{[x_i,x_j]}\equiv g_{ij}}}\{g_{ij} \in B(i,j)\}.
\end{equation*}
This implies that 
\begin{equation}\label{ciStarS}
\{\omega \in \cC_{1-\alpha} (i^{\star}_1,j^{\star}_1,...,i^{\star}_m,j^{\star}_m)\} \supset \{T_n (Y,g)\leq q_n(\alpha)\}
\end{equation}
and therefore it holds uniformly in $g\in \cM^\delta$ that
\begin{equation}\label{ciStar}
\Pp(\omega \in \cC_{1-\alpha} (i^{\star}_1,j^{\star}_1,...,i^{\star}_m,j^{\star}_m)) \geq \Pp(T_n (Y,g)\leq q_n(\alpha)) \geq 1 - \alpha.
\end{equation}
Of course, as the source functions $f$ are unknown, intervals $[x_{i^{\star}_r},x_{j^{\star}_r}]$ which satisfy (\ref{iStar}) are not available immediately and thus, one cannot construct the $m$-box $B^{\star}$ required for (\ref{ciomegacan}) directly.

For this reason, we will describe a strategy to obtain a sub-system of $m$-boxes, that is, a subset $\fB ^{\star}\subset\fB^m$, which covers $B^{\star}$ conditioned on $\{T_n (Y,g)\leq q_n(\alpha)\}$ almost surely. 
To this end, observe that for any random set $\cC^{\star}(Y) \subset \R^m$ with
\begin{equation}\label{condTn}
\Pp\left(\cC^{\star}(Y) \supset \cC_{1-\alpha} (i^{\star}_1,j^{\star}_1,\ldots,i^{\star}_m,j^{\star}_m) \middle| T_n (Y,g)\leq q_n(\alpha)\right) = 1
\end{equation}
(\ref{ciStarS}) and (\ref{ciStar}) imply $\Pp(\omega \in \cC^{\star}(Y)) \geq 1 - \alpha$.
We then define $\cC_{1-\alpha}$ as in (\ref{Comega}). 
To this end, $\fB^{\star}$ is constructed such that the diameter of the resulting $\cC_{1-\alpha}$ is of order $\ln(n)/\sqrt{n}$ (see Corollary \ref{cor: omegaconsi}). 
The construction will be done explicitly by an algorithm which relies on the application of certain reduction rules to $\fB^m$ to be described in the following.

Let $\abbr{\proj_r}{\fB^m}{\fB}$, for $r = 1, \ldots, m$, denote the $r$-th projection (i.e., $\proj_r(B_1 \times \ldots \times B_m) \ZuWeis B_r$) and define the set of boxes on which any signal fulfilling the multiscale constraint is non constant (nc) as
\begin{equation}\label{def: bnc}
\fB_{\text{nc}} \ZuWeis \{B(i,j) \in \fB: \exists [s,t], [u,v] \subset [i,j] \text{ with }B(s,t)\cap B(u,v)=\emptyset \}.
\end{equation}

\begin{rr}\label{notconst}
Delete $B \in \fB^m$ if there exists an $r \in \{1,\ldots,m\}$ such that $B(i,j) \ZuWeis \proj_r(B) \in \fB_{\text{nc}}$ as in (\ref{def: bnc}).
\end{rr}

The reasoning behind R\ref{notconst} is as follows. $g|_{[x_{i_r^{\star}},x_{j_r^{\star}}]}$ is constant for $r=1,\ldots,m$ as $f^1,\ldots,f^m$ are constant on $[x_{i_r^{\star}},x_{j_r^{\star}}]$. Consequently, all $m$-boxes that include a box $B(i,j)\in \fB$ such that $g$ cannot be constant on $[x_i,x_j]$ (conditioned on $T_n (Y,g)\leq q_n(\alpha)$) can be deleted in order to preserve coverage of $B^{\star}$. 
Let $[x_i,x_j]$ be an interval on which $g$ is constant (say $g|_{[x_i,x_j]}\equiv c$) and assume that there exist intervals $[s,t], [u,v] \subset [i,j]$ such that $B(s,t)\cap B(u,v)=\emptyset $. Then by construction of the boxes $B(s,t)$ and $B(u,v)$, $T_n(Y,g)\leq q_n(\alpha)$ implies that $c\in B(s,t)$ and $c\in B(u,v)$, which contradicts $B(s,t)\cap B(u,v)=\emptyset $. 
In other words, $\fB_{\text{nc}}$ (nc $\hat{=}$ non constant ) in (\ref{def: bnc}) includes all boxes $B(i,j)$ such that all function $\tilde{g}\in\cM^\delta$ which fulfill the multiscale constraint $T_n(Y,\tilde{g}) \leq q_n(\alpha)$ cannot be constant on $[x_i,x_j]$. 
Note that, in contrast to the following two reduction rules, the reduction rule R\ref{notconst} does not depend on the specific matrix $A$ in the identifiablity condition in (\ref{separable}).

\begin{rr}\label{validred}
Delete $B \in \fB^m$, with $[\underline{b}_r,\overline{b}_r] \ZuWeis \proj_r(B)$ if at least one of the following statements holds true
\begin{enumerate}
\item $\overline{b}_1 \leq a_1$ or $\underline{b}_1 \geq a_1 + \frac{a_2- a_1}{m}$, \label{val1}
\item for any $2\leq r \leq m$ 
\begin{equation*}
\frac{a_2 + (m-1)a_1 - \sum_{j=1}^{r-1} \underline{b}_j}{m-r+1} \leq \underline{b}_{r} \quad \text{or} \quad \underline{b}_{r-1} \geq \overline{b}_{r},
\end{equation*}
 \label{val2}
\item $\sum_{j = 1}^m \overline{b}_j \leq a_2 + (m-1)a_1$.
\label{val3}
\end{enumerate}
\end{rr}

R\ref{validred} \ref{val1}. comes from the fact that $0 < \omega_1 < 1/m$, R\ref{validred} \ref{val2}. from $\omega_{i-1} < \omega_i < (1- \sum_{j = 1}^{i-1}\omega_j)/(m-i+1)$, and R\ref{validred} \ref{val3}. from $\sum_{j = 1}^{m}\omega_j = 1 $, together with the specific choice of the matrix $A$ in (\ref{Amatrix}). For a different choice of $A$ in (\ref{separable}) the equations in R\ref{validred} can be modified accordingly.

In what follows, define for $k = 1,\ldots,n$
\begin{equation}\label{Ik}
\cJ_k \ZuWeis \{[i,j] : k\in[i,j] \text{ and } B(i,j) \not\in \fB_{\text{nc}} \}.
\end{equation}

\begin{rr}\label{exred}
Delete $B \in \fB^m$, if there exists a $k\in\{1,\ldots,n\}$ such that for all $[i,j] \in \cJ_k$ 
\begin{equation}\label{notex}
\Big[\max_{i \leq u \leq v \leq j} \underline{b}_{uv}, \min_{i\leq u \leq v \leq j} \overline{b}_{uv}\Big]\cap \Big\{\tilde{\omega}^{\top}a: a\in \fA^m\ \text{ and }\tilde{\omega} \in A^{-1} B \Big\}
\end{equation}
is empty, with $ [ \underline{b}_{uv},  \overline{b}_{uv}] \ZuWeis B(u,v) \in \fB$.
\end{rr}
 
Conditioning on $T_n(Y,g)\leq q_n(\alpha)$ implies $\omega \in A^{-1} B^{\star}$, and, in particular, that there exists an $\tilde{\omega} \in  A^{-1} B^{\star}$ such that $\imag (g) \ZuWeis \{ g(x_1),\ldots,g(x_n) \} \subset \{\tilde{\omega}^{\top}a: a\in \fA^m\}.$
Moreover, for every $k\in\{1,\ldots,n\}$ there exists an interval $[x_i,x_j]$ where $g$ is constant with $g|_{[x_i,x_j]}\equiv g(x_k)\in \imag(g)$. So, $T_n(Y,g)\leq q_n(\alpha)$ implies $g(x_k)\in B(u,v)$ for all $[u,v]\subset[i,j]$ and, therefore, for $B = B^\star$ (\ref{notex}) is not empty (conditioned on $T_n (Y,g)\leq q_n(\alpha)$).

\begin{remarkmn}{Incorporating prior knowledge on minimal scales}\label{rem:prior}
\quad 

\begin{enumerate}[a)]
\item If we restrict to a minimal scale $\lambda \in (0,1)$ on which a jump of $g$ may occur, that is, for $\tau_j$, $j= 0,\ldots,K+1$, being the c.p.'s of $g$
\begin{equation}\label{lambda}
\lambda \ZuWeis \min_{j \in \{0,\ldots,K\}}\abs{\tau_{j+1}-\tau_j} > 0,
\end{equation}
we can modify R\ref{exred} with $\cJ_k$ in (\ref{Ik}) replaced by $\cJ_k \cap \{[i,j]: j-i + 1 \geq n\lambda\}$.

\item In many applications (see Section \ref{sec:appgen}), it is very reasonable to assume apriori knowledge of a minimal interval length $\lambda^{\star}$ of $[x_{i_r^{\star}}, x_{j_r^{\star}}]$ in (\ref{iStar}). This means that there exists some interval $I_r \subset [0,1)$ of minimum size $\lambda^{\star}$, where $(f^1,\ldots,f^m)$ take the value $[A]_r$ as in (\ref{Amatrix}) for $r=1,\ldots,m$. This is summarized in the following reduction rule.

\begin{rr}\label{minred}
Knowing that  $j^{\star}_r-i^{\star}_r+1 \geq \lambda^{\star} n$ for $r=1,\ldots,m$ in (\ref{iStar}), delete $B \in \fB^m$ if there exists an $r\in\{1,\ldots,m\}$ such that for $B(i,j)\ZuWeis \proj_r(B)$ $j-i+1< \lambda^{\star} n$.
\end{rr}

\end{enumerate}
\end{remarkmn}

R\ref{notconst} - R\ref{minred} is summarized in Algorithm CRW, in Section \ref{subsec: psCRW} in the supplement, for constructing a confidence region for $\omega$.

\begin{remarkmn}{Noninformative $m$-box}\label{rem: bs}
If $\fB^{\star} = \emptyset$, we formally may set $\cC_{1-\alpha}\ZuWeis \Omega(m)$, the trivial confidence region. As
$\{\fB^{\star} = \emptyset\} \subset \{T_n(Y,g) > q_n(\alpha)\}$,
the probability that this happens can be bounded from above by $\alpha$. This is in general only a very rough bound, simulations show that $\fB^{\star} = \emptyset$ is hardly ever the case when $\alpha$ is reasonably small. For instance, in $10,000$ simulations of Example \ref{example1} with $n = 1280$, $\sigma = 0.1$, $\alpha = 0.1$ it did not happen once. Of course, when $\alpha \nearrow 1$, $\fB^{\star} = \emptyset$ finally, as no mixture $g\in \cM^\delta$ can fulfill the multiscale constrained $T_n(Y,g) \leq q$ for arbitrarily small $q$.
\end{remarkmn}

\begin{remarkmn}{Shape of $\cC_{1-\alpha}$}
The previous construction of the confidence set $\cC_{1-\alpha}$ does not ensure that the confidence set is of $m$-box form
\begin{equation}\label{spcio}
[\underline{\omega}_1,\overline{\omega}_1]\times \ldots \times [\underline{\omega}_m,\overline{\omega}_m].
\end{equation}
In general it is a union of $m$-boxes.
However, we can always take the smallest covering $m$-box of $\cC_{1-\alpha}$, given by
\begin{equation}\label{scmb}
[\inf_{\tilde{\omega} \in \cC_{1 - \alpha}} \tilde{\omega}_1, \sup_{\tilde{\omega} \in \cC_{1 - \alpha}} \tilde{\omega}_1 ]\times \ldots \times [\inf_{\tilde{\omega} \in \cC_{1 - \alpha}} \tilde{\omega}_m, \sup_{\tilde{\omega} \in \cC_{1 - \alpha}} \tilde{\omega}_m ],
\end{equation}
in order to get a confidence set as in (\ref{spcio}). 
Note, that $\dist(\omega,\cC_{1-\alpha}) \WeisZu d$ remains the same when we replace $\cC_{1-\alpha}$ by (\ref{scmb}). To see this, consider $\hat{\cC} \ZuWeis \omega +  [- d , d]^m$, which is a covering $m$-box of $\cC_{1 - \alpha}$, so in particular $\hat{\cC}$ covers (\ref{scmb}), with $\dist(\omega,\hat{\cC}) = d$.
\end{remarkmn}

Summing up, we have now constructed a confidence set $\cC_{1-\alpha}$ for the mixing vector $\omega$ in the SBSSR-model.
Given $\cC_{1-\alpha}$ SLAM estimates $\omega$ as in (\ref{omes}). From this, in the next section we derive estimators for the sources $f^1,\ldots,f^m$. 

\subsection{Estimation of source functions} \label{subsec: estfF}
SLAM estimates $f = (f^1, \ldots, f^m)$ by solving the constraint optimization problem (\ref{def: gEstS}), which admits a solution if and only if
\begin{equation}\label{cHne}
\min_{\tilde{f}\in \cS(\fA)^m} T_n(Y,\hat{\omega}(\alpha)^{\top}\tilde{f}) \leq q_n(\beta).
\end{equation}
(\ref{cHne}) cannot be guaranteed in general but it can be shown that it holds asymptotically with probability one (see Theorem \ref{lem: H} in the supplement), independently of the specific choice of $\hat{\omega} \in \cC_{1-\alpha}(Y)$ in (\ref{omes}). For finite $n$ our simulations show that violation of (\ref{cHne}) is hardly ever the case. For instance, in $10,000$ simulation runs of Example \ref{example1} with $\alpha = \beta = 0.1$ it did not happen once. Therefore, in practice, failure of (\ref{cHne}) might rather indicate that the model assumption is not correct (e.g., due to outliers) and could be treated by pre-processing of the data. Another strategy can be to decrease $\beta$ and hence the constraint in (\ref{cHne}) as for $\beta> \beta^{\prime}$ it holds that $q_n(\beta^{\prime}) > q_n(\beta)$. 

\begin{remarkmn}{Incorporating identifiability conditions in SLAM}
The separability condition in (\ref{separable}) could be incorporated in the estimator (\ref{def: gEstS}), which provides a further restriction on $\cH(\beta)$ in (\ref{Hn}). This may yield a finite sample improvement of SLAM, however, at the expense of being less robust if such a particular identifiability condition is violated (see Section \ref{subsec: simASB} for a simulation study of SLAM when the identifiability conditions in (\ref{def: OmegaD}) and (\ref{separable}) are violated). 
\end{remarkmn}

\subsection{Confidence bands for the source functions} \label{subsec: estfC}
Obviously, uniform confidence sets for $f$ cannot be obtained if we allow for an arbitrarily small distance between two c.p.'s of $g$ (as for any c.p. problem, see \cite{frick}). However, if we restrict to a minimal scale $\lambda$ as in (\ref{lambda}), the SLAM estimation procedure in (\ref{def: gEstS}) leads to asymptotically uniform confidence bands for the source functions $f^1,\ldots,f^m$.
To this end, we introduce
\begin{equation}\label{MAlam}
\cM^{\delta}_\lambda \ZuWeis  \Big\{g \in \cM^\delta : \min_{j \in \{0,\ldots,K(g)\}}\abs{\tau_{j+1}-\tau_j} \geq \lambda \Big \},
\end{equation}
where, as in (\ref{def:SA}), $\tau_0 = 0 < \tau_1 < \ldots < \tau_{K(g)} <  \tau_{K(g) + 1} = 1$ denote c.p.'s of $g$.
Moreover, let $\tilde{T}_n$ be as in (\ref{def: Tn}), but with $pen(j-i+1)$ replaced by $pen(j-i+1) + \left((a_2 - a_1) \ln(n)/m + \sqrt{8\sigma^2 \ln(e/\lambda)/\lambda}\right)\sqrt{(j-i+1)/n}$ and let $\tilde{\cH}(\beta)$ be as in (\ref{Hn}) but with $T_n$ replaced by $\tilde{T}_n$. Then $\tilde{\cH}(\beta)$ constitutes an asymptotically uniform confidence band as the following theorem shows.

\begin{theo}\label{theo: CIf}
Consider the SBSSR-model and
let $\hat{\omega}$ be the SLAM estimator from (\ref{omes}) for $\alpha = \alpha_n$ as in (\ref{alphaN}). Then, for $\tilde{\cH}(\beta)$ as in (\ref{Hn}) with $T_n$ replaced by $\tilde{T}_n$, $\tilde{\cH}(\beta)$ provides an asymptotically uniform confidence region for the sources $f$,
\begin{equation*}
\lim_{n\longrightarrow \infty}\inf_{g \in \cM^\delta_\lambda}\Pp((f^1,...,f^m)\in \tilde{\cH}(\beta))\geq 1-\beta.
\end{equation*}
\end{theo}

For a proof see Section \ref{subsec: app2} in the supplement.

\subsection{Consistency and rates}\label{subsec: consi}

In the following, we investigate further theoretical properties of SLAM. As in Theorem \ref{theo: CIf} our results will be stated uniformly over the space $\cM^\delta_\lambda$ in (\ref{MAlam}), that is, for a given minimal length $\lambda$ of the constant parts of the mixture $g$ and a given minimal ASB $\delta$ as in (\ref{def: ASB}). 
Define the constants 
\begin{equation}\label{c0c1c2}
c_1 = \frac{\delta^2(a_2 - a_1)^2}{48600 \sigma^2 m^2 (a_k - a_1)^2}, 
\quad c_2 = \frac{\delta + \sqrt{2\sigma^2 \ln(e/\lambda)}}{\sqrt{\lambda}}.
\end{equation}

Further, let $N^{\star} \in \N$ be the smallest integer, s.t.
\begin{align}
\sqrt{\frac{2\ln\left( e N^{\star} / \ln^2(N^{\star}) \right)}{\ln^2(N^{\star})}} + \frac{\sqrt{6\ln(3 e / \lambda)}}{\sqrt{N^{\star} \lambda}} &\leq  \frac{\delta}{4\sigma}, \quad \text{ and }\label{Nstar2}\\
\frac{\ln(N^{\star})}{\sqrt{N^{\star} \lambda}} &\leq  \frac{\delta (a_2 - a_1)/ (a_k - a_1) }{2m(\delta + \sqrt{2 \sigma^2 \ln(e/\lambda)}) }. \label{Nstar1}
\end{align}

\begin{remarkmn}{Behavior of $N^{\star}$}\label{rem: Nstar}
Note that the left-hand side in (\ref{Nstar2}) and (\ref{Nstar1}) is decreasing in $N^\star$, respectively.
For fixed $\lambda$ and $\delta/\sigma \searrow 0$, (\ref{Nstar2}) dominates the behavior of $N^{\star}$ as it is essentially of the form $\sigma / \delta \leq c(\lambda) \sqrt{\ln(N^{\star})}$, whereas (\ref{Nstar1}) is of the form $\sigma / \delta \leq c(\lambda, \fA, m) \sqrt{N^{\star}} / \ln(N^{\star})$.
Conversely, for fixed $\delta/\sigma$ and $\lambda \searrow 0$, (\ref{Nstar1}) dominates the behavior of $N^{\star}$ as it is essentially of the form $\lambda^{-1}\ln(\lambda^{-1}) \leq c(\delta / \sigma, \fA, m) N^{\star} / \ln^2(N^{\star})$ whereas (\ref{Nstar2}) is of the form $\lambda^{-1}\ln(\lambda^{-1}) \leq c(\delta / \sigma) N^{\star}$.
\end{remarkmn}

\begin{theo}\label{consimain}
Consider the SBSSR-model with $g \in \cM^\delta_\lambda$.
Let $\hat{\omega}$ and $\hat{f} = (\hat{f}^1,\ldots,\hat{f}^m)$ be the SLAM estimators from (\ref{omes}) and (\ref{def: gEstS}), respectively, with $\alpha = \alpha_n$ and $\beta = \beta_n$ as in (\ref{alphaN}). Further, let $\hat{\tau}^i$ and $\tau^i$ be the vectors of all c.p. locations of $\hat{f}^i$ and $f^i$, respectively, for $i = 1,\ldots,m$.
Then for all $n>N^{\star}$ in (\ref{Nstar2}) and (\ref{Nstar1}) and for all $i = 1,\ldots,m$
\begin{enumerate}
\item $K(\hat{f}^i) = K(f^i)$ ,
\item $\max_{j}|\hat{\tau}^i_j-\tau^i_j|\leq 2 \frac{\ln^2(n)}{n}$,
\item $\max_{j}\abs{\hat{f^i}|_{[\hat{\tau}_j,\hat{\tau}_{j+1})}-f^i|_{[\tau_j,\tau_{j+1})}}=0$, \text{ and}
\item $\abs{\hat{\omega}_i - \omega_i } \leq \frac{c_2}{a_2 - a_1}\frac{\ln(n)}{\sqrt{n}}$
\end{enumerate}
with probability at least $1 - \exp(- c_1 \ln^2(n))$, with $c_1$ and $c_2$ as in (\ref{c0c1c2}).
\end{theo}
From the proof of Theorem \ref{consimain} (see Section \ref{subsec: app1} in the supplement) it also follows that assertions 1. - 4. hold for any $\hat{\omega} \in \cC_{1-\alpha}(Y)$ and we obtain the following.
\begin{coron}\label{cor: omegaconsi}
Consider the SBSSR-model with $g \in \cM^\delta_\lambda$. Let $\cC_{1-\alpha}(Y)$ be as in (\ref{Comega}) and $\alpha_n$ as in (\ref{alphaN}). Further, let $\dist$ be is as in (\ref{def: dist}). Then for all $n > N^{\star}$ in (\ref{Nstar2}) and (\ref{Nstar1})
\begin{equation*}
\dist(\omega, \cC_{1-\alpha_n}(Y)) < \frac{c_2}{a_2 - a_1}\frac{\ln(n)}{\sqrt{n}}
\end{equation*}
with probability at least $1 - \exp(- c_1 \ln^2(n))$, with $c_1$ and $c_2$ as in (\ref{c0c1c2}).
\end{coron}

\begin{remarkmn}{SLAM (almost) attains minimax rates}\label{remarkOptRates}
\quad

\begin{enumerate}[a)]

\item (C.p. locations) Theorem \ref{consimain} states that we can recover the c.p. locations of $f^i$ in probability with rate $\ln^2(n)/n$.
Obviously, the estimation rate of the c.p. locations is bounded from below by the sampling rate $1/n$. Consequently, the rate of Theorem \ref{consimain} differs from the optimal rate only by a $\ln^2(n)$ factor. 

\item (Weights) By the one-to-one correspondence between the weights and the function values of $g$ the weights' detection rate $\ln(n)/\sqrt{n}$ immediately follows from the box height in (\ref{def: box}) with $q_n (\alpha_n)\in \mathcal{O}(\ln(n))$ and coincides with the optimal rate $\mathcal{O}(1/\sqrt{n})$ up to a $\ln(n)$ term.

\item (Dependence on $\lambda$) The minimal scale $\lambda$ in Theorem \ref{consimain} may depend on $n$, i.e., $\lambda = \lambda_n$. In order to ensure consistency of SLAM's estimates $\hat{\omega}$ and $(\hat{f}^1,\ldots,\hat{f}^m)$, Theorem \ref{consimain} requires that (\ref{Nstar2}) and (\ref{Nstar1}) holds (for a sufficiently large $N^{\star}$) and that $c_2 \ln(n)/\sqrt{n}\rightarrow 0$, as $n\rightarrow \infty$. By Remark \ref{rem: Nstar} this is fulfilled whenever $\lambda^{-1} \ln(\lambda^{-1}) \in \co(n / \ln^2(n))$.
This means that the statements 1. - 4. in Theorem \ref{consimain} hold true asymptotically with probability one as the minimal scale $\lambda_n$ of successive jumps in a sequence of mixtures $g_n$ does not asymptotically vanish as fast as of order $\ln^{3}(n) / n$. We stress that no method can recover finer details of a bump signal (including the mixture $g$) below its detection boundary which is of the order $\ln(n)/n$, that is, SLAM achieves this minimax detection rate up to a $\ln^2(n)$ factor (see \cite{dumbgen2008,frick}). \label{rem: lam}

\item (Dependence on $\delta$) Just as the minimal scale $\lambda$, the minimal ASB $\delta$ in Theorem \ref{consimain} may depend on $n$ as well, that is, $\delta = \delta_n$. Analog to \ref{rem: lam}), the SLAM's estimates remain consistent whenever $\delta^{-1} \in \co\left(\sqrt{\ln(n)}\right)$, that is, the statements 1. - 4. in Theorem \ref{consimain} hold true asymptotically with probability one if the minimal ASB $\delta_n$ in a sequence of mixtures $g_n$ does not decrease as fast as of order $1/\sqrt{\ln(n)}$. 
We stress that no method can recover smaller jump heights of the mixture $g$ below its minimax detection rate, which in $1/\ln(n)$. To see this, note that statement 2. in Theorem \ref{consimain} provides asymptotic detection power one for $2\ln(n)^2$ i.i.d. observations with mean $\delta_n$ (recall that the ASB corresponds to the minimal possible jump height of the mixture $g$). Hence, SLAM achieves the minimax rate up to a $\sqrt{\ln(n)}$ factor.

\end{enumerate}
\end{remarkmn}

\begin{remarkmn}{SLAM for known $\omega$}\label{rem: knownO}
If $\omega$ is known in the SBSSR-model, the second part of SLAM can be used separately. We may then directly solve (\ref{Khat}) without pre-estimating $\omega$, that is, in Section \ref{subsec: intSources}, we simply replace $ \hat \omega$ by $\omega$.
Then, Theorem \ref{theo: CIf} is still valid for $\tilde{\cH}(\beta)$ replaced by $\cH(\beta)$. Further, a careful modification of the proof of Theorem \ref{consimain} shows that the assertions in Theorem \ref{consimain} hold for a possibly smaller $N^{\star}$ in (\ref{Nstar2}) and (\ref{Nstar1}) and for $c_1$ replaced by $75 m^2 (a_k - a_1)^2 c_1 / (a_2 - a_1)^2$.
We stress that the finite alphabet assumption  is still required and the corresponding identifiability assumption $ASB(\omega) \geq \delta$ must be valid.
\end{remarkmn}

\section{Computational issues}\label{subsec: Imple}

SLAM is implemented in two steps. 
In the first step, for a given $\alpha \in (0,1)$ a confidence region for the mixing weights $\omega$ is computed as in Algorithm CRW (see Section \ref{subsec: CO} and \ref{subsec: psCRW}). To this end, each of the $n^{2m}$ $m$-boxes in $\fB^m=\{B(i,j): 1\leq i \leq j \leq n\}^m$ needs to be examined with the reduction rules R1 - R4 for validity as a candidate box for the intervals $[i^\star_1, j^\star_1]\times \ldots \times [i^\star_m, j^\star_m]$, which yields the complexity $\cO(n^{2m})$. There are, however, important pruning steps, which can lead to a considerably smaller complexity.

First, note that it suffices to consider $m$-boxes which are maximal elements with respect to the partial order of inclusion, that is, for $B^1 = [\underline{b}_1^1, \overline{b}_1^1]\times \ldots \times [\underline{b}_m^1, \overline{b}_m^1]$, $B^2 = [\underline{b}_1^2, \overline{b}_1^2]\times \ldots \times [\underline{b}_m^2, \overline{b}_m^2] \in \fB^m$
\begin{equation*}
B^1 \preccurlyeq B^2 \Leftrightarrow [\underline{b}_i^1, \overline{b}_i^1] \subseteq [\underline{b}_i^2, \overline{b}_i^2] \quad\text{for all }i = 1,\ldots,m,
\end{equation*}
where an element $a$ of a partially ordered set $P$ is maximal if there is no element $b$ in $P$ such that $b > a$.
To see this, assume that an $m$-box $B$ is not deleted by the reduction rule R\ref{exred} in the second last line of Algorithm CRW, then an $m$-box $B^{\prime} \in \fB^{m}$ with $B^{\prime} \prec B$ does not influence the confidence region $\cC_{1-\alpha}$ (see last line of Algorithm CRW), as $A^{-1}B^{\prime}\subset A^{-1}B$. Conversely, if an $m$-box $B$ is deleted by the reduction rule R\ref{exred} in the second last line of Algorithm CRW, then an $m$-box $B^{\prime} \in \fB^{m}$ with $B^{\prime} \prec B$ will be deleted by R\ref{exred} as well, such that $B^{\prime}$ does not need to be considered either.

Second, note that the parameter $\omega$ which is inferred in Algorithm CRW is global and hence, one can restrict to observations on a subinterval $[x_i,x_j] \subset [0,1)$ as long as $g|_{[x_i, x_j]}$ fulfills the identifiability conditions of $\cM^\delta$. 

The explicit complexity of Algorithm CRW depends on the finial solution $\hat{f}$ itself. Depending on the final $\hat{f}$, the above mentioned pruning steps yield a complexity between $\cO(n^m)$ and $\cO(n^{2m})$. $\hat{\omega}$ is then computed as in (\ref{omes}).

In the second step, for a given $\beta \in (0,1)$ and given $\hat{\omega}$ SLAM solves the constrained optimization problem (\ref{def: gEstS}), which can be done using dynamic programming. 
Frick et al. \cite{frick} provide a pruned dynamic programming algorithm to efficiently solve a one-dimensional version of (\ref{def: gEstS}) without the finite alphabet restriction in (\ref{restH}). As this restriction is crucial for SLAM we outline the details of the necessary modifications in Section \ref{subsec: comFDP} in the supplement. These modifications, however, do not change to complexity of the algorithm. Frick et al. \cite{frick} show that the overall complexity of the dynamic program depends on the final solution $\hat{g}$ and is between $\cO(n)$ and $\cO(n^2)$.

We stress finally that significant speed up (which is, however, not the subject of this paper) can be achieved by restricting the system of intervals in $T_n$ and $\fB$, respectively, to a smaller subsystem, for example, intervals of dyadic length,  which for example reveals the complexity of the second step as $\cO(n\ln(n))$.


\section{Simulations}\label{subsec: Appsim}

In the following we investigate empirically the influence of all parameters and the underlying signal on the performance of the SLAM estimator.
As performance measures we use the mean absolute error, $\mae$, for $\hat{\omega}$ and the mean absolute integrated error, $\miae$, for $\hat{f}$. Further, we report the centered mean, $\meanNcp - K$, the centered median, $\medNcp - K$, of the number of c.p.'s of $\hat{f}$, the frequency of correctly estimated number of c.p.'s for the single source functions $f^i$, $\meanKK_i$, and for the whole source function vector $f$, $\meanKK$. To investigate the accuracy of the c.p. locations of the single estimated source functions $\hat{f}^1, \ldots, \hat{f}^m$ we report the mean of $\max_i \min_j \abs{\tau_i - \hat{\tau}_j}$ and $\max_j \min_i \abs{\tau_i - \hat{\tau}_j}$, where $\tau$ and $\hat{\tau}$ denotes the vector of c.p. locations of the true signal and the estimate, respectively. Furthermore, we report common segmentation evaluation measures for the single estimated source functions $\hat{f}^1, \ldots, \hat{f}^m$, namely the entropy-based $V$-measure, $\vm $, with balancing parameter $1$ of \cite{rosenberg} and the false positive sensitive location error, $\fpsle$, and the false negative sensitive location error, $\fnsle$, of \cite{futschik}.
The $V$-measure, taking values in $[0,1]$, measures whether given clusters include the correct data points of the corresponding class. 
Larger values indicate higher accuracy, $1$ corresponding to a perfect segmentation.
The $\fpsle$ and the $\fnsle$ capture the average distance between true and estimated segmentation boundaries, with FPSLE being larger if a spurious
split is included, while FNSLE getting larger if a true boundary is
not detected (see \cite{futschik} for details).
To investigate the performance of the confidence region $\cC_{1-\alpha}$ for $\omega$, we use $\dist(\omega,\cC_{1-\alpha})$ from (\ref{def: dist}), the mean coverage $\mean(\omega \in \cC_{1 - \alpha})$, and the diameters $\overline{\omega}_i - \underline{\omega}_i$, where $\cC_{1 - \alpha} = [\overline{\omega}_1 , \underline{\omega}_1] \times \ldots \times [\overline{\omega}_m , \underline{\omega}_m]$. Further, we report the mean coverage of the confidence band $\tilde{\cH}(\beta)$, i.e. $\mean(f \in \tilde{\cH}(\beta))$.
In order to reduce computation time, we only considered intervals of dyadic length as explained in Section \ref{subsec: Imple}, possibly at expense of detection power. Simulation runs were always $10,000$.
\subsection{Number of source functions $m$}

In order to illustrate the influence of the number of source functions $m$ on the performance of SLAM we vary $m = 2,\ldots,5 $ while keeping the other parameters in the SBSSR- model fixed.

We investigate a binary alphabet $\fA = \{0,1\}$ and set  $f^i = \indE_{[(i-1)/5, i/5)}$ for $i = 1,\ldots,5$, simple bump functions. For each $m \in \{2,3,4,5\}$, we choose $\omega$ such that $ASB(\omega) = 0.02$ in (\ref{def: ASB}) (see Table \ref{tab: omegaM} in the supplement). For $\sigma = \delta = 0.02$, $n = 1000$, and $\alpha = \beta = 0.1$, we compute $\hat{\omega}$, $\cC_{0.9}$, $\hat{f}^1,\ldots, \hat{f}^m$, and $\tilde{\cH}(0.1)$ for each $m \in \{2,3,4,5\}$, incorporating prior knowledge $\lambda \geq 0.025$ (see (\ref{lambda}) and Remark \ref{rem:prior}) (with truth $\lambda = 0.05$). The results are displayed in Table \ref{tab: depM}. 
A major finding is that as the number of possible mixture values equals $k^m$, the complexity of the SBSSR-model grows exponentially in $m$ such that demixing becomes substantially more difficult with increasing $m$. 

\subsection{Number of alphabet values $k$}\label{subsec: depK}

To illustrate the influence of the number of alphabet values $k$, we consider three different alphabets $\fA_k = \{0,\ldots,k\}$ for $k = 2,3,4$.
For $m = 2$, we set 
\begin{equation}\label{fk}
f^1 = \sum_{i = 0}^{15} \left(i \mod k\right) \indE_{[i, i+1)/16},\quad
f^2 = \sum_{i = 0}^{[15/k]} \left(i \mod k \right) \indE_{k[i, i+1)/16},
\end{equation}
step functions taking successively every alphabet value in $\fA^2$ (see Figure \ref{fig: fk} in the supplement).
Further, we set $\omega = (0.02, 0.98)$ such that $ASB(\omega) = 0.02$ for $k = 2,3,4$.
For $\sigma = 0.05$, $n = 1056$, and $\alpha = \beta = 0.1$ we compute $\hat{\omega}$, $\cC_{0.9}$, $\hat{f}^1,\ldots, \hat{f}^m$, and $\tilde{\cH}(0.1)$ for each $k = 2,3,4$, incorporating prior knowledge $\lambda \geq 1/32$ (see (\ref{lambda}) and Remark \ref{rem:prior}) (with truth $\lambda = 1/16 $). The results are displayed in Table \ref{tab: depK} in the supplement.
From this we find that an increasing $k$ does not influence SLAM's performance for $\hat{\omega}$ and $\cC_{1-\alpha}$ too much. However, the model complexity $k^m$ increases polynomially (for $m=2$ as in Table \ref{tab: depK} quadratically) in $k$, reflected in a decrease of SLAM's performance for the estimate of the source functions $\hat{f}$.

\subsection{Confidence levels $\alpha$ and $\beta$} \label{subsec: stabAB}

We illustrate the influence of the confidence levels $\alpha$ and $\beta$ on SLAM's performance with $f$ and $\omega$ as in Example \ref{example1}, that is, $m = 3$, $\fA = \{0,1,2\}$, $\omega = (0.11, 0.29, 0.6)$, and $f$ as displayed in Figure \ref{fig: example1}.
For $\sigma = 0.02, 0.05, 0.1$ and $n = 1280$, we compute $\hat{\omega}$, $\cC_{1-\alpha}$, $\hat{f}^1,\ldots, \hat{f}^m$, and $\tilde{\cH}(\beta)$ for each $(\alpha, \beta) \in \{0.01, 0.05, 0.1 \}^2$, incorporating prior knowledge $\lambda \geq 0.025$ (see (\ref{lambda}) and Remark \ref{rem:prior}) (with truth $\lambda = 0.05 $). 
Results are displayed in Table \ref{tab: stabABO} and Table \ref{tab: stabABF} in the supplement.
These illustrate that SLAM's estimate $\hat{\omega}$ for the mixing weights is very stable under the choice of $\alpha$. The diameters $\dist(\omega,\cC_{1-\alpha})$ and $\overline{\omega}_i - \underline{\omega}_i$, respectively decrease slightly with increasing $\alpha$, as expected. Further, we found that the coverage $\mean(\omega \in \cC_{1 - \alpha})$ is always bigger than the nominal coverage $1-\alpha$ indicating the conservative nature of the first inequality in (\ref{ciStar}). 
With increasing $\beta$ the multiscale constraint in (\ref{Khat}) becomes stricter leading to an increase of $\hat{K}$. However, as Table \ref{tab: stabABF} illustrates, this effect is remarkably small, resulting also in a high stability of $\hat{f}$ with respect to $\alpha$ and $\beta$. 
In contrast to the uniform coverage of the confidence region $\cC_{1 - \alpha}$ for $\omega$ for finite $n$ (recall (\ref{covComega})), this holds only asymptotically for the confidence band $\tilde{\cH}(\beta)$ (see Theorem \ref{theo: CIf}). This is reflected in Table \ref{tab: stabABF}, where with increasing $\sigma$ the coverage $\mean(f \in \tilde{\cH}(\beta))$ can be smaller than the nominal $1 - \beta$. Nevertheless, the coverage of the single source functions remains reasonably high even for large $\sigma$ (see Table \ref{tab: stabABF}).
In summary, we draw from Table \ref{tab: stabABO} and \ref{tab: stabABF} a high stability of SLAM in the tuning parameters $\alpha$ and $\beta$, for both, the estimation error and the confidence statements, respectively. 

\subsection{Prior information on the minimal scale $\lambda$}

In the previous simulations we always included prior information on the minimal scale $\lambda$ (see (\ref{lambda}) and Remark \ref{rem:prior}). In the following, we demonstrate the influence of this prior information on SLAM's performance in Example \ref{example1}, that is, $m = 3$, $\fA = \{0,1,2\}$, $\omega = (0.11, 0.29, 0.6)$, and $f$ as displayed in Figure \ref{fig: example1}.
For $\sigma = 0.02$, $n = 1280$, and $\alpha = \beta = 0.1$ we compute $\hat{\omega}$, $\cC_{0.9}$, $\hat{f}^1,\ldots, \hat{f}^m$, and $\tilde{\cH}(0.1)$ under prior knowledge 
$\lambda \geq$ $0.05$, $0.04$, $0.025$, $0.015$, $0.005$ (with truth $\lambda = 0.05 $). 
The results in Table \ref{tab: infP} in the supplement show a certain stability for a wide range of prior information on $\lambda$. Only when the prior assumptions on $\lambda$ is of order $0.1\lambda$ (or smaller) SLAM's performance gets significantly worse.

\subsection{Robustness of SLAM}\label{subsec: simRob}
Finally, we want to analyze SLAM's robustness against violations of model assumptions.

\subsubsection{Robustness against non-identifiability}\label{subsec: simASB}
Throughout this work, we assumed $g\in \cM^\delta$, that is, $\omega \in \Omega^\delta(m)$ as in (\ref{def: OmegaD}) and $f\in \cS(\fA)^m$ separable as in (\ref{separable}), in order to ensure identifiability.  
In the following, we briefly investigate SLAM's behavior if these conditions are close to be, or even violated.

\paragraph{Alphabet separation boundary $\delta$} 
We start with the identifiability condition $\omega \in \Omega^\delta(m)$, i.e., $ASB(\omega) \geq  \delta > 0$ as in (\ref{def: ASB}).
We reconsider Example \ref{example1}, that is, $m = 3$, $\fA = \{0,1,2\}$, and $f$ as displayed in Figure \ref{fig: example1}, but with $\omega$ chosen randomly, uniformly distributed on $\Omega(3)$.
For $\sigma = 0.05$, $n = 1280$, and $\alpha = \beta = 0.1$ we compute $\hat{\omega}$, $\cC_{1-\alpha}$, $\hat{f}^1,\hat{f}^2, \hat{f}^3$, and $\tilde{\cH}(\beta)$, incorporating prior knowledge $\lambda \geq 0.025$ (see (\ref{lambda}) and Remark \ref{rem:prior}) (with truth $\lambda = 0.05 $). 
Consequently, for each run we get a different $\omega$ and $ASB(\omega)$, respectively.

We found that SLAM's performance of $\hat{\omega}$ and $\cC_{1-\alpha}$, respectively, is not much influenced by $ASB(\omega)$ (see Table \ref{tab: vioIOC}, where the average mean squared error of $\hat{\omega}$ and $\dist(\omega,\cC_{1-\alpha})$ remain stable when $ASB(\omega)$ becomes small). 
The situation changes of course, when it comes to estimation of $f$ itself. $ASB(\omega) = 0$ in (\ref{def: ASB}) implies non-identifiability of $f$, that is, it is not possible to recover $f$ uniquely. Therefore, it is expected that small $ASB(\omega)$ will lead to a bad performance of any estimator of $f$. 
This is also reflected in Theorem \ref{consimain} where $\delta$, with $ASB(\omega)  \geq \delta$, appears as a ``conditioning number'' of the SBSSR-problem.
The results in Table \ref{tab: vioIFH} in the supplement confirm the strong influence of $ASB(\omega)$ on the performance of SLAM's estimate for $f$. However, as SLAM does not only give an estimate of $f$ but also a confidence band $\tilde{\cH}(\beta)$ this (unavoidable) uncertainty is also reflected in its coverage. To illustrate this define a local version of $ASB(\omega)$ as
$ASB_x(\omega) \ZuWeis \min _{a \neq f(x)\in \fA^m } \abs{\omega^{\top}a - \omega^{\top}f(x)}$.
Intuitively, $ASB_x(\omega)$ determines the difficulty to discriminate between the source functions at a certain location $x \in [0,1)$.
Now, define the local size of $\tilde{\cH}(\beta)$ as
$|\tilde{\cH}_x(\beta)| \ZuWeis \#\{a \in \fA^m: \exists f \in \tilde{\cH}(\beta) \text{ s.t. } f(x) = a\}$.
Table \ref{tab: vioIFH} in the supplement shows that the uncertainty in $|\tilde{\cH}_x(\beta)|$ increases in non-identifiable regions, that is, when $ASB_x(\omega)$ is small.

\paragraph{Violation of separability condition}
Next, we consider the separability condition in (\ref{separable}).
We consider a modification of Example \ref{example1}, that is, $m = 3$, $\fA = \{0,1,2\}$, where we modified the source function $f^1$ in such a way, that it violates the separability condition in (\ref{separable}) for $r = 1$ (see Figure \ref{fig: fvioF} in the supplement). 
For $\sigma = 0.05$, $n = 1280$, and $\alpha = \beta = 0.1$, we compute $\hat{\omega}$ and $\hat{f}^1,\hat{f}^2, \hat{f}^3$ incorporating prior knowledge $\lambda \geq 0.025$ (see (\ref{lambda}) and Remark \ref{rem:prior}) (with truth $\lambda = 0.05 $). The results are shown in Table \ref{tab: rob} in the supplement.
The violation of the separability condition in (\ref{separable}) leads to non-identifiabilty of $\omega$, which is naturally reflected in a worse performance of SLAM's estimate of $\omega$. 
As the condition is violated for $r=1$ this has a particular impact on $\hat{\omega}_1$. The performance for $\hat{\omega}_2$ and $\hat{\omega}_3$ remains relatively stable.
The same holds true for $\hat{f}$ itself, where the estimation error of $\hat{\omega}_1$ propagates to a certain degree to the estimation of $\hat{f}^1$. The performance of $\hat{f}^2$ and $\hat{f}^3$, however, is not much influenced.

\subsubsection{Violation of normality assumption}
In the SBSSR-model we assume that the error distribution is normal, that is, $\epsilon = (\epsilon_1, \ldots, \epsilon_n)^{\top} \sim \cN (0,I_n)$. In the following we study SLAM's performance for $t$-(heavy tails) and $\chi^2$-(skewed) distributed errors.
Again, we reconsider Example \ref{example1}, that is, $m = 3$, $\fA = \{0,1,2\}$, and $f$ as displayed in Figure \ref{fig: example1}. We add to $g$ now $t$-distributed and $\chi^2$-distributed errors, respectively, with $3$ degrees of freedom, re-scaled to a standard deviation of $\sigma = 0.05$.
For $n = 1280$ and $\alpha = 0.1$, we compute $\hat{\omega}$ and $\hat{f}^1,\hat{f}^2, \hat{f}^3$, incorporating prior knowledge $\lambda \geq 0.025$ (see (\ref{lambda}) and Remark \ref{rem:prior}) (with truth $\lambda = 0.05 $). 
We simulated the statistic $T_n$ for $t$- and $\chi^2$- distributed errors, respectively, and choose $q(\beta)$ to be the corresponding $90 \%$ quantile. For $t$-distributed errors this gave $q(\beta) = 13.03$ and for $\chi^2$-distributed errors $q(\beta) = 3.73$.
The results (see Table \ref{tab: rob} in the supplement) indicate a certain robustness to misspecification of the error distribution, provided the quantiles for $T_n$ are adjusted accordingly.

\subsection{Selection of $q_n(\alpha)$ and $q_n(\beta)$}\label{subsec: chQ}
On the one hand, for given $\alpha$ and $\beta$ SLAM yields confidence statements for the weights $\omega$ and the source functions $f$ at level $1 - \alpha$ and $1 - \beta$, respectively. This suggests the choice of these parameters as confidence levels.
On the other hand, when we target to estimate $\omega$ and $f$ $q_n(\alpha)$ and $q_n(\beta)$ can be seen as tuning parameters for the estimates $\hat{\omega}$ and $\hat{f}$.
Although, we found in Section \ref{subsec: stabAB} that SLAM's estimates are quite stable for a range of $\alpha$'s and $\beta$'s, a fine tuning of these parameters improves estimation accuracy, of course. In the following, we suggest a possible strategy for this.
First, we discuss $q_n(\alpha)$ for tuning the estimate $\hat{\omega}_q \ZuWeis \hat{\omega}(Y, q)$ . Recall that for estimating $\omega$, $q_n(\beta)$ is not required.

\paragraph{Minimal valid threshold (MVT)}

Theorem \ref{consimain} yields $\ln(n)/\sqrt{n}$-consistency of $\hat{\omega}$ when $q_n(\alpha) = q_n(\alpha_n)$ with $\alpha_n$ as in (\ref{alphaN}), independently of the specific choice of $\hat{\omega} \in \cC_{1 - \alpha_n}$. Further, for $\alpha^{\prime}$ (and $q_n(\alpha^{\prime})$, respectively) with $\alpha^{\prime} \geq \alpha_n$ (and $q_n(\alpha^{\prime}) \leq q_n(\alpha_n)$ , respectively) $\cC_{1 - \alpha^{\prime}} \subseteq \cC_{1 - \alpha_n}$ whenever $\fB^{\star} = \fB^{\star}_{q_n(\alpha^{\prime})} \neq \emptyset$ in (\ref{Comega}). Thus, choosing the threshold $q$, for any discrete set $Q = \{q_1,q_2,\ldots,q_N = q_n(\alpha_n)\}$, as
$q^{\star} \ZuWeis \min\left( q \in Q: \fB^{\star}_q \neq \emptyset \right)$
guarantees the convergence rates of Theorem \ref{consimain} for the corresponding estimate $\hat{\omega}(Y, q^{\star})$. In practice, we found $Q = \{-1.0,-0.9,...,1.9, 2.0 \}$ to be a sufficiently rich candidate set.

\paragraph{Sample splitting (SST)}

Alternatively, we can choose $q$ such that a given performance measure $h(q)\ZuWeis \E\lbrack L(\hat{\omega}_q - \omega)\rbrack $ for estimating $\omega$, for example, the MSE with $L = \norm{\cdot}^2_2$, is minimized. 
As $\omega$ is unknown, we have to estimate $h(q)$, for which we suggest a simple sample splitting procedure. Details are given in Section \ref{sec: chqSupp}, in the supplement.
Simulations indicate that, especially for high noise level, the MVT-selection method outperforms the SST-selection method in terms of standard performance measures like MSE and MAE. However, in contrast to the SST-selection method, the MVT-selection method cannot be tailored for a specific performance measure $h$.

\vspace{2ex}

It remains to select $q_n(\beta)$ (and $\beta$, respectively), which is required additionally for $\hat{f}$, recall (\ref{def: gEstS}) and (\ref{Hn}). Theorem \ref{consimain} suggests to choose $q_n(\beta) = q_n(\beta_n)$ with $\beta_n$ as in (\ref{alphaN}), i.e., $q_n(\beta) \rightarrow \infty$ with rate $\mathcal{O}(\log(n))$. For finite $n$, there exist several methods for selection of $q_n(\beta)$ in c.p. regression (see e.g., \cite{mBIC}), which might be used here as well. However, due to the high stability of $\hat{f}$ in $q$ (see Section \ref{subsec: stabAB} and Figure \ref{fig: chQmisemiae} in the supplement) we simply suggest to choose $\beta = 0.1$, which we have used here for our data analysis. This choice controls the probability of overestimating the number of jumps in $g$, $\Pp(K(\hat{g}) > K(g)) \leq 0.1$ asymptotically. In general, it depends on the application. A large $q_n(\beta)$ (hence small $\beta$) has been selected in the subsequent application  to remove spurious changes in the signal which appear biologically not as of much relevance.

\section{Genetic sequencing data}\label{sec:appgen}
Recall from Section \ref{subsec: intapp} that a tumor often consists of a few distinct sub-populations , so called clones, of DNA with distinct copy-number profiles arising from duplication and deletion of genetic material groups. The copy number profiles of the underlying clones in a sample measurement correspond to the functions $f^1,\ldots,f^m$, the weights $\omega_1,\ldots,\omega_m$ correspond to their proportion in the tumor, and the measurements correspond to the mixture $g$ with some additive noise. 

The most common method for tumor DNA profiling is via whole genome sequencing, which roughly involves the following steps:
\begin{enumerate}
\item Tumor cells are isolated, and the pooled DNA is extracted, amplified and fragmented through shearing into single-strand pieces.
\item Sequencing of the single pieces takes place using short ``reads'' (at time of writing of around $10^2$ base-pairs long).
\item Reads are aligned and mapped to a reference genome (or the patient germline genome if available) with the help of a computer.
\end{enumerate}

Although, the observed total reads are discrete (each observation corresponds to an integer number of reads at a certain locus), for a sufficiently high sequencing coverage, as it is the case in our example with around $55$ average stretches of DNA mapped to a locus, it is well established to approximate this binomial by a normal variate (see \cite{liu} and references there).

In the following, SLAM is applied to the cell line LS411, which comes from colorectal cancer and a paired lymphoblastoid cell line. Sequencing was done through a collaboration of Complete Genomics with the Wellcome Trust Center for Human Genetics at the University of Oxford. This data has the special feature of being generated under a designed experiment using radiation of the cell line ({\em{``in vitro''}}), designed to produce CNAs that mimic real world copy-number events. In this case therefore, the mixing weights and sequencing data for the individual clones are known, allowing for validation of SLAM's results, something that is not feasible for patient cancer samples.

The data comes from a mixture of three different types of DNA, relating to  a normal (germline) DNA and two different clones. Tumor samples, even from micro-dissection, often contain high proportion of  normal cells, which for our purposes are a nuisance, this is known as ``stromal contamination'' of germline genomes in the cancer literature. 
The true mixing weights in our sample are $\omega^{\top} =(\omega_{\text{Normal}},\omega_{\text{Clone1}},\omega_{\text{Clone2}})=(0.2,0.35,0.45).$

SLAM will be, in the following, applied only to the mixture data without knowledge of $\omega$ and the sequenced individual clones and germline. The latter (which serve as ground truth) will then be used only for validation of SLAM's reconstruction.
We restricted attention to regions of chromosome $4,5,6,18$ and $20$, as detailed below. Figure \ref{fig:raw_data} in the supplement shows the raw data. Sequencing produces some spatial artefacts in the data, and waviness related to the sequencing chemistry and local GC-content, corresponding to the relative frequency of the DNA bases $\{$C, G$\}$ relative to $\{$A, T$\}$. This violates the modeling assumptions. To alleviate this we preprocess the data with a smoothing filter  using local polynomial kernel regression on normal data, baseline correction, and binning. We used the local polynomial kernel estimator from the R package {\tt KernSmooth}, with bandwidth chosen by visual inspection. We selected the chromosomal regions above as those showing reasonable denoising, and take the average of every 10th data point to make the computation manageable resulting in $n=7480$ data points spanning the genome. The resulting data is displayed in Figure \ref{fig:prepro-data}, in the supplement, where we can see that the data is much cleaned in comparison with Figure \ref{fig:raw_data} although clearly some artefacts and local drift of the signal remain.

With $\sigma = 0.21$ pre-estimated as in \cite{davies}, SLAM yields the confidence region  for $\alpha = 0.1$ $C_{0.9} = [0.00, 0.31]\times [0.28, 0.50]\times[0.33, 0.72]$.
With $q_n(\alpha) = -0.15$ selected with the MVT-method from Section \ref{subsec: chQ} we obtain $\hat{\omega} = (0.12, 0.35, 0.53)$.
\begin{figure}[h!]
\includegraphics[width=\textwidth]{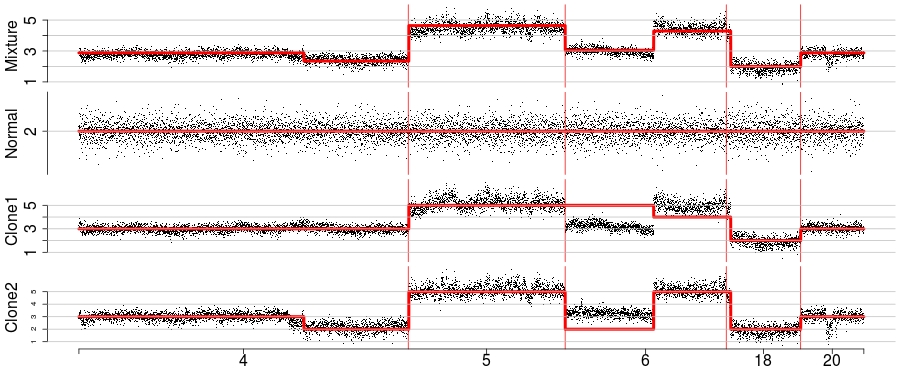}
\caption{SLAM's estimates (red lines) for $q_n(\alpha) = -0.15$ (selected with MVT-method from Section \ref{subsec: chQ}) and $q_n(\beta) = 20$. 
Top row: total copy-number estimates across the genome. Rows 2-4: estimates of the CN profiles of the germline and clones. }\label{fig:est2}
\end{figure}
Figure \ref{fig:est1} in the supplement shows SLAM's estimates for $q_n(\beta) = 2.2$ (which corresponds to $\beta = 0.01$). The top row shows the estimate for total copy number $\sum_j \hat{w}_j \hat{f}^j$ and rows 2-4 show $\hat{f}^1, \hat{f}^2$, and $\hat{f}^3$. We stress that the data for the single clones are only used for validation purposes and do not enter the estimation process.
Inspection of Figure \ref{fig:est1} shows that artefacts and local drifts of the signal result in an overestimation of the number of jumps. However, the overall appearance of the estimated CNA profile remains quite accurate.
This over-fitting effect caused by these artifacts can be avoided by increasing SLAM's tuning parameter $q_n(\beta)$ at the (unavoidable) cost of loosing detection power on small scales (see Figure \ref{fig:est2}, which shows SLAM's estimate for $q_n(\beta) = 20$).
In summary, Figure \ref{fig:est2} (and \ref{fig:est1}) show that SLAM can yield highly accurate estimation of the total CNA profile in this example, as well as reasonable CNA profiles and their mixing proportions for the clones, something which has not been obtainable prior to now. The analysis takes around 1 minute to run on a desktop computer with an intel core i7 processor. In future work we aim to speed up the algorithm and explore association between the CNA patient profiles and clinical outcome data such as time-to-relapse and response to therapy. 
\section{Conclusion and discussion}\label{sec: con}

In this paper, we have established a new approach for separating linear mixtures of step functions with a known finite alphabet for additive Gaussian noise.
This is of major interest for cancer genetics, but appears in other applications as well, for instance, in digital communications. We are not aware of any other method that deals with this problem in such a rigorous and general way.
However, there are still some further generalizations and extensions to be studied.
 
Although we obtained a certain robustness of SLAM to misspecification of the error distribution in our simulation study, it is natural to ask how the results of this work can be extended to other types of error distributions than the normal distribution. 
\cite{dumbgen2001, dumbgen2006, frick} give several results about the multiscale statistic $T_n$, its limit distribution, and its geometric interpretation - which leads to the definition of the boxes $\fB$ (see (\ref{def: box})) for general one-dimensional exponential families.
Combining this with the results of this work should yield extensions for such distributions.  

In contrast to the noiseless case, $\epsilon \equiv 0$ in (\ref{def:Y}), where the weights can be reconstructed in  $\cO(k^m)$ (independent of $n$) steps \cite{diamantaras2006, behr}, SLAM's estimation for the weights requires between $\cO(n^m)$ and $\cO(n^{2m})$ steps. Without further parallelization, this restricts the applicability of the algorithm to small number of mixtures $m$. 
Significant speed up can also be achieved when a smaller
system of intervals in $T_n$ is used (at the possible expense of finite sample detection power), for example, all intervals of dyadic length, in which case the worst case complexity reduces to $\cO((n\ln(n))^m)$.

A further important issue is an extension for unknown number of source functions $m$. Clearly, this is a model selection problem, which might be approached with standard methods like the BIC or AIC criterion in conjunction with SLAM, a topic for further research. 

One may also ask the question, whether the SBSSR-model can be treated for infinite alphabets $\fA$. The condition $ASB(\omega) > 0$ in (\ref{def: OmegaD}) remains necessary in order to guarantee identifiability, that is, different mixture values must be well separated. This condition, however, becomes significantly more restrictive when the size of the alphabet increases. Even for the most simple (infinite) alphabet $\fA = \N$ there exists no $m\geq 2$, $\omega \in \Omega(m)$ which fulfills $ASB(\omega) > 0$, that is, no method can be valid in this situation. 
To see this, fix some $\omega \in \Omega(m)$ and w.l.o.g. assume that $\omega_1 \in \Q$, i.e., $\omega_1 = n/d$ with $n,d \in \N$ and $d > n$. Then, $\tilde{d} \ZuWeis (d - n) d \in \N$, $n\cdot d \in \N$, and 
$ASB(\omega) \leq \abs{\left(\tilde{d}\omega_1 + 0 \cdot \left(1 -  \omega_1 \right) \right) - \left(0 \cdot \omega_1  + nd \left(1- \omega_1 \right) \right) } = 0$.
Hence, finiteness of the alphabet $\fA$ is fundamental for identifiability in the SBSSR-model. 

Another issue is the extension to unknown (but finite) alphabets. If only certain parameters of the alphabet are unknown, for example, an unknown scaling constant, the alphabet is of the form $\fA = \{L a_1,..., L a_k \}$ with $a_i$'s known but $L$ unknown, we speculate that generalizations should be possible and will rely on corresponding identifiability conditions, which are unknown so far. An arbitrary unknown alphabet, however, clearly leads to an unidentifiable model.
This raises challenging issues, which we plan to address in the future.

\section*{Acknowledgements}
Helpful comments of Hannes Sieling, Christopher Yau, and Philippe Rigollet are gratefully acknowledged.
We are also grateful to two referees and one editor for their constructive comments which led to an improved version of this paper.

\begin{supplement}
\sname{Supplement}\label{supp}
\stitle{Supplement to Multiscale Blind Source Separation}
\slink[url]{}
\sdescription{Proofs of Theorem \ref{lem: epsilon}, Theorem \ref{theo: CIf}, and  Theorem \ref{consimain} (Section \ref{sec: proofs}); additional details on algorithms (Section \ref{sec: algoS}); additional figures and tables from Section \ref{subsec: Appsim} and \ref{sec:appgen} (Section \ref{sec: aftS}); details on the SST-method (Section \ref{sec: chqSupp}).}
\end{supplement}

\bibliographystyle{imsart-number}

\clearpage
\setattribute{journal}{name}{}
\arxiv{}
\renewcommand*\footnoterule{}
\renewcommand{\thesection}{S\arabic{section}}
\setcounter{section}{0}
\setcounter{page}{1}
\setcounter{equation}{43}

\begin{frontmatter}
\title{Supplement to \\  multiscale blind source separation}
\runtitle{Supplement}

\begin{aug}
\author{\fnms{Merle} \snm{Behr}\thanksref{m1}\ead[label=e1]{behr@math.uni-goettingen.de}},
\author{\fnms{Chris} \snm{Holmes}\thanksref{m2}\ead[label=e3]{cholmes@stats.ox.ac.uk}},
\and
\author{\fnms{Axel} \snm{Munk}\thanksref{m1,m3}\ead[label=e2]{munk@math.uni-goettingen.de}}

\runauthor{M. Behr et al.}

\affiliation{University of Goettingen \thanksmark{m1}, University of Oxford \thanksmark{m2},\\ and {Max Planck Institute for Biophysical Chemistry \thanksmark{m3}}}

\address{University of Goettingen \\
Institute for Mathematical Stochastics\\
Goldschmidtstr. 7\\
37077 G\"ottingen\\
Germany\\
\printead{e1}\\
\printead{e2}}

\address{University of Oxford\\
Department of Statistics\\
24-29 St Giles'\\
Oxford. OX1 3LB\\
United Kingdom\\
\printead{e3}}

\end{aug}

\end{frontmatter}

\counterwithin{figure}{section}

\section{Additional poofs}\label{sec: proofs}

\subsection{Proof of Theorem \ref{lem: epsilon}} \label{subsec: appExactR}

\begin{proof}
As $g, \tilde{g} \in \cM^\delta$, (\ref{supghg}) implies that there exist $a^1, \ldots, a^m$, $\tilde{a}^1, \ldots,\tilde{a}^m \in \fA^m$ such that
\begin{align}\label{ata}
\begin{aligned}
\abs{\omega^{\top}a^i - \tilde{\omega}^{\top}[A]_i} &< \epsilon\quad\text{for } i = 1,\ldots,m,\\
\abs{\tilde{\omega}^{\top}\tilde{a}^i - \omega^{\top}[A]_i} &< \epsilon\quad \text{for }i = 1,\ldots,m,
\end{aligned}
\end{align}
with $A$ as in (\ref{Amatrix}).

First, we show by induction that (\ref{ata}) implies  \ref{a}..

W.l.o.g. let $\tilde{\omega}_1 > \omega_1$. Assume that $\tilde{\omega}^{\top}\tilde{a}^1 < \tilde{\omega}^{\top}[A]_1 = a_1 + (a_2 - a_1)\tilde{\omega}_1$, i.e.,
\begin{equation}\label{proofappExactEq1}
\sum_{i = 2}^{m} \tilde{\omega}_i(\tilde{a}^1_i - a_1) < \tilde{\omega}_1 (a_2 - \tilde{a}^1_1).
\end{equation}
As $\tilde{\omega}_1$ denotes the smallest mixing weight (recall $\tilde{\omega}_1 \leq \ldots \leq \tilde{\omega}_m$ in (\ref{def: Omega})) and $a_1$ and $a_2$ denote the smallest and second smallest, respectively, alphabet values (recall $a_1 < \ldots < a_k$ in (\ref{def:SA})), it holds for any alphabet value $e \in \fA \setminus \{a_1\} = \{a_2,\ldots,a_k\}$ and $i = 1,\ldots,m$ that 
\begin{equation}\label{proofappExactEq2}
\tilde{\omega}_i ( e - a_1) \geq \tilde{\omega}_1(a_2 - a_1) \geq \tilde{\omega}_1 (a_2 - \tilde{a}^1_1).
\end{equation}
(\ref{proofappExactEq1}) and (\ref{proofappExactEq2}) imply that $\tilde{a}^1 = (a_1, \ldots,a_1)^\top$, i.e., $\tilde{\omega}^{\top}\tilde{a}^1 = a_1$. In particular, (\ref{ata}) yields $\abs{a_1- \omega^{\top}[A]_1} < \epsilon < \delta$, which contradicts $ASB(\omega) \geq \delta$. Consequently, 
\begin{align*}
\tilde{\omega}^{\top}\tilde{a}^1 \geq \tilde{\omega}^{\top}[A]_1 = a_1 + (a_2 - a_1)\tilde{\omega}_1 > a_1 + (a_2 - a_1)\omega_1  = \omega^{\top}[A]_1
\end{align*}
and therefore, by (\ref{ata})
\begin{align*}
(a_2 - a_1)\abs{\tilde{\omega}_1 - \omega_1} = \abs{\tilde{\omega}^{\top}[A]_1 - \omega^{\top}[A]_1} < \epsilon.
\end{align*}

Now, assume that $(a_2 - a_1)\abs{\tilde{\omega}_i - \omega_i} < \epsilon$ for $i = 1,\ldots,l-1$.

W.l.o.g., let $\tilde{\omega}_l > \omega_l$. Assume that $\tilde{\omega}^{\top}\tilde{a}^l < \tilde{\omega}^{\top}[A]_l = a_1 + (a_2 - a_1)\tilde{\omega}_l$, i.e.,
\begin{equation}\label{proofappExactEq3}
\sum_{i = 1,\; i \neq l}^{m} \tilde{\omega}_i(\tilde{a}^l_i - a_1) < \tilde{\omega}_l (a_2 - \tilde{a}^l_l).
\end{equation}
Again, as $\tilde{\omega}_1 \leq \ldots \leq \tilde{\omega}_m$ and $a_1 < \ldots < a_k$, it holds for any alphabet value $e \in \fA \setminus \{a_1\} = \{a_2,\ldots,a_k\}$ and $i\geq l$ that 
\begin{equation}\label{proofappExactEq4}
\tilde{\omega}_i ( e - a_1) \geq \tilde{\omega}_l (a_2 - a_1) \geq \tilde{\omega}_l (a_2 - \tilde{a}^l_l).
\end{equation}
(\ref{proofappExactEq3}) and (\ref{proofappExactEq4}) imply that $\tilde{a}^l_l = \ldots = \tilde{a}^l_m = a_1$ and therefore,
\begin{align*}
\abs{\omega^{\top}[A]_l - \omega^{\top}\tilde{a}^l} &\leq \abs{\omega^{\top}[A]_l - \tilde{\omega}^{\top}\tilde{a}^l} + \abs{\tilde{\omega}^{\top}\tilde{a}^l - \omega^{\top}\tilde{a}^l}\\
&< \epsilon + \abs{\sum_{i = 1}^{l-1} (\tilde{a}_i^l - a_1)(\tilde{\omega}_i - \omega_i)}\\
&\leq \epsilon + (m-1)\frac{a_k - a_1}{a_2 - a_1}\epsilon \leq m\frac{a_k - a_1}{a_2 - a_1}\epsilon < \delta,
\end{align*}
which contradicts $ASB(\omega) \geq \delta$.
Consequently, $\tilde{\omega}^{\top}\tilde{a}^l \geq \tilde{\omega}^{\top}[A]_l > \omega^{\top}[A]_l$ and therefore,
\begin{align*}
(a_2 - a_1)\abs{\tilde{\omega}_l - \omega_l} = \abs{\tilde{\omega}^{\top}[A]_l - \omega^{\top}[A]_l} < \epsilon.
\end{align*}

By induction \ref{a}. follows.

To prove \ref{b}., assume the contrary. Then there exist $a \neq \tilde{a} \in \fA^m$ such that 
\begin{align*}
\epsilon > \abs{\omega^{\top}a - \tilde{\omega}^{\top}\tilde{a}}
\geq \abs{\omega^{\top}a - \omega^{\top}\tilde{a}} - \abs{\omega^{\top}\tilde{a} - \tilde{\omega}^{\top}\tilde{a}}
\end{align*}
and by \ref{a}.
\begin{align*}
\abs{\omega^{\top}\tilde{a} - \tilde{\omega}^{\top}\tilde{a}} 
= \abs{\sum_{i = 1}^m (\omega_i (\tilde{a}_i - a_1) - \tilde{\omega}_i(\tilde{a_i}- a_1)) } \leq m \frac{a_k - a_1}{a_2 - a_1}\epsilon.
\end{align*}
The last two inequalities give $\epsilon > \delta - m \epsilon (a_k - a_1)/(a_2 - a_1)$, which contradicts $2 m (a_k - a_1) \epsilon < \delta (a_2 - a_1)$ as $m (a_k - a_1) > (a_2 - a_1)$.

\end{proof}

\subsection{Proof of Theorem \ref{consimain}} \label{subsec: app1}
The following Theorem is needed for the proof of Theorem \ref{consimain} and shows that SLAM admits a solution with probability converging to one at a superpolynomial rate. 

Let $N_1^{\star}$ be such that
\begin{align}\label{N1S}
\frac{\delta}{\sigma} \ln(N_1^\star) \geq
 139 \left( 1 + 2 m \frac{a_k - a_1}{a_2 -a_1} \right) \sqrt{2 \ln(e/ \lambda^\star)} +  70
\end{align}
and $\lambda^{\star} \geq \lambda$ as in R\ref{minred}. Analog to $\cM^\delta_\lambda$ in (\ref{MAlam}) define
\begin{align}\label{SfAl}
\cS(\fA)^m_\lambda \ZuWeis \{f \in \cS(\fA)^m \text{ separable}:\; \min_{j \in \{0,,\ldots,K(f)\}} \abs{\tau_{j+1} - \tau_j} \geq \lambda\},
\end{align}
where $\tau_j$ denote the change points of $f$, that is, at least one of the $f^i$'s jumps, and $K(f)$ the number of change points of $f$.

\begin{theo}\label{lem: H}
Consider the SBSSR-model with $g\in \cM_\lambda^\delta$. Let $\alpha_n$ and $\beta_n$ be as in (\ref{alphaN}).
Further, let $\cC_{1-\alpha}(Y)$ be as in (\ref{Comega}) and let $\hat{\omega}$ be any weight vector in $\cC_{1-\alpha_n}(Y)$.
Then for all $n\geq N_1^{\star}$ in (\ref{N1S})
\begin{align*}
\Pp\left(\min_{\tilde{f}\in \cS(\fA)^m_\lambda} T_n(Y,\hat{\omega}^{\top}\tilde{f}) \leq q_n(\beta_n) \middle| \;  T_n(Y,g) \leq q_n(\alpha_n)\right) = 1.
\end{align*}
\end{theo}

\begin{proof}
Let $\tilde{\omega} \in \Omega(m)$ and $\alpha \in (0,1)$ be fixed. Define the set 
\begin{align*}
\cN(\tilde{\omega}) \ZuWeis \Bigg\{\check{\omega}^{\top} a:  a\in \fA^m \text{ and } \norm{\check{\omega} - \tilde{\omega}}_{\infty} \leq 2\sigma \frac{q_n(\alpha) + \sqrt{2\ln(e / \lambda^{\star})}}{\sqrt{n\lambda^{\star}}(a_2 - a_1)}\Bigg\}
\end{align*}
and, analog to $\cS(\fA)_\lambda^m$ in (\ref{SfAl}),
\begin{align*}
&\cS(\cN(\tilde{\omega}))_\lambda \ZuWeis\\
&\Big\{ g\in \cS(\cN(\tilde{\omega})):\;  \min_{j \in \{0,,\ldots,K(g)\}} \abs{\tau_{j+1} - \tau_j} \geq \lambda \text{ and } a_1 + (a_2-a_1)\tilde{\omega}_i \in \imag(g)    \Big\},
\end{align*}
where $\imag(g) \ZuWeis \{g(x):\; x \in [0,1)\}$ denotes the image of $g$.
Then it follows from R\ref{notconst}, R\ref{exred}, R\ref{minred}, (\ref{def: box}), and Remark \ref{rem: bs} that conditioned on $\{\tilde{\omega} \in \cC_{1-\alpha}(Y)\}$ and $\{T_n(Y,g)\leq q_n(\alpha)\}$
\begin{align}\label{infTSNleqq}
\inf_{\tilde{g} \in \cS(\cN(\tilde{\omega}))_\lambda} T_n(Y,\tilde{g}) \leq q_n(\alpha) \quad \text{a.s..} 
\end{align}
Further, for $\epsilon_n \ZuWeis 2 m \sigma \frac{a_k - a_1}{a_2 - a_1} \left(q_n(\alpha) + \sqrt{2 \ln(e/ \lambda^{\star})}\right)/\sqrt{n \lambda^\star}$ we have that
\begin{align}\label{subnormEpsilon}
\sup_{\tilde{g} \in \cS(\cN(\tilde{\omega}))_\lambda} \min_{\tilde{f} \in \cS(\fA)^m_\lambda} \norm{\tilde{g} - \tilde{\omega}^{\top}\tilde{f}}_{\infty} \leq \epsilon_n.
\end{align}

Let $(y_n)_{n\in\N}$ be a fixed sequence in $\R$, and denote $y^n \ZuWeis (y_1,\ldots,y_n)$. Let $\epsilon > 0$, and $g,g^{\prime}\in \cM_\lambda$ be such that $\sup_{x\in [0,1)}\abs{g(x)- g^{\prime}(x)} \leq \epsilon$. Then by the reverse triangle inequality
\begin{align*}
\abs{T_n(y^n, g) - T_n(y^n, g^{\prime})} &\leq \max_{\substack{1\leq i \leq j \leq n \\ j-i+1\geq n\lambda }}  \abs{\frac{\abs{ \sum_{l = i}^j y_l - g(x_l)} - \abs{\sum_{l = i}^j y_l - g^{\prime}(x_l)}}{\sigma \sqrt{j-i+1}}}\\
&\leq \max_{\substack{1\leq i \leq j \leq n \\ j-i+1\geq n\lambda }}  \frac{\abs{\sum_{l = i}^j g(x_l) - g^\prime(x_l) }}{\sigma \sqrt{j-i+1}} \leq  \frac{\sqrt{n \lambda}}{\sigma} \epsilon.
\end{align*} 
This, together with (\ref{infTSNleqq}) and (\ref{subnormEpsilon}), implies that conditioned on $\{\tilde{\omega}\in \cC_{1-\alpha}(Y)\}$ and $\{T_n(Y,g)\leq q_n(\alpha)\}$ 
\begin{align}\label{infAE}
\begin{aligned}
&\inf_{\tilde{\omega}\in \Omega(m)}\Pp\left(\min_{\tilde{f}\in \cS(\fA)^m_\lambda} T_n(Y,\tilde{\omega}^{\top}\tilde{f}) \leq q_n(\alpha) + \frac{\sqrt{n \lambda}}{\sigma} \epsilon_n \right) \\
\geq &\inf_{\tilde{\omega}\in \Omega(m)}\Pp\left(\inf_{\tilde{g} \in \cS(\cN(\tilde{\omega}))_\lambda} T_n(Y,\tilde{g}) \leq q_n(\alpha)\right) = 1,
\end{aligned}
\end{align}
where the inequality results from
\begin{align*}
&\min_{\tilde{f}\in \cS(\fA)^m_\lambda} T_n(Y,\tilde{\omega}^{\top}\tilde{f})\\
= &\inf_{\tilde{g} \in \cS(\cN(\tilde{\omega}))_\lambda} T_n(Y,\tilde{g}) + \left( \min_{\tilde{f}\in \cS(\fA)^m_\lambda} T_n(Y,\tilde{\omega}^{\top}\tilde{f}) - \inf_{\tilde{g} \in \cS(\cN(\tilde{\omega}))_\lambda} T_n(Y,\tilde{g}) \right) \\
\leq & \inf_{\tilde{g} \in \cS(\cN(\tilde{\omega}))_\lambda} T_n(Y,\tilde{g}) + \sup_{\tilde{g} \in \cS(\cN(\tilde{\omega}))_\lambda} \min_{\tilde{f}\in \cS(\fA)^m_\lambda} \abs{T_n(Y,\tilde{\omega}^{\top}\tilde{f}) -  T_n(Y,\tilde{g})}.
\end{align*}
It remains to show that for all $n \geq N_1^{\star}$
\begin{align}\label{qaqb}
q_n(\alpha_n) + \frac{\sqrt{n \lambda}}{\sigma} \epsilon_n  \leq q_n(\beta_n).
\end{align}
To this end, we need some results about the quantile function of the multiscale statistic $T_n$ from (\ref{def: Tn}).
Easy calculations and Mill's ratio give for all $n \in \N$
\begin{align*}
\Pp(T_n > q) \geq \sqrt{\frac{2}{\pi}} \left( \frac{1}{\tilde{q}} - \frac{1}{\tilde{q}^3}\right) \exp\left( - \tilde{q}^2 / 2 \right), \quad \text{with }\tilde{q} \ZuWeis q + \sqrt{2\ln(e/\lambda^{\star})},
\end{align*}
which implies 
\begin{align}\label{theoBoundDF2}
q_n(\alpha) \geq \sqrt{\abs{- \ln(\alpha \sqrt{\pi / 2})}} - \sqrt{2 \ln(e/\lambda^{\star})}.
\end{align}
Further, a slight modification of \cite[Corollary 4]{hannes} gives for all $n\in \N$ and $q > C$, for some constant $C< \infty$, that
\begin{align}\label{theoBoundDF}
\Pp(T_n > q)\leq  \exp(-q^2/8),
\end{align}
which implies
\begin{align}\label{theoBoundDF3}
q_n(\alpha) \leq \sqrt{-8 \ln(\alpha)}.
\end{align}

From (\ref{theoBoundDF3}) and (\ref{alphaN}) we follow that
\begin{align}\label{qau}
\begin{aligned}
&q_n(\alpha_n) + \frac{\sqrt{n \lambda}}{\sigma} \epsilon_n
= q_n(\alpha_n) + 2 m \frac{a_k - a_1}{a_2 - a_1} \left(q_n(\alpha_n) + \sqrt{2 \ln(e/\lambda^\star} \right)\\
 \leq &\left(\sqrt{8 c_1} + 2 m \frac{a_k - a_1}{a_2 - a_1} \sqrt{8 c_1}\right) \ln(n) + 2 m \frac{a_k - a_1}{a_2 - a_1} \sqrt{2\ln(e/\lambda^\star)}
 \end{aligned}
\end{align}
and from (\ref{theoBoundDF2}) and (\ref{alphaN}) that
\begin{align}\label{qbl}
q_n(\beta_n)\geq \sqrt{75 m^2 \left(\frac{a_k - a_1}{a_2 - a_1} \right)^2 c_1}\ln(n) - \sqrt{\ln(\sqrt{\pi / 2})} - \sqrt{2\ln(e/\lambda^{\star})}.
\end{align}
(\ref{N1S}) yields that the right hand side of (\ref{qau}) is smaller than the right hand side of (\ref{qbl}) for all $n \geq N_1^{\star}$, which yields (\ref{qaqb}) and, thus, together with (\ref{infAE}), that conditioned on $\{\tilde{\omega}\in \cC_{1-\alpha_n}(Y)\}$ and $\{T_n(Y,g) \leq q_n(\alpha_n)\}$
\begin{align*}
\inf_{\tilde{\omega}\in \Omega(m)}\Pp\left(\min_{\tilde{f}\in \cS(\fA)^m_\lambda} T_n(Y,\tilde{\omega}^{\top}\tilde{f}) \leq q_n(\beta_n)\right) = 1.
\end{align*}
As $\hat{\omega} \in \cC_{1-\alpha_n}$ a.s., this yields the assertion.
\end{proof}

The following theorem is a slight variation of Theorem \ref{consimain}, from which, together with Theorem \ref{lem: H}, Theorem \ref{consimain} will follow easily.

\begin{theo}\label{theoconsi}
Consider the SBSSR-model with $g\in \cM_\lambda^\delta$.
Let $q_n(\alpha)$ be as in (\ref{def: qalpha}), $\alpha_n$ as in (\ref{alphaN}), and $\beta_n$ such that 
\begin{align}\label{qnlog}
q_n(\alpha_n) < q_n(\beta_n)<  \frac{\delta}{9 \sigma} \ln(n).
\end{align}
Let $\hat{g} = \hat{\omega}^{\top}\hat{f}\in \cM$ be the SLAM estimator of $g$ with $\alpha = \alpha_n$, $\beta = \beta_n$, and $T_n(Y,\hat{g})\leq q_n(\beta_n)$. Further, let $\hat{\tau}$ and $\tau$ be the vectors of all change points of $\hat{g}$ and $g$, respectively. Define
\begin{align*}
A_n\ZuWeis&\Big\{\max_{j}|\hat{\tau_j}-\tau_j|\leq 2 \frac{\ln(n)^2}{n}\Big\}\cap \Big\{ K(\hat{g}) = K(g) \Big\}\\
\cap &\Big\{\max_{j}\max_{i}\abs{\hat{f^i}|_{[\hat{\tau}_j,\hat{\tau}_{j+1})}-f^i|_{[\tau_j,\tau_{j+1})}}=0\Big\}\\
\cap  &\Big\{ \max_{i} \abs{\hat{\omega}_i - \omega_i} < \frac{\delta + \sqrt{2 \sigma^2 \ln(e/\lambda)}}{\sqrt{\lambda}(a_2 - a_1)} \; \frac{\ln(n)}{\sqrt{n}}  \Big\}.
\end{align*}
Then for all $n>N^{\star}$ in (\ref{Nstar2}) and (\ref{Nstar1}) $\Pp\left(A_n  \middle| \; T_n(Y,g) \leq q_n(\alpha_n)\right) = 1$.
\end{theo}

\begin{proof}
Let $d_n \ZuWeis \ln^2(n)/n$ and 
\begin{align*}
\cI \ZuWeis \{[x_i,x_j]: 1\leq i \leq j \leq n \text{ and } j-i+1\geq n\lambda\}.
\end{align*}

We define a partition $\cI=\cI_1\cup \cI_2 \cup \cI_3$ as follows. 
\begin{align*}
&\cI_1 \ZuWeis \{I \in \cI : \text{ $I$ contains more than two change points of $g$}\},\\
&\cI_2 \ZuWeis \{I \in \cI : g|_I = g_1^I \indE_{I_1} + g_2^I \indE_{I_2} +g_3^I \indE_{I_3}, \text{ with } \abs{I_1} \geq \abs{I_2} \geq \abs{I_3},\\ &\abs{I_2}\leq d_n, \text{ and } g_1^I,g_2^I,g_3^I \in\imag(g)\text{ pairwise different} \}\\
&\cI_3 \ZuWeis \{I \in \cI : g|_I = g_1^I \indE_{I_1} + g_2^I \indE_{I_2} +g_3^I \indE_{I_3}, \text{ with } \abs{I_1} \geq \abs{I_2} \geq \abs{I_3},\\ &\abs{I_2} > d_n, \text{ and } g_1^I,g_2^I,g_3^I \in\imag(g)\text{ pairwise different} \}.
\end{align*}

Moreover, let $\fB \ZuWeis \{B(I) = B(i,j) \;: \; I = [x_i,x_j] \in \cI\}$ be as in (\ref{def: box}) with $q = q_n(\beta_n)$ and define $\norm{B(I)} \ZuWeis \overline{b}-\underline{b}$ with $B(I)=[\underline{b},\overline{b}]$. Furthermore,  let $\fB_{\text{nc}}$ be as in (\ref{def: bnc}) and define
\begin{align}\label{epsilonN}
\epsilon_n \ZuWeis \frac{\delta + \sqrt{2 \sigma^2 \ln(e/\lambda)}}{\sqrt{\lambda}} \frac{\ln(n)}{\sqrt{n}}
\end{align}
and
\begin{align*}
&E_1\ZuWeis \bigcap_{I \in \cI_1 \cup \cI_3}\{B(I)\in \fB_{\text{nc}}\}, \\
&E_2 \ZuWeis \bigcap_{I \in \cI_2}  \{B(I)\subset [g_1^I-\epsilon_n, g_1^I +\epsilon_n]\},\\
&E_3 \ZuWeis \{K(\hat{g}) =  K(g)\}\cap  \{\max_{j}|\hat{\tau_j}-\tau_j|\leq 2 d_n\} \cap \{\max_j \abs{ \hat{g}(\hat{\tau}_j) - g(\tau_j)} < \epsilon_n\}.
\end{align*}

First, we show that
\begin{align}\label{E1E2E3}
E_1 \cap E_2 \subset E_3.
\end{align}
To this end, consider Figure \ref{fig: proof} and note that (conditioned on $\{T_n (Y,g)\leq q_n(\alpha_n)\}$) by Theorem \ref{lem: H} and (\ref{Khat}) $\hat{g}$ has minimal scale $\lambda$ for all $n > N^\star$.

If $B(I)\in \fB_{\text{nc}}$, then $\hat{g}$ is not constant on $I$. Therefore, it follows from $E_1$ that $\hat{g}$ is constant only on intervals $I \in \cI_2$.

Conversely, if $\hat{g}$ is constant on $I\in \cI_2$ then $\hat{g}|_{I} \in B(I)$ (see orange bars in Figure \ref{fig: proof}) as $T_n(Y,\hat{g})\leq q_n(\beta_n)$ by assumption. 

Now, consider a change point of $\hat{g}$. Let $I, I^{\prime} \in \cI_2$ be the constant parts of $\hat{g}$ left and right of this change point and $I_1, I_1^\prime$ be those sub-intervals which include the largest constant piece of $g$ (see green lines in Figure \ref{fig: proof}), with $g|_{I_1} \equiv g_1^I$ and $g|_{I_1^\prime} \equiv g_1^{I^\prime}$.

As $\epsilon_n < \delta/2$ for all $n > N^{\star}$ (see (\ref{Nstar1})) $\abs{g_1^I - g_1^{I^\prime}}  > 0$ (see the vertical distance between the left and the right green line in Figure \ref{fig: proof}), such that $g$ has at least one jump in a $2d_n$-neighborhood of a jump of $\hat{g}$. Conversely, as $2d_n < \lambda $ for all $n > N^{\star}$ (see (\ref{Nstar1})) $g$ has at most one jump in a $2d_n$-neighborhood of a jump of $\hat{g}$. Consequently, (\ref{E1E2E3}) follows.

\begin{figure}[!h]
\includegraphics[width= 5cm, height= 5cm]{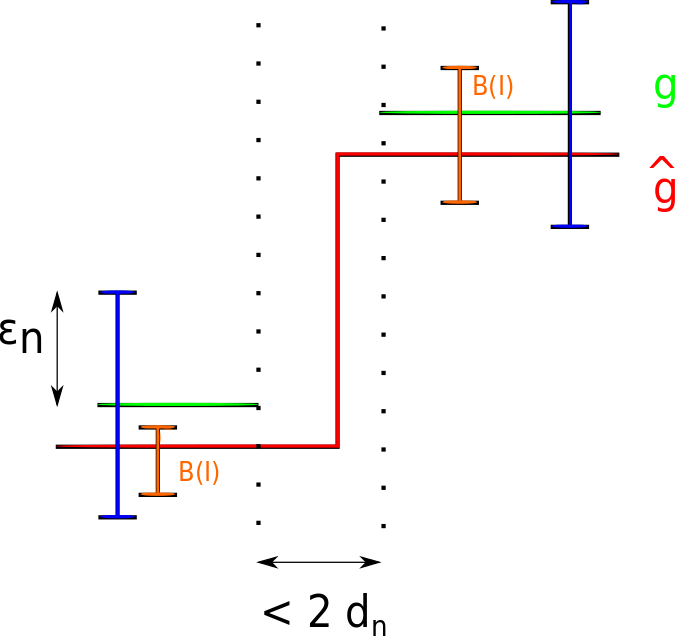}
\caption{The key argument underlying $E_1 \cap E_2 \subset E_3$.}\label{fig: proof}
\centering
\end{figure}

Furthermore, as $\epsilon_n < \delta(a_2 - a_1) / (2m(a_k - a_1))$ for all $n > N^{\star}$ (see (\ref{Nstar1})), Theorem \ref{lem: epsilon} implies that 
\begin{align}\label{E3}
E_3 \subset A_n.
\end{align}

In the following we write $q_n\ZuWeis q_n(\beta_n)$.

(\ref{E1E2E3}) and (\ref{E3}) implies that for all $n > N^{\star}$ 
\begin{align*}
\Pp\left(A_n \middle|\; T_n (Y,g)\leq q_n(\alpha_n)\right)\geq \Pp\left(E_1\cap E_2\middle| \; T_n (Y,g)\leq q_n(\alpha_n) \right).
\end{align*} 

First, consider $E_1$ conditioned on $\{T_n (Y,g)\leq q_n(\alpha_n)\}$:

Every interval $I \in \cI_1$ includes a sub-interval $I^{\prime}$, which is the union of two constant pieces of $g$ and, as $2d_n < \lambda $ for all $n > N^{\star}$ (see (\ref{Nstar1})), $I^{\prime} \in \cI_3$. 

Consequently, conditioned on $\{T_n (Y,g)\leq q_n(\alpha_n)\}$ we have that for all $n> N^{\star}$
\begin{align*}
E_1 \supseteq \bigcap_{I \in \cI_3}\{B(I)\in \fB_{\text{nc}}\}
\supseteq \bigcap_{I\in \cI_3}\{\delta > \norm{B(I_1)}+ \norm{B(I_2)}\},
\end{align*}
where $I_1$ and $I_2$ are the sub-intervals of $I\in \cI_3$ such that $g|_{I_i}\equiv g_i^I$ for $i= {1,2}$ (as in the definition of $\cI_3$).

By the definition of $\cI_3$ it follows that $\abs{I_1} \geq \lambda - 2 d_n \geq \lambda /3$ for all $n > N^\star$ and $\abs{I_2} > d_n$ and hence, (\ref{def: box}) implies
\begin{align*}
&\norm{B(I_1)}+ \norm{B(I_2)} \leq 2\left( \frac{q_n +\sqrt{2\ln(3 e / \lambda)}}{\sqrt{n\lambda/3}/\sigma}+\frac{q_n+\sqrt{2\ln(e/d_n)}}{\sqrt{n d_n}\sigma}\right) \\
= &\frac{2 \sigma}{\sqrt{n}}\left(\sqrt{\frac{3}{\lambda}}\left(q_n +\sqrt{2\ln(3 e / \lambda)}\right)+ \sqrt{\frac{1}{d_n}}\left(q_n +\sqrt{2\ln(e/d_n)}\right)\right).
\end{align*}

In summary we obtain that conditioned on $\{T_n (Y,g)\leq q_n(\alpha_n)\}$ for all $n > N^{\star}$
\begin{align}\label{A1cond}
\begin{split}
E_1
&\supseteq \left\{\delta > \frac{2 \sigma}{\sqrt{n}}\left(\sqrt{\frac{3}{\lambda}}\left(q_n +\sqrt{2\ln(3 e / \lambda)}\right)+ \sqrt{\frac{1}{d_n}}\left(q_n +\sqrt{2\ln(e/( d_n))}\right)\right) \right\}\\
&= \left\{  q_n < \left( \frac{\sqrt{n}\delta}{2 \sigma} - \sqrt{\frac{6\ln(3 e / \lambda)}{\lambda}} -  \sqrt{\frac{2\ln(e/( d_n))}{d_n}}  \right) \left(\sqrt{\frac{3}{\lambda}} + \sqrt{\frac{1}{d_n}}\right)^{-1} \right\}\\
&\supseteq \left\{  q_n <  \frac{\sqrt{n}\delta}{4 \sigma}  \left(\sqrt{\frac{3}{\lambda}} + \frac{\sqrt{n}}{\ln(n)} \right)^{-1} \right\}\\
&\supseteq \left\{  q_n <  \frac{\delta}{9 \sigma} \ln(n) \right\},
\end{split}
\end{align}
where the second inclusion results from (\ref{Nstar2}) and the last inclusion from $2d_n < \lambda $ for all $n > N^{\star}$ (see (\ref{Nstar1})).

In particular, (\ref{A1cond}) and (\ref{qnlog}) yield $\Pp(E_1|T_n (Y,g)\leq q_n(\alpha_n)) = 1$ for all $n > N^{\star}$. 

Second, consider $E_2$ conditioned on $\{T_n (Y,g)\leq q_n(\alpha_n)\}$:

By (\ref{qnlog}), (\ref{epsilonN}), and (\ref{c0c1c2}) it holds for all $I = [x_i, x_j] \in \cI$ that
\begin{align*}
\norm{B(I)} = 2\sigma \frac{q_n + \sqrt{2\ln(\frac{en}{j-i+1})}}{\sqrt{j-i+1}} \leq 2 \sigma \frac{\frac{\delta}{9 \sigma} \ln(n) + \sqrt{2\ln(\frac{e}{\lambda})}}{\sqrt{n \lambda}} < \epsilon_n / 2
\end{align*}
and as $\bar{g}_I \ZuWeis \sum_{l \in I} g(x_l) / (n \abs{I}) \in B(I)$,
\begin{align*}
E_2 &\supseteq  \bigcap_{I \in \cI_2}\{ |\bar{g}_I - g_1^{I}|\leq \epsilon_n - \norm{B(I)} \}\\
&\supseteq  \bigcap_{I \in \cI_2}\{ |\bar{g}_I - g_1^{I}|\leq \epsilon_n/2 \}.
\end{align*}
Moreover, for $I \in \cI_2$ 
\begin{align}\label{gijg1}
\begin{aligned}
\abs{\bar{g}_I - g_1^{I}} &= \abs{(g_2^I - g_1^I)\frac{\abs{I_2}}{\abs{I}} + (g_3^I - g_1^I)\frac{\abs{I_3}}{\abs{I}}}\\
&\leq \frac{\abs{I_2} + \abs{I_3}}{\abs{I}}(a_k - a_1)
\leq \frac{2d_n}{\lambda} (a_k - a_1).
\end{aligned}
\end{align}
Summarizing, conditioned on $\{T_n (Y,g)\leq q_n(\alpha_n) \}$
\begin{align}\label{A2cond}
E_2 &\supseteq
\Big\{ \frac{2d_n}{\lambda} (a_k - a_1) \leq \frac{\delta + \sqrt{2 \sigma^2 \ln(e/\lambda)}}{2\sqrt{\lambda}} \frac{\ln(n)}{\sqrt{n}} \Big \}\\
&=
\Big\{ \frac{\ln(n)}{\sqrt{n}} \leq \sqrt{\lambda}\; \frac{\delta + \sqrt{2 \sigma^2 \ln(e/\lambda)}}{4 (a_k - a_1)}\Big \}.
\end{align}
(\ref{Nstar1}) implies that the right hand side of (\ref{A2cond}) holds for all $n \geq N^{\star}$ and in particular, $\Pp(E_2|T_n (Y,g)\leq q_n(\alpha_n)) = 1$ for all $n \geq N^{\star}$.

Together with (\ref{A1cond}) this gives $\Pp(E_1 \cap E_2 | T_n (Y,g)\leq q_n(\alpha_n)) = 1$ for all $n > N^{\star}$. This proves the assertion.
\end{proof} 

With Theorem \ref{theoconsi} and Theorem \ref{lem: H} the proof of Theorem \ref{consimain} is straight forward.

\begin{proof}[Proof of Theorem \ref{consimain}]
Let $A_n$ be as in Theorem \ref{theoconsi},
\begin{align*}
\fT_\alpha \ZuWeis \{T_n(Y,g) \leq q_n(\alpha)\}, \quad \text{and}\quad
\hat{\fT}_{\alpha} \ZuWeis \{T_n(Y,\hat{g}) \leq q_n(\alpha)\}.
\end{align*}
Theorem \ref{lem: H} implies that 
\begin{align}\label{theoconsiE}
\Pp\left(\hat{\fT}_{\beta_n} \middle| \; \fT_{\alpha_n}\right) = 1.
\end{align}

From  (\ref{theoBoundDF}) we deduce that for $\beta_n$ as in (\ref{alphaN}) $q_n(\beta_n) < \delta / (9 \sigma) \ln(n)$. 
Thus, Theorem \ref{theoconsi} yields
\begin{align}\label{lemHE}
\Pp\left(A_n \middle| \; \fT_{\alpha_n} \cap \hat{\fT}_{\beta_n} \right) = 1.
\end{align}
(\ref{theoconsiE}) and (\ref{lemHE}) give
\begin{align*}
\Pp\left( A_n \right) &\geq \Pp\left( A_n \middle| \; \fT_{\alpha_n} \cap \hat{\fT}_{\beta_n} \right) \Pp\left(\fT_{\alpha_n} \cap \hat{\fT}_{\beta_n} \right)\\
&\geq \Pp\left(\fT_{\alpha_n} \cap \hat{\fT}_{\beta_n} \right) = \Pp(\fT_{\alpha_n}) \geq 1-\alpha_n.
\end{align*}
Finally, remember that the identfiability condition $ASB(\omega)\geq \delta > 0$ implies that $g$ jumps if and only if $f$ jumps. Hence, when $f^i$ and $\hat{f}^i$ take the same function values on constant pieces, results about change points of $g$ directly translate to results about change points of $f^1,\ldots,f^m$.
\end{proof}

\subsection{Proof of Theorem \ref{theo: CIf}} \label{subsec: app2}
\begin{proof}
It follows from the proof of Theorem \ref{theoconsi} that conditioned on $\{T_n(Y,g) \leq q_n(\alpha_n)\}$
\begin{align}\label{oa1}
\max_{a \in \imag(f)}\abs{\omega^{\top}a - \hat{\omega}^{\top}a} \leq  \left(\delta \frac{\ln(n)}{\sqrt{n}} + \sqrt{\frac{8\sigma^2 \ln(e/\lambda)}{n\lambda}}\right)
\end{align}
and
\begin{align}\label{k1}
K(\hat{\omega}^{\top} f) = K(\hat{g}).
\end{align}
Let $B(i,j) = [\underline{b}_{ij}, \overline{b}_{ij}]$ be as in (\ref{def: box}) and 
\begin{align*}
\tilde{B}(i,j)\ZuWeis \Bigg [\underline{b}_{ij} - \left(\delta \frac{\ln(n)}{\sqrt{n}} + \sqrt{\frac{8\sigma^2 \ln(e/\lambda)}{n\lambda}}\right), \overline{b}_{ij} + \left(\delta \frac{\ln(n)}{\sqrt{n}} + \sqrt{\frac{8\sigma^2 \ln(e/\lambda)}{n\lambda}}\right)\Bigg],
\end{align*}
with $q = q_n(\beta)$ as in (\ref{alphaN}), then
\begin{align*}
&\Pp\left(f=(f^1,...,f^m)^\top\in \tilde{\cH}(\beta)\right)\\ = &\Pp\left(\bigcap_{\substack{1\leq i \leq j \leq n \\ (\hat{\omega}^{\top}f)|_{[i,j]}\equiv (\hat{\omega}^{\top}f)_{ij}  }} (\hat{\omega}^{\top}f)_{ij} \in \tilde{B}(i,j) \text{ and } K\left(\hat{\omega}^{\top}f\right) = K(\hat{g})\right)\\
\geq &\Pp\left( \bigcap_{\substack{1\leq i \leq j \leq n \\ g|_{[ij]}\equiv g_{ij} }} g_{ij} \in B(i,j)\text{ and } T_n(Y,g) \leq q_n(\alpha_n)\right) \\
= &\Pp\left(T_n(Y,g)\leq q_n(\beta) \right) + \co(1),
\end{align*}
where the inequality in the third line follows from (\ref{oa1}) and (\ref{k1}). Finally, the assertion follows from the fact that $\delta \leq (a_2 - a_1)/m $.
\end{proof}

\clearpage

\section{Algorithms} \label{sec: algoS}
\subsection{Pseudocode for Algorithm CRW} \label{subsec: psCRW}
\quad \\

\begin{algorithm}[h]
\renewcommand\thealgorithm{}
\caption{CRW (Confidence region for weights)}
\begin{algorithmic}[1]
\Input $Y$, $m$, $\fA$, $\alpha$, $\lambda$, $\lambda^{\star}$  \Comment{see the SBSSR-model and Remark \ref{rem:prior}}
\State $\overline{\fB} \gets \{B(i,j)\in\fB\setminus\fB_{\text{nc}}: j - i + 1 \geq \lambda^{\star}n \}$ \Comment{see R\ref{notconst} and  R\ref{minred}}
\State $\fB^{\star} \gets \{[\underline{b},\overline{b}] \in \overline{\fB}: \overline{b} \geq a_1 \text{ and } \underline{b} \leq a_1 + \frac{a_2- a_1}{m} \}$ \Comment{see R\ref{validred} }
\For{i=2\ldots m}
\State $\fB^{\star} \gets$ \begin{align*}
\Big\{[\underline{b}_1,\overline{b}_1] \times ... \times  [\underline{b}_i,\overline{b}_i] \in  \fB^{\star} \times \overline{\fB}:\\
\frac{a_2 + (m-1)a_1 - \sum_{k=1}^{r-1} \underline{b}_k}{m-r+1} \geq \underline{b}_{r} \text{ and } \underline{b}_{r-1} \leq \overline{b}_{r}  \Big \}
\end{align*} 
\Comment{see R\ref{validred} }
\EndFor
\State $\fB^{\star} \gets 
\Big\{[\underline{b}_1,\overline{b}_1]\times ... \times [\underline{b}_m,\overline{b}_m] \in \fB^{\star}:
\sum_{j = 1}^m \overline{b}_r \geq a_2 + (m-1) a_1 \Big\}$ \Comment{see R\ref{validred}}
\State $\fB^{\star} \gets \text{ R\ref{exred} applied to $\fB^{\star}$ }$ \Comment{see Remark \ref{rem:prior}}\\
\Return $\bigcup_{B \in \fB^{\star}} A^{-1}B$
\end{algorithmic}
\end{algorithm}

\subsection{Computation of $(\hat{f}^1,\ldots,\hat{f}^m)$}\label{subsec: comFDP}
For a given $\beta \in (0,1)$ SLAM solves the constrained optimization problem (\ref{def: gEstS}). 

Note that $\hat{f}^1, \ldots, \hat{f}^m$ are the unique source functions such that $\sum_{i=1}^m \hat{\omega}_i \hat{f}^i = \hat{g}$ for
\begin{align}\label{gEstSS}
\hat{g} \ZuWeis \argmax_{\tilde{g} \in \fH(\beta)} \sum_{i = 1}^n \phi_{\tilde{g}(x_i)}(Y_i),
\end{align}
with
\begin{align}\label{restH}
\fH(\beta)\ZuWeis\{\tilde{g}\in \cS(\{\hat{\omega}^{\top}a: a\in \fA^m\}) : T_n\left(Y, \tilde{g}\right) \leq q_n(\beta) \text{ and } K\left(\tilde{g}\right) = \hat{K}\}
\end{align}
and $\hat{K}$ as in (\ref{Khat}).
Frick et al. \cite{frick} provide a pruned dynamic programming algorithm how to efficently solve (\ref{gEstSS}) without the restriction that $\hat{g}$ can only attain values in $\{\hat{\omega}^{\top} a : a\in \fA^m\}$ as it is the case here, see (\ref{restH}). As this restriction is crucial for SLAM we outline the details of the necessary modifications below.

To this end, it is necessary for a finite set $\fL = \{l_1,\ldots,l_k\}$ of possible function values to check finiteness of their \textit{minimal cost} $d_{[i,j]}^{\star} = \min_{\theta \in \R}d_{[i,j]}$ (see \cite[eq. 30]{frick}) with $\R$ replaced by $\fL$.

In \cite{frick} finiteness of $d_{[i,j]}^{\star} = \min_{\theta \in \R}d_{[i,j]}$ is examined by the relation
\begin{align}\label{minCostCond}
\min_{\theta \in \R}d_{[i,j]} = \infty \quad \Leftrightarrow \quad \max_{i \leq u \leq v \leq j}\underline{b}_{uv} > \min_{i \leq u \leq v \leq j}\overline{b}_{uv},
\end{align}
with $\{B(i,j) = [\underline{b}_{ij}, \overline{b}_{ij}] : 1\leq i \leq j \leq n \}$ as in (\ref{def: box}).

Let $L$ be any number such that $L> \max(\fL)$ and define $Q(i,j) = $
\begin{align}
[\underline{q}_{ij}, \overline{q}_{ij}] \ZuWeis \begin{cases} [\max(\fL \cap B(i,j)), \min(\fL \cap B(i,j))] & \text{ if } \fL \cap B(i,j) \neq \emptyset \\
[L, L] & \text{ else} \end{cases}.
\end{align}
Then we observe, as in (\ref{minCostCond}), that
\begin{align}
\min_{\theta \in \fL}d_{[i,j]} = \infty \quad \Leftrightarrow \quad \max_{i \leq u \leq v \leq j}\underline{q}_{uv} > \min_{i \leq u \leq v \leq j}\overline{q}_{uv}.
\end{align}

This allows to adapt the dynamic program from \cite{frick}.

Again, in order to reduce computation time, one can only consider subintervals, e.g., of dyadic length, possibly at the expense of deletion power.

\FloatBarrier

\clearpage

\section{Additional figures and tables} \label{sec: aftS}

\subsection{Additional tables and figure from Section \ref{subsec: Appsim}}

\quad \\

\begin{table}[ht!]
\caption{Weight vector $\omega$ for $m = 2,3,4,5$ such that the  $ASB(\omega) = 0.02$.}\label{tab: omegaM}
\begin{tabular}{|c|c|c|c|c|}
\hline
& $m = 2$ & $m = 3$ & $m = 4$ & $m = 5$\\
\hline \hline 
$\omega$ & $(0.02, 0.98)$ & $(0.02, 0.04, 0.94)$ & $(0.04, 0.06, 0.12, 0.78)$ & $(0.06, 0.08, 0.12, 0.16, 0.58)$\\
\hline
\end{tabular}
\end{table}

\begin{table}[ht]
\tiny
\caption{Influence of the number of source functions $m$ for $m = 2,3,4,5$.}\label{tab: depM}
\begin{tabular}{|c|c|c|c|c|}
\hline
 & $m = 2$ & $m = 3$ & $m = 4$ & $m = 5$\\
\hline
$\mae(\hat{\omega})$ $[10^{-4}]$ & $(1 , 1)$ & $(11 , 18, 24)$ & $(90, 154, 62, 69)$ & $(91, 68, 81, 196, 54)$ \\
$\dist(\omega,\cC_{1-\alpha})$ $[10^{-3}]$ & $11$ & $23$ & $63$ & $54$ \\
$\mean(\omega \in \cC_{1 - \alpha})$ $[\%]$ & $100$ & $99.99$ & $99.96$ & $100$ \\
$\overline{\omega}_i - \underline{\omega}_i$ $[10^{-3}]$ & $(21,21)$ & $(37, 33, 23)$ & $(68, 93, 35, 23)$ & $(40, 55, 84, 63, 23)$ \\
$\miae(\hat{f}^i)$ $[10^{-3}]$ & $(0.2, 0.0)$ & $(26, 9, 0.0)$ & $(115, 103, 67, 0.0)$ & $(315, 317, 49, 183, 0.0)$ \\
$\meanNcp - K$ & $(0, 0)$ & $(0.22, -0.03, 0)$ & $(3.7, 2.6, -0.6, 0)$ & $(2.75, 2.28, 0.75, -1.61, 0)$ \\
$\medNcp - K$ & $(0,0)$ & $(0,0,0)$ & $(4,2,0,0)$ & $(2,2,0,-2,0)$ \\
$\meanKK_i$ $[\%]$ & $(99.8, 99.8)$ & $(88.5, 98, 100)$ & $(15.9, 31, 69.4, 100)$ & $(7.1, 30.4, 63.8, 12, 99.9)$ \\
$\meanKK$ $[\%]$ & $99.8$ & $87.2$ & $15.8$ & $1$ \\
$\max_i \min_j \abs{\tau_i - \hat{\tau}_j}$ & $(0.37, 0.02)$ & $(33.82, 4.77, 0.00)$ & $(245.49, 95.75, 2.52, 0.00)$ & $(374.38, 208.32, 40.12, 7.41, 0.02)$ \\
$\max_j \min_i \abs{\tau_i - \hat{\tau}_j}$ & $(0.03, 0.00)$ & $(18.59, 12.53, 0.000)$ & $(9.61, 18.66, 126.33, 0.00)$ & $(83.09, 117.17, 61.13, 348.89, 0.00)$ \\
$\vm $ $[\%]$ & $(99.9, 100)$ & $(88.3, 96.2, 100)$ & $(60.9, 83.4, 68.6, 100)$ & $(37.5, 54.1, 82.8, 12.6, 100)$ \\
$\fpsle$ & $(0.07, 0.00)$ & $(8.98, 6.05, 0.00)$ & $(51.52, 21.36, 78.23, 0.00)$ & $(110.3, 92.21, 34.98, 216.82, 0.00 )$ \\
$\fnsle$ & $(0.3, 0.02)$ & $(24.04, 3.22, 0.00)$ & $(168.04, 45.09, 62.15, 0.00)$ & $(205.75, 137.64, 41.29, 90.02, 0.02)$ \\
 $\mean(f \in \tilde{\cH}(\beta))$ $[\%]$ & $99.93$ & $99.49$ & $98.77$ & $91.08$ \\
\hline
\end{tabular}
\end{table}

\begin{table}[hb]

\caption{Influence of the number of alphabet values $k$ for $k = 2,3,4$.}\label{tab: depK}
\begin{tabular}{|c|c|c|c|}
\hline
 & $k = 2$ & $k = 3$ & $k = 4$\\
\hline
$\mae(\hat{\omega})$ $[10^{-3}]$ & $(19 , 12)$ & $(18 , 12)$ & $(15, 11)$  \\
$\dist(\omega,\cC_{1-\alpha})$ $[10^{-3}]$ & $51$ & $51$ & $47$  \\
$\mean(\omega \in \cC_{1 - \alpha})$ $[\%]$ & $100$ & $100$ & $100$ \\
$\overline{\omega}_i - \underline{\omega}_i$ $[10^{-3}]$ & $(71,71)$ & $(71, 71)$ & $(67, 67)$ \\
$\miae(\hat{f}^i)$ $[10^{-2}]$ & $(29, 0)$ & $(49, 0)$ & $(60,0)$ \\
$\meanNcp - K$ & $(-6.65, 0)$ & $(-7.42, 0)$ & $(-7.04, 0)$ \\
$\medNcp - K$ & $(-6,0)$ & $(-7,0)$ & $(-7,0)$ \\
$\meanKK_i$ $[\%]$ & $(0.39, 99.99)$ & $(0, 100)$ & $(0, 100)$ \\
$\meanKK$ $[\%]$ & $0.39$ & $0$ & $0$ \\
$\max_i \min_j \abs{\tau_i - \hat{\tau}_j}$ & $(17.5, 0.0)$ & $(22.0, 0.0)$ & $(23.31, 0.00)$ \\
$\max_j \min_i \abs{\tau_i - \hat{\tau}_j}$ & $(96.0, 0.0)$ & $(134.4, 0.0)$ & $(79.8, 0.0)$ \\
$\vm $ $[\%]$ & $(81.7, 100)$ & $(78, 100)$ & $(81.5, 100)$ \\
$\fpsle$ & $(0.4, 0.0)$ & $(58.3, 0.0)$ & $(37.2, 0.0)$ \\
$\fnsle$ & $(25.7, 0.0)$ & $(29.3, 0.0)$ & $(25.2, 0.0)$ \\
 $\mean(f \in \tilde{\cH}(\beta))$ $[\%]$ & $94.60$ & $98.49$ & $98.60$ \\
\hline
\end{tabular}
\end{table}

\begin{figure}[hb!]
\includegraphics[width=\textwidth]{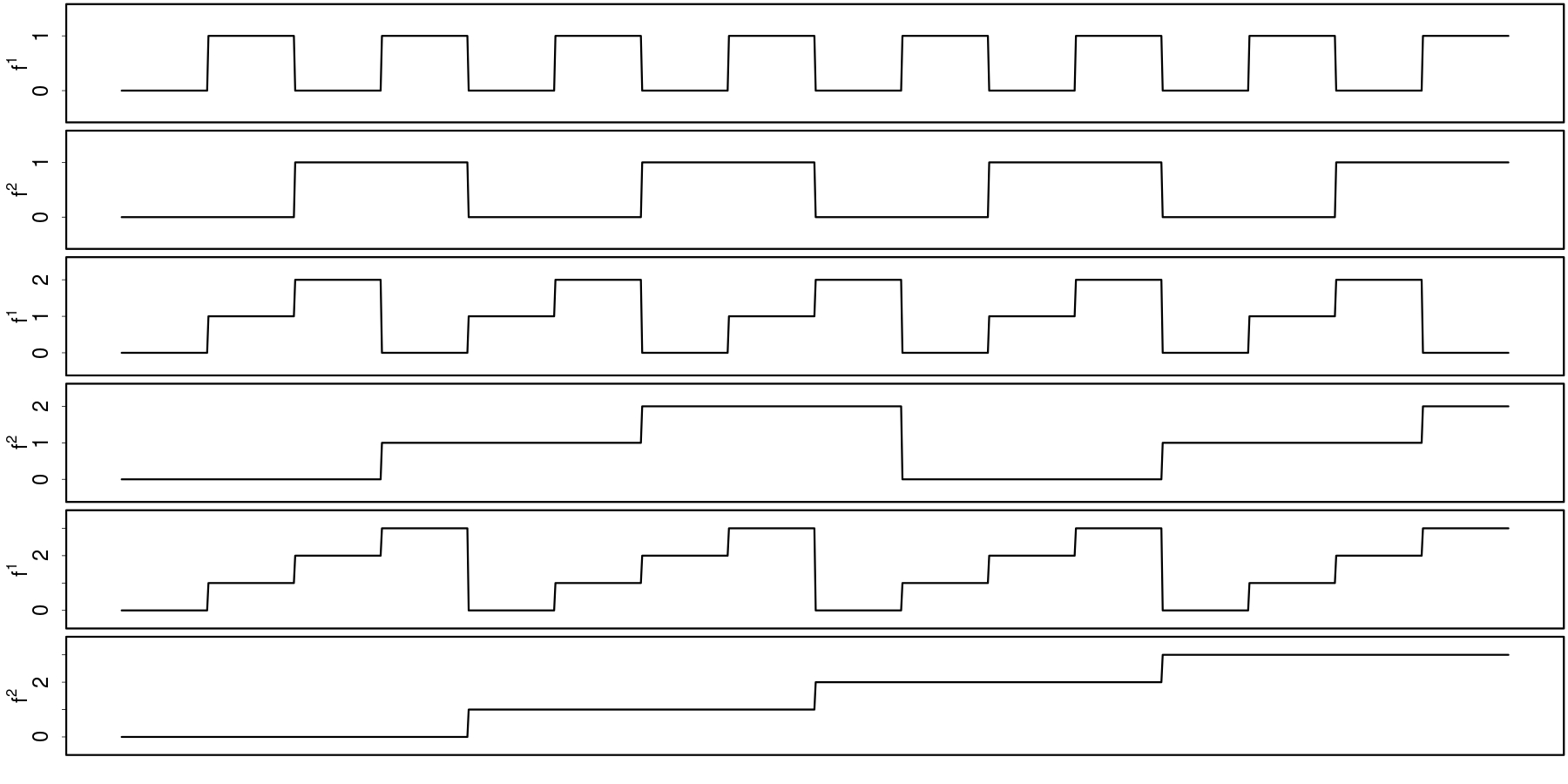}
\caption{$f^1$ and $f^2$ from (\ref{fk}) in Section \ref{subsec: depK} for $\fA = \{0,1\}, \{0,1,2\}$, and $\{0,1,2,3\}$ (from top to bottom).}\label{fig: fk}
\end{figure}

\begin{table}[ht]
\caption{Influence of the confidence level $\alpha$ on $\hat{\omega}$ and $\cC_{1-\alpha}$ for $\alpha = 0.01, 0.05, 0.1$.}\label{tab: stabABO}
\begin{tabular}{|c|c|c|c|}
\hline
\multicolumn{4}{|c|}{ $\sigma = 0.02$ }\\
\hline
 & $\alpha = 0.01$ & $\alpha = 0.05$ & $\alpha = 0.1$\\
\hline
$\mae(\hat{\omega})$ $[10^{-3}]$ & $(2 , 2, 2)$ & $(1 , 1, 1)$ & $(1 , 1, 1)$ \\
$\dist(\omega,\cC_{1-\alpha})$ $[10^{-3}]$ & $29$ & $25$ & $24$  \\
$\mean(\omega \in \cC_{1 - \alpha})$ $[\%]$ & $100$ & $100$ & $100$ \\
$\overline{\omega}_i - \underline{\omega}_i$ $[10^{-3}]$ & $(48, 46, 44)$ & $(43, 42, 42)$ & $(42, 42, 42)$\\
\hline

\multicolumn{4}{|c|}{ $\sigma = 0.05$ }\\
\hline
 & $\alpha = 0.01$ & $\alpha = 0.05$ & $\alpha = 0.1$\\
\hline
$\mae(\hat{\omega})$ $[10^{-3}]$ & $(22 , 7, 16)$ & $(23 , 7, 16)$ & $(22 , 7, 16)$ \\
$\dist(\omega,\cC_{1-\alpha})$ $[10^{-3}]$ & $109$ & $105$ & $102$  \\
$\mean(\omega \in \cC_{1 - \alpha})$ $[\%]$ & $100$ & $100$ & $99$ \\
$\overline{\omega}_i - \underline{\omega}_i$ $[10^{-3}]$ & $(168 ,123, 115)$ & $(160, 112, 106)$ & $(155, 107, 102)$\\
\hline


\multicolumn{4}{|c|}{ $\sigma = 0.1$ } \\
\hline
 & $\alpha = 0.01$ & $\alpha = 0.05$ & $\alpha = 0.1$\\
\hline
$\mae(\hat{\omega})$ $[10^{-3}]$ & $(59, 51 ,13)$ & $( 45 ,48, 13)$ & $(32 ,43 ,18)$ \\
$\dist(\omega,\cC_{1-\alpha})$ $[10^{-3}]$ & $231$ & $218$ & $210$  \\
$\mean(\omega \in \cC_{1 - \alpha})$ $[\%]$ & $100$ & $100$ & $100$ \\
$\overline{\omega}_i - \underline{\omega}_i$ $[10^{-3}]$ & $( 329, 344 ,282)$ & $(305, 323, 226)$ & $(276, 312, 212)$\\
\hline
\end{tabular}
\end{table}

\begin{table}[ht]
\caption{Influence of the confidence levels $\alpha$ and $\beta$ on $\hat{f}$ and $\tilde{\cH}(\beta)$ for each $(\alpha, \beta) \in \{0.01, 0.05, 0.1\}^2$, for $\sigma = 0.02, 0.05, 0.1$. In the displayed matrices $\alpha$ increases within a column and $\beta$ increases within a row.}\label{tab: stabABF}
\tiny
\begin{tabular}{|c|c|c|c|}
\hline
\multicolumn{4}{|c|}{ $\sigma = 0.02$ } \\
\hline
& $f^1$ & $f^2$ & $f^3$ \\
\hline
$\miae(\hat{f}^i)$ $[10^{-4}]$  & $\begin{pmatrix} 0  &  2 &   10 \\  0 &   2 &   10  \\  0 &   2 &   10 \end{pmatrix}$ &
$\begin{pmatrix} 6 & 3 & 11 \\ 9 & 5 & 12   \\ 11 & 7  & 13 \end{pmatrix}$
 &
$\begin{pmatrix} 3 &  1 &  4  \\  5  & 2  & 4 \\  6 &  3  & 5  \end{pmatrix}$ 
  \\
\hline
$\medNcp - K$ &  
$ \begin{pmatrix} 0  &  0  &  0 \\   0&    0  &  0 \\   0  &  0  &  0
 \end{pmatrix} $ &
$ \begin{pmatrix} 0  &  0  &  0 \\   0&    0  &  0 \\   0  &  0  &  0
 \end{pmatrix} $
&
$ \begin{pmatrix} 0  &  0  &  0 \\   0&    0  &  0 \\   0  &  0  &  0
 \end{pmatrix} $ \\
\hline
$\meanKK_i$ $[\%]$ & $\begin{pmatrix}
100  & 100 &  100 \\  100 &  100 &  100  \\ 100  & 100  & 100 \end{pmatrix}$
&
$\begin{pmatrix}
98 & 100 & 100\\  97 &  99 &  99 \\  96 &  98 &  99 \end{pmatrix}$
&
$\begin{pmatrix}
 99 & 100 & 100 \\   98  & 99 &  99 \\  97 &  99 &  99 \end{pmatrix}$ \\
\hline
$\meanKK$ $[\%]$ & \multicolumn{3}{c|}{$\begin{pmatrix}
98  & 99 & 100\\ 97  & 99  & 99 \\ 96  & 98  & 99  \end{pmatrix}$} \\
\hline
$\vm $ $[\%]$ & $ \begin{pmatrix}
 100 &  100 &  100 \\  100 &  100 &  100 \\  100 &  100  & 100 \end{pmatrix} $
&
 $ \begin{pmatrix}
 100 &  100 &  100 \\  100 &  100 &  100 \\  100 &  100  & 100 \end{pmatrix} $
&
 $ \begin{pmatrix}
 100 &  100 &  100 \\  100 &  100 &  100 \\  100 &  100  & 100 \end{pmatrix} $ \\
\hline

$\mean(f \in \tilde{\cH}(\beta))$ $[\%]$ & \multicolumn{3}{c|}{
$ \begin{pmatrix}
95.8 & 93.3 & 92.3\\ 99.0 & 97.7 & 97.0 \\99.2 & 98.6 & 98.1\end{pmatrix} $
} \\

\hline

$\mean(f^i \in \tilde{\cH}(\beta)_ i)$ $[\%]$ & $ 
\begin{pmatrix}
 99.90 & 99.74 & 99.34 \\  99.94 & 99.78 & 99.64 \\  99.90 & 99.70 & 99.68 \end{pmatrix} 
 $
&
 $ 
\begin{pmatrix}
 99.84 & 99.60 & 99.38 \\  99.92 & 99.84 & 99.74 \\  99.90 & 99.82 & 99.74 \end{pmatrix} 
 $
&
 $ 
\begin{pmatrix}
 96.68 & 95.46 & 94.92 \\  99.18 & 98.34 & 98.10 \\  99.42 & 99.02 & 98.64 \end{pmatrix} 
 $   \\
\hline


\multicolumn{4}{|c|}{ $\sigma = 0.05$ } \\
\hline
& $f^1$ & $f^2$ & $f^3$ \\
\hline
$\miae(\hat{f}^i)$ $[10^{-3}]$  & $\begin{pmatrix} 6  &  7 &   8  \\  6  &  8  &  9   \\  6  &  8  &  9 \end{pmatrix}$ &
$\begin{pmatrix} 160 & 161 & 160 \\ 164 & 165 & 164   \\ 160 & 161  &161 \end{pmatrix}$
 &
$\begin{pmatrix} 80 &  80 &  80  \\  82  & 83  & 82 \\  80 &  80  & 80  \end{pmatrix}$ 
  \\
\hline
$\medNcp - K$ &  
$ \begin{pmatrix} 0  &  0  &  0 \\   0&    0  &  0 \\   0  &  0  &  0
 \end{pmatrix} $ &
$ \begin{pmatrix} 2 &   2  &  2 \\   2  &  2 &   2 \\   2  &  2  &  2 \end{pmatrix} $
&
$ \begin{pmatrix} 
-2 &  -2 &  -2 \\  -2  & -2  & -2  \\ -2 &  -2 &  -2
\end{pmatrix} $ \\
\hline
$\meanKK_i$ $[\%]$ & $\begin{pmatrix}
96  & 90 &  85 \\  93 &  86 &  80  \\ 93  & 85  & 80 \end{pmatrix}$
&
$\begin{pmatrix}
21 &  19 &  17 \\  19  & 16 &  15 \\  21  & 19  & 17 \end{pmatrix}$
&
$\begin{pmatrix}
 24 &  25 &  27 \\  21 &  23 &  24 \\  24 &  25 &  26 \end{pmatrix}$ \\
\hline
$\meanKK$ $[\%]$ & \multicolumn{3}{c|}{$\begin{pmatrix}
19 & 16 &14 \\ 17& 14 &12 \\ 19 &16 &14  \end{pmatrix}$} \\
\hline
$\vm $ $[\%]$ & $ \begin{pmatrix}
 99 &  99 &  99 \\  99 &  99 &  99 \\  99 &  99  & 99 \end{pmatrix} $
&
$ \begin{pmatrix}
 92 &  92 &  92 \\  92 &  92 &  92 \\  92 &  92  & 92 \end{pmatrix} $
&
$ \begin{pmatrix}
91  & 91  & 91 \\  91 &  91 &  91 \\  91  & 91  & 91 \end{pmatrix} $ \\
\hline

$\mean(f \in \tilde{\cH}(\beta))$ $[\%]$ & \multicolumn{3}{c|}{$ 
 \begin{pmatrix}
83.1 & 76.7 & 74.0 \\ 81.3 & 75.6 & 73.4 \\ 81.7 & 76.4 & 74.5 \end{pmatrix}
$} \\

\hline

$\mean(f^i \in \tilde{\cH}(\beta)_ i)$ $[\%]$ & $ 
\begin{pmatrix}
 100 &  100 &  100 \\  100 &  100 &  99.98 \\  100 &  100 &  99.98 \end{pmatrix} 
 $
&
 $  
\begin{pmatrix}
 89.34 & 84.78 & 82.82 \\  86.60 & 83.04 & 83.18 \\  87.24 & 84.16 & 83.18 \end{pmatrix} 
 $
&
 $ 
\begin{pmatrix}
 85.80 & 80.56 & 78.34 \\  83.14 & 78.48 & 77.14 \\  83.58 & 79.48 & 78.16 \end{pmatrix} 
 $   \\


\hline
\multicolumn{4}{|c|}{ $\sigma = 0.1$ } \\
\hline
& $f^1$ & $f^2$ & $f^3$ \\
\hline
$\miae(\hat{f}^i)$ $[10^{-3}]$  & $\begin{pmatrix}  327 & 327 & 327 \\ 297 & 296  & 296  \\ 255 & 254 & 253 \end{pmatrix}$ &
$\begin{pmatrix} 245 & 246 & 246 \\ 233 & 234 & 234 \\ 231 &  232 & 232  \end{pmatrix}$
 &
$\begin{pmatrix} 90 &  91 &  91 \\ 67 &  68 &  68 \\ 75 &  76 &  76 \end{pmatrix}$ 
  \\
\hline
$\medNcp - K$ &  
$ \begin{pmatrix}
2 &   3  &  3  \\  1  &  2 &   2 \\   1  &  1  &  1 \end{pmatrix} $ &
$ \begin{pmatrix}
1 &   1  &  1  \\  0  &  0 &   0 \\   0  &  0  &  0
 \end{pmatrix} $
&
$ \begin{pmatrix}
0 &    0   & 0  \\  0  &  0 &   0 \\   0  &  0  &  0
\end{pmatrix} $ \\
\hline
$\meanKK_i$ $[\%]$ & $\begin{pmatrix}
12 &   9  &  7 \\  22 &  19 &  17 \\  36 &  32 &  29 \end{pmatrix}$
&
$\begin{pmatrix}
15 &  12 &  11 \\  24 &  22 &  21 \\  35 &  33  & 32 \end{pmatrix}$
&
$\begin{pmatrix}
44 &  37 &  34 \\  62 &  53 &  49 \\  59 &  52 &  48 \end{pmatrix}$ \\
\hline
$\meanKK$ $[\%]$ & \multicolumn{3}{c|}{$\begin{pmatrix}
4& 2& 1 \\ 7& 5& 4 \\ 8& 7& 6 \end{pmatrix}$} \\
\hline
$\vm $ $[\%]$ & $ \begin{pmatrix}
 85 &  85 &  85 \\  86 &  86 &  86 \\  88 &  87 &  87 \end{pmatrix} $
&
$ \begin{pmatrix}
74  & 74  & 75 \\  73  & 74 &  74 \\  75 &  76 &  76 \end{pmatrix} $
&
$ \begin{pmatrix}
95 &  95 &  95 \\  97 &  97 &  97 \\  96  & 96 &  96 \end{pmatrix} $ \\
\hline

$\mean(f \in \tilde{\cH}(\beta))$ $[\%]$ & \multicolumn{3}{c|}{$  
\begin{pmatrix}
60.7 & 58.6 & 55.7 \\ 71.0 & 63.5 & 63.2 \\ 80.2 & 71.0 & 66.9 \end{pmatrix}
$} \\

\hline

$\mean(f^i \in \tilde{\cH}(\beta)_ i)$ $[\%]$ & $ 
\begin{pmatrix}
 90.4 & 89.6 & 89.3 \\   99.0 & 98.8 & 98.8 \\  99.7 & 99.6 & 99.6 \end{pmatrix} 
 $
&
 $ 
\begin{pmatrix}
 96.7 & 91.5 & 86.0\\  97.8 & 95.0 & 94.3 \\  97.9 & 95.2 & 92.9 \end{pmatrix} 
 $
&
 $ 
\begin{pmatrix}
 72.8 & 74.6 & 77.0\\  83.5 & 80.2 & 79.4 \\  90.1  & 86.2 & 85.6 \end{pmatrix} 
 $   \\
 
 \hline

\end{tabular}
\end{table}

\FloatBarrier

\begin{figure}[ht!]
\includegraphics[width=\textwidth]{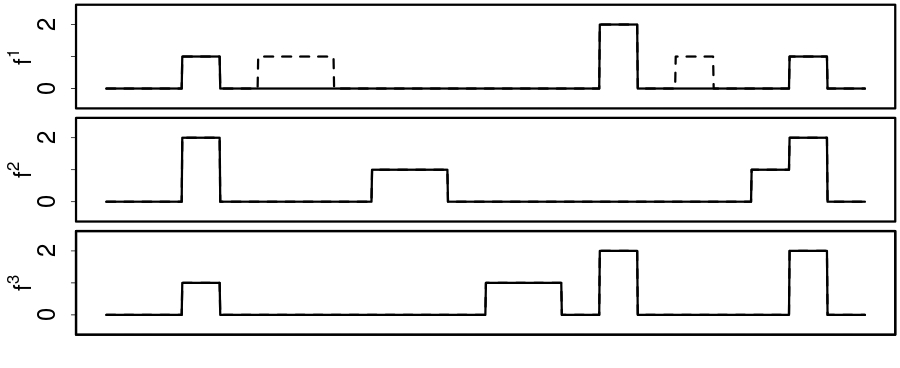}
\caption{Source functions $f$ from Example \ref{example1} modified such they violate the separability condition in (\ref{separable}) for $r = 1$ (solid line). The dotted lines indicate the removed jumps. }\label{fig: fvioF}
\end{figure}

\begin{table}[h]
\caption{Result illustrating robustness. (1): Setting as in Example \ref{example1} but with $f$ modified such it violates the separability condition in (\ref{separable}) (see Figure \ref{fig: fvioF}). (2): Setting as in Example \ref{example1}, but with $t$-distributed errors with $3$ degrees of freedom. (3): Setting as in Example \ref{example1}, but with $\chi^2$-distributed errors with $3$ degrees of freedom. } \label{tab: rob}
\begin{tabular}{|c|c|c|c|}
\hline
 & (1) & (2) &  (3) \\
\hline
$\mae(\hat{\omega})$ $[10^{-3}]$ & $( 73, 36 ,39)$ & $(43, 58, 16)$ & $( 42, 59, 17)$ \\
$\miae(\hat{f}^i)$ $[10^{-3}]$ & $(123 ,181 , 84)$ & $(447, 435, 137)$ & $( 563, 279,  99)$\\
$\medNcp - K$ & $(-4,  2,  0)$ & $(4,  1, -2)$ & $( 11,  4, -2)$ \\
$\meanKK_i$ $[\%]$ & $(10, 10, 19)$ & $( 5,  0, 33)$ & $(2, 1, 4)$ \\
$\vm $ $[\%]$ & $(71, 85, 96)$ & $( 84, 72, 88)$ & $( 78, 82, 89)$  \\
\hline
\end{tabular}
\end{table}

\begin{table}
\caption{Results illustrating the influence of the alphabet separation boundary $ASB = ASB(\omega)$ on $\hat{\omega}$ with $\omega \sim \cU(\Omega(m))$.}\label{tab: vioIOC}
\begin{tabular}{|c|c|c|c|c|c|}
\hline
& $\mae(\hat{\omega})$ $[10^{-3}]$  & $\dist(\omega,\cC_{1-\alpha})$ $[10^{-3}]$  \\
\hline
$ 0 \leq ASB \leq 0.0001$ & $( 6, 4, 5)$ & $29 $ \\
$ 0.0001 \leq ASB \leq 0.01$ & $(7, 4, 7)$ & $34 $\\
$ 0.01 \leq ASB \leq 0.02$ & $( 4, 4, 4 )$ & $30 $\\
$ 0.02 \leq ASB \leq 0.03$ & $(4, 4, 4 )$ & $29 $\\
$ 0.03 \leq ASB \leq 0.04$ & $(4, 3, 4 )$ & $31 $\\
$ 0.04 \leq ASB \leq 0.05$ & $(4, 3, 4 )$ & $31 $\\
$ 0.05 \leq ASB \leq 0.06$ & $(4, 3, 5 )$ & $31 $\\
$ 0.06 \leq ASB \leq 0.07$ & $(3, 3, 4 )$ & $31 $\\
\hline
\end{tabular}
\end{table}

\begin{table}[h]
\caption{Influence of the alphabet separation boundary $ASB = ASB(\omega)$ on $\hat{f}$ with $\omega \sim \cU(\Omega(m))$.}\label{tab: vioIFH}
\begin{tabular}{|c|c|c|c|c|c|c|}
\hline
& $\miae(\hat{f}^i)$ $[10^{-4}]$  & \multicolumn{2}{|c|}{$|\tilde{\cH}_x(0.1)|$} &  \\
& & mean & median & \\
\hline
$ 0 \leq ASB \leq 0.0001$ & $(1916, 1067,  483)$ & $2.71$ & $3$ & $ 0 \leq ASB_x \leq 0.001$ \\
$ 0.0001 \leq ASB \leq 0.01$ & $(1536,  923,  354)$ & $2.68$ & $3$ & $ 0.001 \leq ASB_x \leq 0.01$ \\
$ 0.01 \leq ASB \leq 0.02$ & $(671, 474 ,147)$ & $2.67$ & $3$ & $ 0.01 \leq ASB_x \leq 0.02$ \\
$ 0.02 \leq ASB \leq 0.03$ & $(236, 164,  40)$ & $2.66$ & $3$ & $ 0.02 \leq ASB_x \leq 0.03$ \\
$ 0.03 \leq ASB \leq 0.04$ & $(96, 37,  7)$ & $2.53$ & $2$ & $ 0.03 \leq ASB_x \leq 0.04$\\
$ 0.04 \leq ASB \leq 0.05$ & $(100, 7, 2)$ & $2.49$ & $2$ & $ 0.04 \leq ASB_x \leq 0.05$\\
$ 0.05 \leq ASB \leq 0.06$ & $(42,1, 0)$ & $2.36$ & $2$ & $ 0.05 \leq ASB_x \leq 0.1$\\
$ 0.06 \leq ASB \leq 0.07$ & $(16,4, 0)$ & $1.97$ & $1$ & $ 0.1 \leq ASB_x $ \\
\hline
\end{tabular}
\end{table}

\FloatBarrier

\begin{table}
\tiny
\caption{Influence of prior information on $\lambda$ for prior knowledge $\lambda \geq 0.05, 0.04, 0.025, 0.015, 0.005$.}\label{tab: infP}
\begin{tabular}{|c|c|c|c|c|c|}
\hline
Prior knowledge $\lambda \geq$ & $0.05$ & $0.04$ & $0.025$ & $0.015$ & $0.005$\\
\hline
$\mae(\hat{\omega})$ $[10^{-3}]$ & $(6,5,3)$ & $(2,2,1)$ & $(2,2,1)$ & $(5,5,6)$ & $(159, 126, 186)$ \\
$\dist(\omega,\cC_{1-\alpha})$ $[10^{-3}]$ & $17$ & $23$ & $23$ & $37$ & $123$ \\
$\mean(\omega \in \cC_{1 - \alpha})$ $[\%]$ & $100$ & $100$ & $100$  & $100$ & $100$\\
$\overline{\omega}_i - \underline{\omega}_i$ $[10^{-3}]$ & $(24,25,25)$ & $(42,42,42)$ & $(42,42,42)$ & $(65,64,63)$ & $(183,171,144)$ \\
$\miae(\hat{f}^i)$ $[10^{-3}]$ & $(3,13,6)$ & $(1,4,2)$ & $(1,4,2)$ & $(1,23,11)$ & $(40,175,88)$ \\
$\meanNcp - K$ & $(0.1, 0.2, 0.0)$ & $(0.1, 0.1, 0.0)$ & $(0.1, 0.1, 0.0)$ & $(0.0, 0.3, -0.1)$ & $(2.4, 2.5, -0.2)$ \\
$\medNcp - K$ & $(0,0,0)$ & $(0,0,0)$ & $(0,0,0)$ & $(0,0,0)$ & $(0,-2,-2)$ \\
$\meanKK_i$ $[\%]$ & $(99,93,97)$ & $(100,98,99)$ & $(100, 98, 99)$ & $(99,87,93)$ & $(54,24,16)$ \\
$\meanKK$ $[\%]$ & $93$ & $98$ & $98$ & $86$ & $6$\\
$\max_i \min_j \abs{\tau_i - \hat{\tau}_j}$ $[10^{-1}]$& $(13, 148, 4)$ & $(6,40,2)$ & $(6,40,2)$ & $(7,299,9)$ & $(508, 1794, 122)$ \\
$\max_j \min_i \abs{\tau_i - \hat{\tau}_j}$ $[10^{-1}]$ & $(2, 41, 50)$ & $(1,11,15)$ & $(1,11,15)$ & $(1,45,91)$ & $(223, 331, 1343)$ \\
$\vm $ $[\%]$ & $(100, 99, 100)$ & $(100, 100, 100)$ & $(100, 100, 100)$ & $(100,98,99)$ & $(96,89,91)$\\
$\fpsle$ $[10^{-2}]$ & $(16,246,167)$ & $(8,67,51)$ & $(8, 67, 51)$ & $(5, 398, 304)$ & $(708, 1994, 4491)$ \\
$\fnsle$ $[10^{-2}]$ & $(34, 407, 41)$ & $(17, 113, 14)$ & $(17, 113, 14)$ & $(16, 785, 71)$ & $(1610, 5786, 1168)$ \\
 $\mean(f \in \tilde{\cH}(\beta))$ $[\%]$ & $96.01$ & $98.96$ & $98.95$ & $94.78$ & $56.65$  \\
\hline
\end{tabular}
\end{table}


\FloatBarrier


\subsection{Additional figures from Section \ref{sec:appgen}}

\quad \\

\begin{figure}[hb!]
\includegraphics[width= \textwidth]{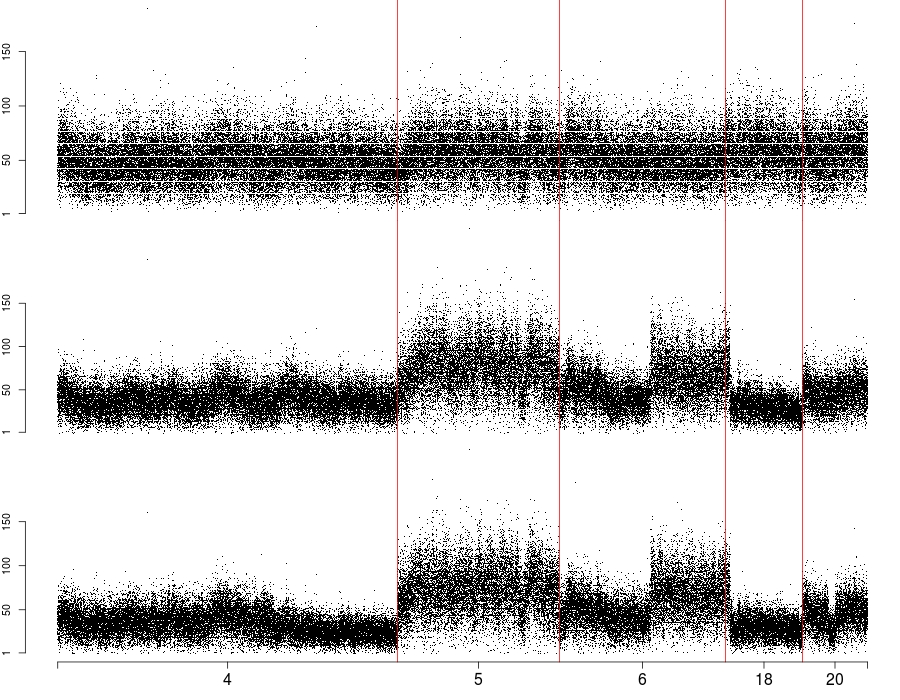}
\caption{Raw whole genome sequencing data from cell line LS411}\label{fig:raw_data}
\end{figure}

\begin{figure}[ht!]
\includegraphics[width= \textwidth]{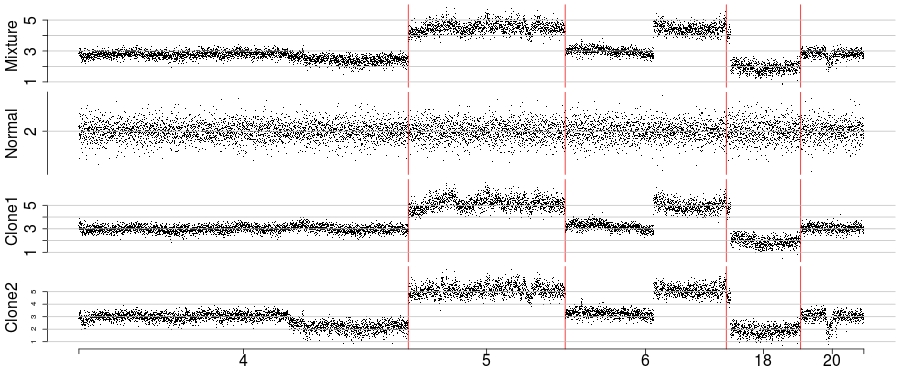}
\caption{Preprocessed whole genome sequencing data from cell line LS411}\label{fig:prepro-data}
\end{figure}

\begin{figure}[h!]
\includegraphics[width= \textwidth]{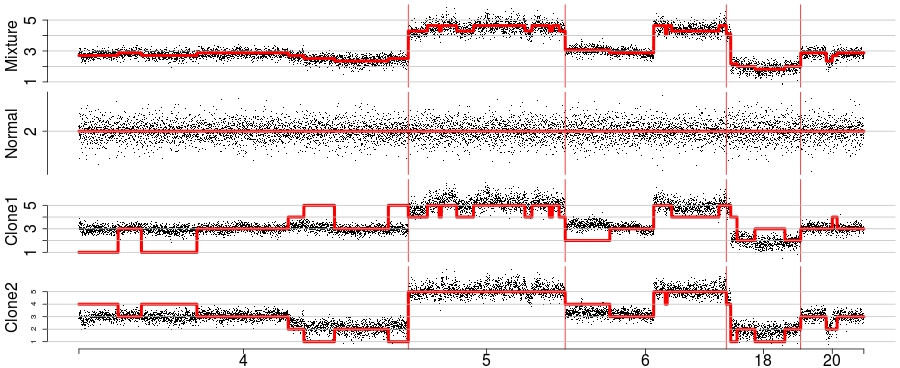}
\caption{
SLAM's estimates (red lines) for $q_n(\alpha) = -0.15$ (selected with MVT-method from Section \ref{subsec: chQ}) and $q_n(\beta) = 2.2$. 
Top row: total copy-number estimates across the genome. Rows 2-4: estimates of the CN profiles of the germline and clones. }\label{fig:est1}
\end{figure}

\section{Data driven selection of $q_n(\alpha)$} \label{sec: chqSupp}

In the following we give further details on the SST-method for selection of $q_n(\alpha)$ introduced in Section \ref{subsec: chQ}.
To simplify notation let $n$ be even. Then $Y^1 \ZuWeis (Y_1, Y_3,\ldots,Y_{n-1})$ and $Y^2 \ZuWeis (Y_2, Y_4,\ldots,Y_n) $ are both samples of size $n/2$ from the same underlying mixture $g$, with corresponding estimates $\hat{\omega}^1_q \ZuWeis \hat{\omega}(Y^1,q)$ and $\hat{\omega}^2_q \ZuWeis \hat{\omega}(Y^2,q)$, respectively.
Let $L$ be a loss function and $h(q)\ZuWeis \E\lbrack L(\hat{\omega}_q - \omega)\rbrack $ its corresponding performance measure  for estimating $\omega$, e.g., the MSE with $L = \norm{\cdot}^2_2$, which is to be minimized. 
As $\omega$ is unknown, $h(q)$ has to be estimated. This is done by
\begin{align*}
\hat{h}(q) \ZuWeis \frac{1}{2}\left(  L(\hat{\omega}_q - \hat{\omega}^1_q) + L(\hat{\omega}_q - \hat{\omega}^2_q)  \right)
\end{align*}
and we estimate the minimizing $q$ of $h$ as
\begin{align}\label{hatH}
\hat{q} \ZuWeis \argmin_{q \leq q_0} \hat{h}(q).
\end{align}
Bounding $q$ from above by $q_0$ is necessary as for $q \rightarrow \infty$, i.e. $\alpha \rightarrow 0$, the corresponding confidence region $\cC_{1-\alpha}$ converges to the entire domain $\Omega(m)$, hence $h(q) \rightarrow 0$ as $q \rightarrow \infty$. We found empirically that $q_0 \ZuWeis q_n(0.01)$ serves as a good bound (as statements with higher confidence as $0.99$ are rarely demanded), also to reduce computation time for the optimization of (\ref{hatH}).

\begin{figure}[h]
\includegraphics[width=0.7\textwidth]{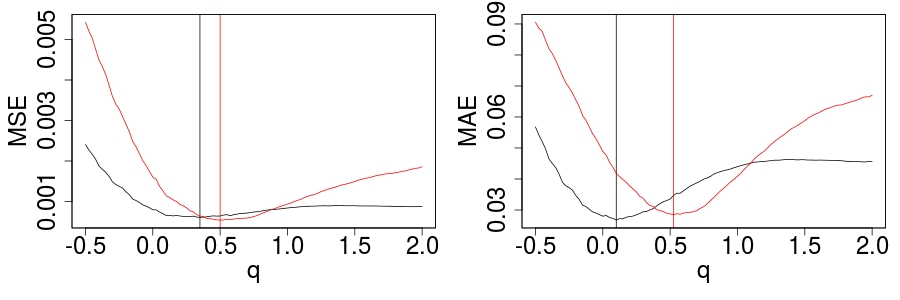}
\caption{Pointwise mean of $1,000$ replications of $\norm{\hat{\omega}_q - \omega}_2$, $\norm{\hat{\omega}_q - \omega}_1$, respectively (black) and of $\left( \norm{\hat{\omega}^1_q - \hat{\omega}_q }_2 + \norm{\hat{\omega}^2_q - \hat{\omega}_q}_2 \right)$, $\left( \norm{\hat{\omega}^1_q - \hat{\omega}_q }_1 + \norm{\hat{\omega}^2_q - \hat{\omega}_q}_1 \right)$, respectively (red) (from left to right), for the setting as in Example \ref{example1} with $n = 1280$ and $\sigma = 0.05$. The vertical lines indicate the corresponding minima.}\label{fig: chQmsemae}
\end{figure}

\begin{figure}[h]
\includegraphics[width=0.7\textwidth]{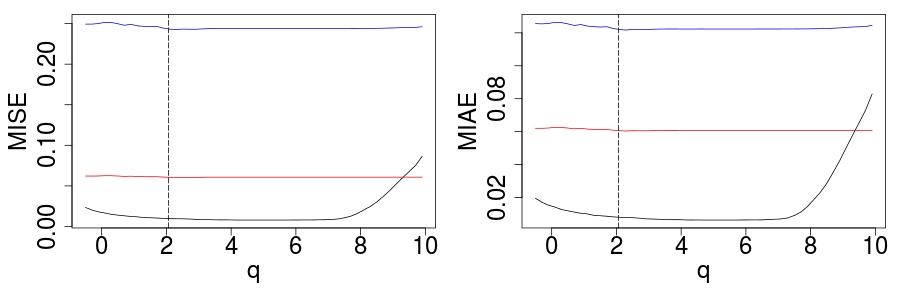}
\caption{MISE and MIAE of $\hat{f}^1_q$ (black), $\hat{f}^2_q$ (blue), and $\hat{f}^3_q$ (red) observed from $1,000$ realizations for the setting as in Example \ref{example1} with $n = 1280$ and $\sigma = 0.05$. The vertical dotted line indicates $q_n(0.01) = 2.07$. }\label{fig: chQmisemiae}
\end{figure}

The performance of the selector in (\ref{hatH}) is illustrated for the setting of Example \ref{example1} (with $n = 1280$ and $\sigma = 0.05$) in Figure \ref{fig: chQmsemae} for the MSE($q$) ($L = \norm{\cdot}_2^2$) and the MAE($q$) ($L = \norm{\cdot}_1$), respectively. 
From this we find that the optimal $q$ (the minimizer of the black line) is quite well approximated by its estimate $\hat{q} \approx 0.5$ (the minimizer of the red line). 
Simulations for different $n$ and $\sigma$ with $\sigma/\sqrt{n}$ in the order of our application example (see Section \ref{subsec: Appsim}) show the same. 
Recall from the previous Section \ref{subsec: stabAB} that $\omega$ is estimated quite stable for a range of $q$. In Figure \ref{fig: chQmsemae} $q \approx 0.5$ corresponds to $\alpha \approx 0.69$. The optimal $q$ for the MSE is $q \approx 0.35$, corresponding to $\alpha \approx 0.81$ and for the MAE $q \approx 0.1$, corresponding to $\alpha \approx 0.95$. 

For large noise levels, however, we found that the SST-selection method is outperformed by the MVT-method from Section \ref{subsec: chQ} illustrated for the setting of Example \ref{example1} with $n = 1280$ and $\sigma = 0.05, 0.08, 0.1, 0.2$ in Table \ref{tab: chQ}.

\begin{table}[ht!]
\caption{MSE and MAE for the SST-method and the MVT-method for the setting of Example \ref{example1} with $n = 1280$ and $\sigma = 0.05, 0.08, 0.1, 0.2$ obtained from $2,000$ replications.}\label{tab: chQ}
\begin{tabular}{|c|c|c|c|c|}
\hline
&\multicolumn{2}{|c}{MSE $[10^{-4}]$} & \multicolumn{2}{c|}{MAE $[10^{-3}]$}\\
\hline
& SST & MVT & SST & MVT \\
\hline
$\sigma = 0.05$ & $4$ & $4$ & $27$ & $18$\\
$\sigma = 0.08$ & $26$ & $34$ & $73$ & $81$ \\
$\sigma = 0.1$ & $56$ & $30$ & $110$ & $78$\\
$\sigma = 0.2$ & $166$ & $44$ & $206$ & $95$\\
\hline
\end{tabular}
\end{table}

\FloatBarrier

\end{document}